\documentclass[11pt,a4paper]{article}

\usepackage{jheppub}
\usepackage{graphicx}				
\usepackage[sort&compress]{natbib}
\usepackage{url}
\usepackage{hyperref}
\usepackage{amsfonts}
\usepackage{epsfig}
\usepackage{dcolumn}
\usepackage{bm}
\usepackage{slashed}
\usepackage{color}

\usepackage{amssymb}
\usepackage[caption=false]{subfig}
\usepackage{tikz}

\newcommand{\softsusy}[0]{{{\tt SOFTSUSY}}}
\newcommand{\spheno}[0]{{\tt SPheno}}
\newcommand{\feynhiggs}[0]{{\tt FeynHiggs}}
\newcommand{\susyhd}[0]{{\tt SusyHD}}
\newcommand{\micromegas}[0]{{\tt micrOMEGAs}}
\newcommand{\darksusy}[0]{{\tt DarkSUSY}}

\newcommand{\be}{\begin{equation}}
\newcommand{\ee}{\end{equation}}
\newcommand{\bea}{\begin{eqnarray}}
\newcommand{\eea}{\end{eqnarray}}
\newcommand{\mnt}[1]   {m_{\tilde\chi^0_{#1}}}

\newcommand{\neu}[1]{\ensuremath{\tilde{\chi}_{#1}^0}}

\newcommand{\gsim}{\lower.7ex\hbox{$\;\stackrel{\textstyle>}{\sim}\;$}}
\newcommand{\lsim}{\lower.7ex\hbox{$\;\stackrel{\textstyle<}{\sim}\;$}}

\title{{\color{blue} Theoretical Uncertainties in the Calculation of Supersymmetric Dark Matter Observables}}
\date{\today}
\author{Paul Bergeron$^1$, Pearl Sandick$^1$, Kuver Sinha$^2$}
\affiliation{$^1$Department of Physics \& Astronomy,  University of Utah,  Salt Lake City, UT 84112, USA\\
$^2$Dept. of Physics and Astronomy, 
University of Oklahoma, Norman, OK 73019, USA}
\date{\today}

\abstract{We estimate the current theoretical uncertainty in supersymmetric dark matter predictions by comparing several state-of-the-art calculations within the minimal supersymmetric standard model (MSSM). We consider standard neutralino dark matter scenarios -- coannihilation, well-tempering, pseudoscalar resonance -- and benchmark models both in the pMSSM framework and in frameworks with Grand Unified Theory (GUT)-scale unification of supersymmetric mass parameters.  The pipelines we consider are constructed from the publicly available software packages \softsusy{}, \spheno{}, \feynhiggs{}, \susyhd{}, \micromegas{}, and \darksusy{}. We find that the theoretical uncertainty in the relic density as calculated by different pipelines, in general, far exceeds the statistical errors reported by the Planck collaboration. In GUT models, in particular, the relative discrepancies in the results reported by different pipelines can be as much as a few orders of magnitude. We find that these discrepancies are especially pronounced for for cases where the dark matter physics relies critically on calculations related to electroweak symmetry breaking, which we investigate in detail, and for coannihilation models, where there is heightened sensitivity to the sparticle spectrum. The dark matter annihilation cross section today and the scattering cross section with nuclei also suffer appreciable theoretical uncertainties, which, as experiments reach the relevant sensitivities, could lead to uncertainty in conclusions regarding the viability or exclusion of particular models.}

\begin{document}
\maketitle

\section{Introduction}

The search for supersymmetry and its connection to dark matter physics have been prominent areas of research in particle phenomenology, both theoretical and experimental, over the last few decades. Experimental results have provided significant constraints on the Minimal Supersymmetric Standard Model (MSSM) via the discovery of the Higgs boson~\cite{Aad:2012tfa,Chatrchyan:2012xdj}, as well as via null results from collider searches for new particles~\cite{Sirunyan:2017kqq, ATLAS:2017msx} and dark matter direct and indirect detection experiments. The LHC in particular has pushed the limits on squark masses to roughly the TeV range, ruling out much of the constrained MSSM (CMSSM)~(eg.~\cite{Olive:2016xmw}). 
Nevertheless, within the full MSSM as well as more minimal frameworks, much parameter space remains in which the thermal relic abundance of the lightest neutralino explains the astrophysical cold dark matter and a Higgs boson consistent with that discovered at the LHC is predicted~\cite{Baer:2011ab,Feng:2011aa,Heinemeyer:2011aa,Arbey:2011ab,Draper:2011aa,Buchmueller:2011ab,Kadastik:2011aa,Strege:2011pk,Aparicio:2012iw,Ellis:2012aa,Baer:2012uya,Bechtle:2012zk,Balazs:2013qva,Ghosh:2012dh,Fowlie:2012im,Buchmueller:2012hv,Kowalska:2012gs,Strege:2012bt,Cabrera:2012vu,Draper:2013cka,Cohen:2013kna,Henrot-Versille:2013yma,Bechtle:2013mda,Buchmueller:2013psa,Buchmueller:2013rsa,Roszkowski:2014wqa,Buchmueller:2015uqa,Bagnaschi:2015eha}. 
Furthermore, it can be argued that, despite not yet having discovered any new supersymmetric (SUSY) partners, the verdict is still out on (even weak-scale) supersymmetry (eg.~\cite{Dutta:2017nqv} and \cite{Baer:2017uvn}).

As the mechanism of supersymmetry breaking is far from clear, models may be defined either in the UV, near the so-called Grand Unification (GUT) scale, or in the IR, for example within the well-studied phenomenological MSSM (pMSSM) framework.  In either case, one would like to calculate the superpartner spectrum at the weak scale and the masses and couplings in the Higgs sector.  Finally, once the Lagrangian is known at the weak scale, it can be used to calculate the dark matter observables, including the relic density, the current annihilation cross section, and the scattering rates, all of which can be compared with experimental results.

A plethora of public software packages have been developed to facilitate the analysis of various SUSY models and compare their predictions with experimental results~(for example,~\cite{Paige:2003mg,Allanach:2001kg,Porod:2003um,Djouadi:2002ze,Hahn:2013ria,Vega:2015fna,Belanger:2014vza,Gondolo:2004sc}). 
Our confidence in the accuracy of the calculations done by any package is based on the continual improvements made by its authors and on the agreement of results from different packages.
Comparative studies of spectrum generators and Higgs sector calculators
have been undertaken before (eg.~\cite{Allanach:2002pz,Allanach:2003jw,Belanger:2005jk,Allanach:2004rh,Bagnaschi:2014rsa,Draper:2016pys}).  Differences in the renormalization group running and predictions for sparticle masses within the same supersymmetric model have been observed~\cite{Allanach:2002pz,Allanach:2003jw,Belanger:2005jk}, and sensitivities of the Higgs sector have also been explored~\cite{Allanach:2004rh,Bagnaschi:2014rsa,Draper:2016pys}. However, previous studies have focused primarily on models with relatively light ($\mathcal{O}(100)$ GeV) sparticles, and the most recent comparison of the full sparticle spectrum was undertaken more than a decade ago~\cite{Belanger:2005jk}. 

There are several publicly-available software packages that calculate quantities that can be observed at dark matter direct and indirect detection experiments as well as the relic abundance of dark matter within a particular model.   \micromegas{}~\cite{Belanger:2014vza} and \darksusy{}~\cite{Gondolo:2004sc} are two examples.
Studies have been carried out regarding the accuracy with which a single observable is calculated by an individual software package~\cite{Allanach:2004xn,Catena:2015uha,Harz:2015qva}, though there are relatively few studies that compare the calculation of dark matter observables by different software packages (eg.~\cite{Gomez:2002tj}), and none of which we are aware that address LHC-era supersymmetric benchmarks.

In this report, we embark on a comparison study designed with three advancements over previous studies: First, we compare calculations for the sparticle spectrum, the Higgs sector, and the dark matter observables, discussing, when possible, differences in the implementations of the underlying physics of the calculations in each case.  Second, we incorporate the various calculators into comprehensive pipelines to study not only the effects of the choice of an individual calculator, but also all downstream effects of those choices on subsequent calculations.  Finally, we analyze the above choices and observables in the context of several SUSY benchmark models chosen as representative of models that are interesting in the light of LHC Run-1 and null results from recent dark matter searches as described below.

This study is conducted in two parts: To begin, we investigate a set of pMSSM models from Ref.~\cite{Cahill-Rowley:2013gca}. The dark matter scenarios we consider are coannihilation (bino-stop and bino-squark), $A$-funnel, well-tempered neutralinos, and pure higgsinos. We will see that the spectrum generators can differ by up to 1 - 2 \% in their predicted masses for the stop and the first two generations of squarks, and by up to 20\% in the gauge composition of the lightest neutralino, for a given pMSSM model. As for the dark matter observables, differences of up to a factor of $\sim 3 - 5$ in the relic density and current annihilation cross section, and up to a factor of $\sim 10$ in the predicted scattering cross section are possible for the different pipelines.  The  theoretical uncertainty in the relic density of neutralino dark matter already far exceeds the statistical errors reported by the Planck collaboration, while the dark matter annihilation cross section today and the scattering cross section with nuclei also suffer appreciable theoretical uncertainties,  which,  as experiments reach the relevant sensitivities, could lead to uncertainty in conclusions regarding the viability or exclusion of particular models.

In the second part of our study, we consider four benchmark models defined at the GUT scale -- two CMSSM points and two points from models with non-universal Higgs masses (NUHM)~\cite{Buchmueller:2013rsa, Cohen:2013kna}.  For GUT-scale models, we will find that discrepancies among the various pipelines are often amplified by the renormalization group running.  For our CMSSM and NUHM benchmarks, we will see that the spectrum generators can give low energy values of the higgsino mass parameter $\mu$ and the pseudoscalar Higgs mass $m_A$ that differ by up to 150\% - 200\% (though the differences can be much greater at larger $m_0$ than the values condidered here). This leads to dramatic differences in the annihilation and scattering cross sections computed by the dark matter calculators.

 Before proceeding, we would like to reflect on whether a study like this is anachronistic at this juncture. With the LHC failing to find new physics yet, and supersymmetric WIMP searches yielding null results, one might ask whether it makes sense to go back to benchmark SUSY scenarios yet again. We remind the reader that the connection of supersymmetry to dark matter physics, while robust from a high level perspective due to the WIMP miracle, was always fragile at the model-building level, at least under the assumption of a standard cosmological history. The dark matter relic density is often obtained in fine-tuned regions of parameter space, which either exhibit compressed spectra, or have suppressed  interactions with nuclei. Many of these scenarios are difficult to probe at colliders or direct detection experiments, and also have small annihilation rates in the current Universe. Moreover, their fine-tuned nature means that detailed predictions for physical quantities in these scenarios are particularly sensitive to the approximations used. It is not a surprise that theoretical uncertainties in these scenarios can substantially outweigh experimental uncertainties.  Given the current precision of the measurement of the dark matter abundance and the dramatically improving sensitivities to dark matter-nucleon scattering in the era of ton-scale experiments, it could be argued that it is more important than ever to examine the precision and accuracy of the predicted values for observable quantities in supersymmetric models.  We also note that our findings may be relevant for some more general (non-supersymmetric) models of dark matter, so long as they share particular characteristics with the benchmarks considered here, for example, models in which the relic abundance is achieved via resonant annihilations.

The paper is organized as follows. 
In Section~\ref{sec:methodology} we present the benchmark MSSM points considered here and discuss the calculators and pipelines we study. 
In Section~\ref{sec:DMphysics} we discuss relevant aspects of the physics of neutralino dark matter.
In Sections~\ref{sec:pMSSM} and~\ref{sec:GUT} we present our results for the pMSSM and GUT-scale benchmarks, respectively.  
Finally, in Section~\ref{sec:conclusions} we summarize the conclusions of our study.  Numerical results for all benchmarks are compiled in Appendices~\ref{app:pmssm} and~\ref{app:msugra}.

\section{Methodology of the Comparison}
\label{sec:methodology}

In this section, we present our benchmark points, the calculator pipelines we study, and the details of the calculations undertaken by each of the calculators.

\subsection{Benchmark Points} %
\label{sec:benchmarks}

We consider two sets of supersymmetric benchmark points, all of which assume that the lightest supersymmetric particle (LSP) is the lightest neutralino, which is therefore the dark matter candidate.  For future reference, we describe our notation here. We choose the bino -- wino -- higgsino basis to write the neutralino mass matrix as
\begin{equation}
  {\cal M}_N =
  \left( \begin{array}{cccc}
  M_1 & 0 & -m_Z s_W c_\beta  & m_Z s_W s_\beta \\
  0 & M_2 &  m_Z c_W c_\beta  & -m_Z c_W s_\beta  \\
  -m_Z s_W c_\beta & m_Z c_W c_\beta   & 0 & -\mu \\
   m_Z s_W s_\beta & - m_Z c_W s_\beta & -\mu & 0
  \end{array}\right),
\label{neutmassmatrix}
\end{equation}
where we follow the standard notation: $s_W=\sin\theta_W$, $c_W=\cos\theta_W$, $s_\beta=\sin\beta$,
$c_\beta=\cos\beta$ and $\tan\beta = v_2/v_1$, with $v_{1,2}$ being the vacuum expectation values of the two Higgs
fields $H_{1,2}$.

The mass matrix can be diagonalized by a unitary mixing matrix $N$,
\begin{equation}
  N^*{{\cal M}_N} N^\dagger =
  {\rm diag}(\mnt{1},\,\mnt{2},\,\mnt{3},\,\mnt{4}),
\end{equation}
where the eigenvalues are the neutralino masses. The lightest neutralino mass eigenstate can be written as
\be
  \tilde{\chi}_1^0= N_{11}\tilde{B}+ N_{12} \tilde{W} +N_{13}\tilde{H_1}+
  N_{14}\tilde{H_2} .
  \label{eq:neuteigenstate}
\ee
As the composition of the neutralino LSP determines much of the dark matter physics, we will often refer to the bino fraction, $|N_{11}|^2$, and the higgsino fraction, $|N_{13}|^2+|N_{14}|^2$. 

The first set of benchmark points consists of 5 pMSSM points from the Snowmass 2013 white paper~\cite{Cahill-Rowley:2013gca}. These points are representative of the pMSSM landscape and of the primary mechanisms by which the correct relic density of neutralino dark matter is achieved: sfermion coannihilation, rapid annihilation via a pseudo-scalar Higgs resonance, pure higgsino content, and the so-called well-tempered neutralino.  We will discuss each of these in Section~\ref{sec:DMphysics}. The spectrum for each point can be found at~\cite{pMSSMspectra}. 

The second set consists of 4 points of interest defined at the GUT scale.  The defining parameters of these 4 points may be found in Table~\ref{tab:gutpoints}. Three of these points are based on the MasterCode Collaboration's post-LHC Run I best fit points from the CMSSM and NUHM~1 (hereafter, ``NUHM~1'' one will simply be referred to as ``NUHM'')~\cite{Buchmueller:2013rsa}. The MasterCode analysis also includes constraints from dark matter direct detection experiments and the observed dark matter abundance.  The MasterCode CMSSM best fit point will be denoted CMSSM 1.  The final CMSSM point, denoted CMSSM 2, is inspired by the stau coannihilation benchmark point from~\cite{Cohen:2013kna}.

Both the MasterCode best fit NUHM point and the CMSSM 2 point are in  coannihilation regions of parameter space, the former by virtue of having nearly pure higgsino dark matter, and are therefore extremely sensitive to variations in the RGE running.  Significant variations in the running can occur between different spectrum calculators as well as between different versions of the same spectrum calculator.  For example, a point that yields the correct dark matter abundance via stau coannihilation may end up with a stau LSP if a different calculator or version is employed.
Since the publication of~\cite{Buchmueller:2013rsa} and~\cite{Cohen:2013kna}, there have been several updates to \softsusy{}, which was used to calculate the sparticle spectrum in both studies.  
Here, we consider two NUHM points inspired by the best fit point in~\cite{Buchmueller:2013rsa}, denoted ``NUHM~A'' and ``NUHM~B'', chosen with the requirement that a valid relic density would be achieved by NUHM~A  via our \spheno{} pipelines and NUHM~B via our \softsusy{} pipelines. The original MasterCode NUHM point is included in Table~\ref{tab:gutpoints} for reference.
Furthermore, the original stau coannihilation benchmark from~\cite{Cohen:2013kna}, calculated with \softsusy{} {\tt 3.3.7}, yields a stau LSP in the more contemporary version \softsusy{} {\tt 3.7.3}. As such, we consider a similar point where $m_0$ has been increased by about $20$~GeV over the original value to avoid a stau LSP.  

\begin{table}
  \centering
   \includegraphics[scale=0.9]{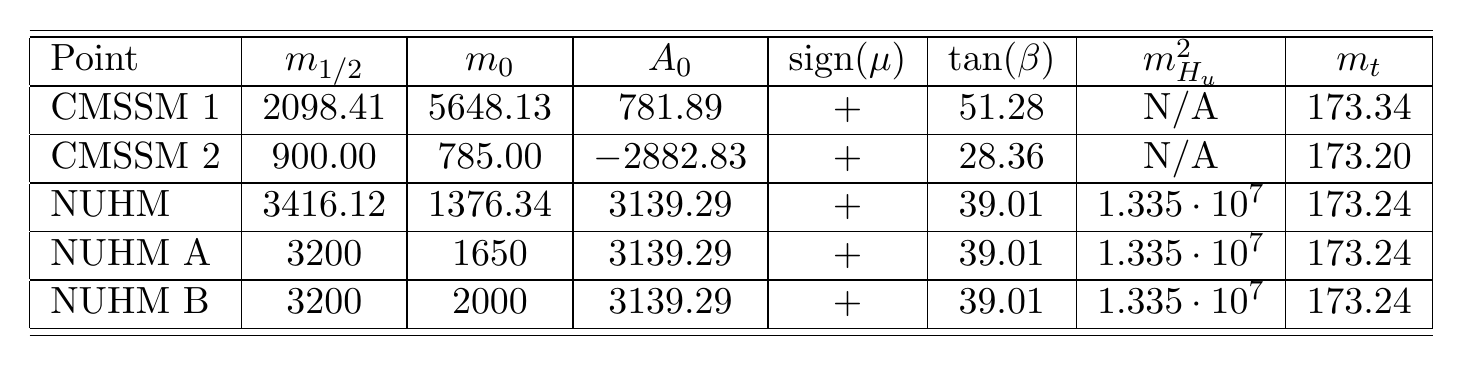}
  \caption{Parameters defining the CMSSM and NUHM benchmarks considered in Section~\ref{sec:GUT}. }
 \label{tab:gutpoints}
\end{table}

\subsection{Pipeline Structure and Nomenclature} \label{sec:pipeandnames}

The pipelines considered here are comprised of a selection of the many publicly-available calculators on the market for studying supersymmetry and dark matter physics. We will refer to the different calculators considered here in terms of their primary functions: 
\begin{itemize}
\item {\it mass spectrum generators}, 
\softsusy{}~\cite{Allanach:2001kg} and \spheno{}~\cite{Porod:2003um,Porod:2011nf}; 
\item {\it Higgs sector calculators}, \feynhiggs{}~\cite{Hahn:2013ria,Frank:2006yh,Degrassi:2002fi,Heinemeyer:1998np,Heinemeyer:1998yj} and \susyhd{}~\cite{Vega:2015fna}; and 
\item {\it dark matter observable calculators}, \micromegas{}~\cite{Belanger:2001fz,Belanger:2004yn,Belanger:2010gh,Belanger:2014vza} and \darksusy{}~\cite{Gondolo:2004sc,Gondolo:1990dk,Edsjo:1997bg,Edsjo:2003us,Bergstrom:1995cz}.
\end{itemize}

\begin{figure}
  \centering
  \includegraphics[scale=0.9]{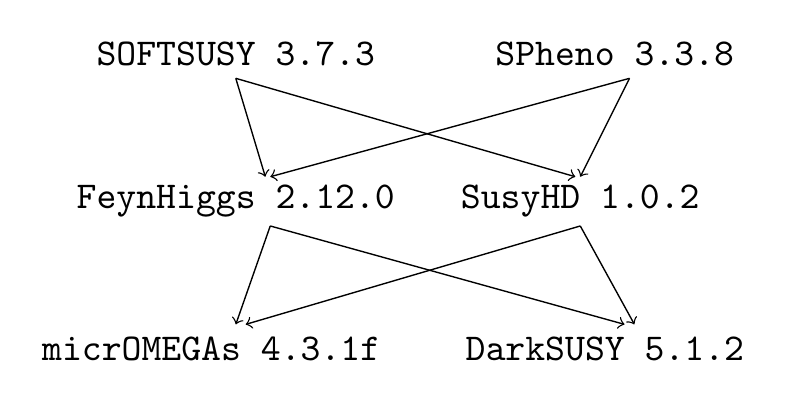}
 \caption{Depiction of the 8 pipelines used in this study. From top to bottom, we show the supersymmetric sparticle mass spectrum generators, the Higgs sector calculators, and the programs that calculate the dark matter observables.}
 \label{fig:flow}
\end{figure}

\begin{table}
  \centering
    \includegraphics[scale=0.9]{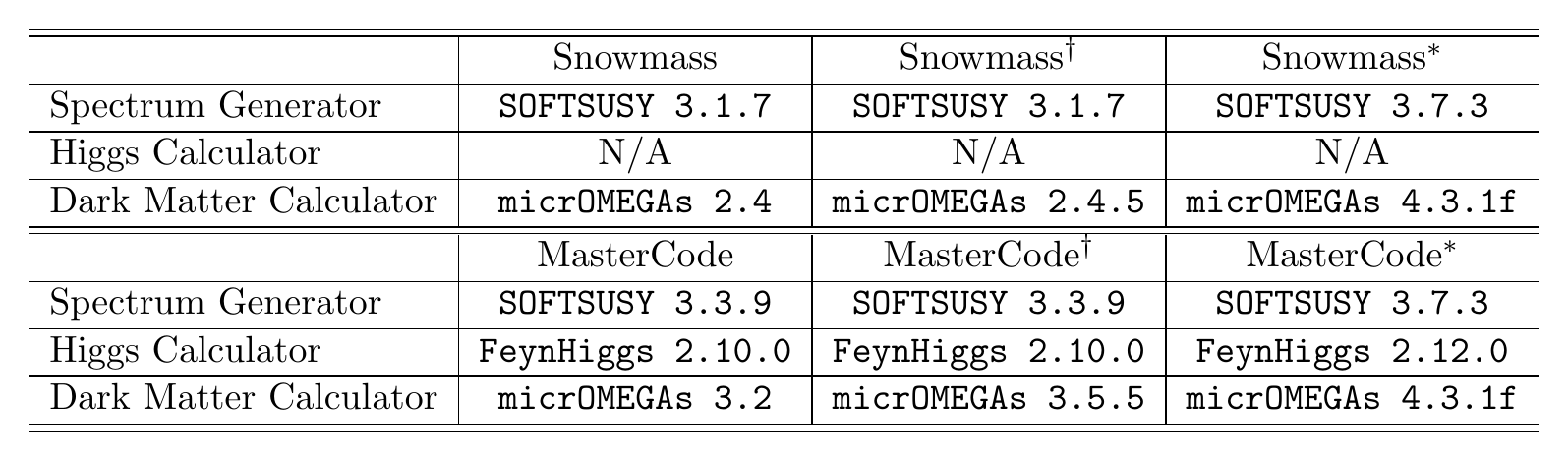}
  \caption{Summary of the calculators used in the Snowmass-type and MasterCode-type pipelines used herein. Unadorned names refer to the original work whose results are quoted in our analysis. Daggers ($\dag$) denote our reproductions of original results with our implementations of the packages; these are the same versions as the original with the exception of \micromegas{} {\tt 2.4.5} for Snowmass$^\dag$. Asterisks ($*$) denote our updated versions of the pipelines using the same contemporary versions as our pipelines.}
  \label{tab:pipelines}
\end{table}

As demonstrated in Figure~\ref{fig:flow}, each pipeline is composed of 3 calculators, one of each type -- spectrum, Higgs, and dark matter. In this way, we consider 8 different pipelines, such as \softsusy{}-\feynhiggs{}-\micromegas{} or \softsusy{}-\susyhd{}-\darksusy{}. The inclusion of the Higgs calculator is to ensure that details for the Higgs sector are achieved before computing dark matter observables. Files in SLHA~\cite{Allanach:2008qq} format are used to pass information between each calculator, with the input and output being retained at each stage. We note that two separate input files are necessary for \softsusy{} and \spheno{}, as there are minor differences in the expected format of the SLHA input files for the two calculators.

An important caveat in the passage of information between the programs is the handling of the branching ratios. While SLHA formatted files do include blocks for detailing particle decays, they are not universally utilized by all spectrum calculators.  For example, \spheno{} does write the decay blocks for its SLHA output files, while \softsusy{} does not.  For \softsusy{} pipelines, if \feynhiggs{} is used, Higgs decay widths will be written, but if \susyhd{} is used, since it only calculates the $CP$-even Higgs mass, no Higgs decay widths will be recorded.  This means that there are no recorded widths in the \softsusy{}-\susyhd{} pipelines, which can lead to discrepancies in the calculation of the dark matter abundance, for example, if dark matter annihilates primarily via the psuedoscalar resonance.

For our analysis, the versions of the calculators implemented (unless otherwise noted) are \softsusy{} {\tt 3.7.3}, \spheno{} {\tt 3.3.8}, \feynhiggs{} {\tt 2.12.0}, \susyhd{} {\tt 1.0.2}, \micromegas{} {\tt 4.3.1f}, and \darksusy{} {\tt 5.1.2}. Since all of the calculators studied here are continuously updated and improved, specifically since the publication of~\cite{Cahill-Rowley:2013gca,Buchmueller:2013rsa,Cohen:2013kna}, we also include the versions of the pipelines used in the Snowmass and MasterCode studies for proper comparisons with their results, as summarized in Table~\ref{tab:pipelines}. The Snowmass pipeline uses \softsusy{} {\tt 3.1.7} and \micromegas{} {\tt 2.4} and is denoted as ``Snowmass'', and the updated pipeline (still without \feynhiggs{}) is denoted as ``Snowmass$^*$,'' i.e. \softsusy{} {\tt 3.7.3} and \micromegas{} {\tt 4.3.1f}. Alternatively, the MasterCode pipeline utilized\footnote{In addition to \micromegas{}, MasterCode's calculation of the relic density is verified by the private code {\tt SSARD}~\cite{SSARD, Bagnaschi:2016xfg}, which is also used to calculate the SI scattering cross sections~\cite{Buchmueller:2013rsa}.}
 \softsusy{} {\tt 3.3.9}, \feynhiggs{}  {\tt 2.10.0}, and \micromegas{} {\tt 3.2}. Since the updated MasterCode pipeline (MasterCode$^*$) is identical to that of our \softsusy{}-\feynhiggs{}-\micromegas{} pipeline we do not denote it separately from here forward.

Furthermore we denote pipelines with a dagger (Snowmass$^\dag$/MasterCode$^\dag$) to indicate when \emph{we have reproduced} the calculation of the original pipeline; otherwise the result is quoted as published. That said, not all versions of \micromegas{} are currently available, in which case we use the closest available version.

As will be discussed below, there can be substantial variation in results and calculational techniques between different versions of the same software package. Indeed, the version numbers are critical to the interpretation of the results presented here. In the remainder of this paper, however, for the sake of brevity, we will suppress the version numbers for the packages that compose the pipelines unless otherwise specified, and refer the reader to Fig.~\ref{fig:flow} and Table~\ref{tab:pipelines}.

\subsection{Details of the Calculations}

Here we discuss the details of the calculations performed by each software package. In particular we focus on the contrasting choices underlying the differences between packages, taking each tier of the pipeline in turn. Unless otherwise specified, we take the default settings for each calculator throughout the following analysis.

\subsubsection{Spectrum Calculators}

For the evaluation of the sparticle mass spectrum, we consider \softsusy{} and \spheno{}. First, they evaluate the gauge and Yukawa couplings at the electroweak scale before running them to the high scale and applying the soft SUSY breaking boundary conditions. After running back down to the electroweak scale, $m_Z$, the initial tree-level values of the sparticle and Higgs masses are calculated. These masses are used as input for the iterative loop that comprises the calculation. In the iterative step, the current mass spectrum is evolved to a high scale $M_x$, defined for GUT models to be the scale at which $g_1(M_x) = g_2(M_x)$ and defined for the pMSSM to be some low scale near $M_{\text{SUSY}} = \sqrt{m_{\tilde{t}_1} m_{\tilde{t}_2}}$, where the soft SUSY breaking parameters are set from the specified boundary conditions. 
GUT models are then evolved down to $M_{\text{SUSY}}$, where the two cases proceed in the same manner. At $M_{\text{SUSY}}$, electroweak boundary conditions are applied and the sparticle and Higgs pole masses are evaluated at the loop level.
These are now input for the next iteration of the loop -- beginning with a new, and more accurate, calculation of the gauge and Yukawa couplings. When a stable solution of a given accuracy is reached, the iteration terminates and the spectrum is run down to $m_Z$.

\begin{table}
  \centering
  \includegraphics[scale=0.9]{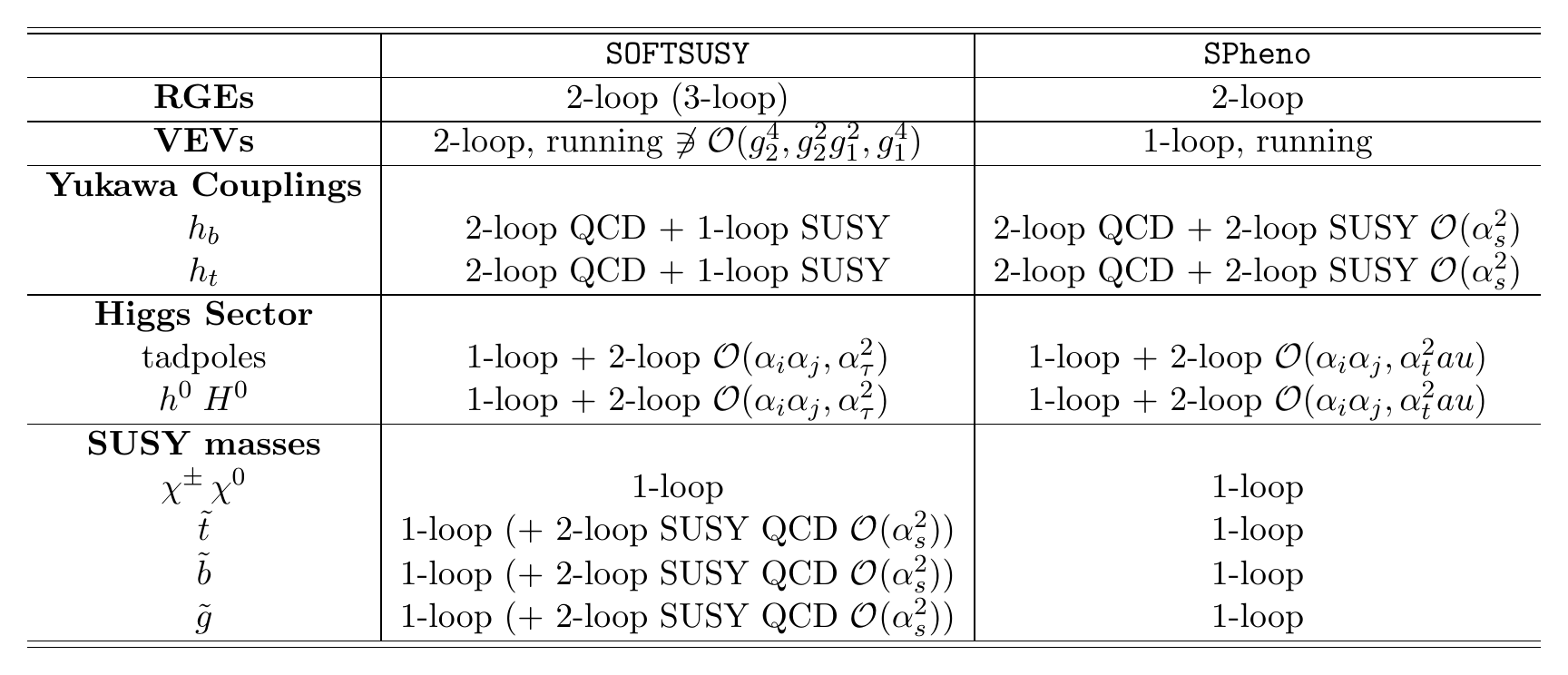}
  \caption{Orders of RGE and radiative corrections employed by \softsusy{} and \spheno{}. The common terms between both \softsusy{} and \spheno{} for the Higgs sector, denoted $\alpha_i \alpha_j$, are the members of the set $\{\alpha_t \alpha_s, \alpha_t^2, \alpha_t \alpha_b, \alpha_b^2, \alpha_b \alpha_s, \alpha_b \alpha_{\tau}\}$. \softsusy{}'s optional modes are detailed in parentheses: a ``high order mode'' for 2-loop radiative corrections to the squark and gluino pole masses~\cite{Allanach:2016rxd} and a ``high accuracy mode'' for 3-loop RGEs (requires {\tt CLN} and {\tt GiNaC} interfaces)~\cite{Allanach:2014nba}. The default modes are used in our analysis.}
  \label{tab:specdetails}
\end{table}

The programs, however, do differ in the details of their calculations, as summarized in Table~\ref{tab:specdetails} (see also Table~1 from Reference~\cite{Allanach:2003jw}). It is worth reminding that, even when the quoted loop level is the same, the scheme in which the calculation is handled can lead to important differences. This manifests in the choice between $\overline{MS}$ and $\overline{DR}$ schemes, where the latter amounts to a higher order correction to the former. While \softsusy{} and \spheno{} both employ running $\overline{DR}$ masses in their calculations, their methods of calculating the $\overline{DR}$ corrections are different. 

This important difference enters in determination of the Yukawa couplings, which are calculated from the quark masses at scale $Q$:
\begin{equation}
  h_t(Q) = \frac{m_t(Q)}{v_2}\sqrt{2},  \quad   h_{b,\tau}(Q) =  \frac{m_{b,\tau}(Q)}{v_1}\sqrt{2}.
\end{equation}
In running to the high scale, the quark masses must be shifted from  $\overline{MS}$ to $\overline{DR}$, and both programs ultimately follow Reference~\cite{Pierce:1996zz} for the 2-loop QCD corrections and Reference~\cite{Avdeev:1997sz} for the 1-loop SUSY contributions. For the bottom mass the $\overline{DR}$ value is arrived at in both \softsusy{} and \spheno{} by
\begin{equation}
  m_b(m_Z)^{\overline{DR}}_{\text{SM}} = m_b(m_Z)^{\overline{MS}}_{\text{SM}}\left(1 - \frac{\alpha_s}{3\pi} - \frac{23\alpha_s^2}{72\pi^2} + \frac{3g_2^2}{128\pi^2} - \frac{13g_1^2}{1152\pi^2} \right),
\end{equation}
and then resummed with the SUSY corrections via
\begin{equation}
  m_b(m_Z)^{\overline{DR}}_{\text{MSSM}} = \frac{m_b(m_Z)^{\overline{DR}}_{\text{SM}}}{1 - \Delta^b_{\text{SUSY}}(m_Z)}.
\end{equation}
For the top mass, \softsusy{} employs a similar correction, with the 2-loop QCD corrections as
\begin{equation}
  m_t(m_Z)^{\overline{DR}}_{\text{SM}} = m_t(m_Z)^{\overline{MS}}_{\text{SM}}\left[ 1 - \frac{\alpha_s}{3\pi}(5 - 3L) - 
                                                                                                          \alpha_s^2 \left( 0.538 - \frac{43}{24\pi^2}L + \frac{3}{8\pi^2}L^2 \right) \right],
\end{equation}
where $L=\ln(m_t^2(m_Z)/m_Z^2)$ for the top mass. But \spheno{} uses a modified $\alpha_s^2$ term according to the large quark mass expansion in~\cite{Bednyakov:2002sf}, which results in an $\alpha_s^2$ coefficient of
\begin{equation*}
  - \left( \frac{8}{9} + \frac{2011}{18\pi^2} + \frac{16}{9}\ln(2) - \frac{8\zeta(3)}{3\pi^2} + \frac{246}{3\pi^2}L + \frac{22}{\pi^2}L^2 \right).
\end{equation*}

\subsubsection{Higgs Calculators}

The two Higgs mass calculators studied here are \feynhiggs{} and \susyhd{}.
Prior \feynhiggs{} {\tt 2.11.3}, \feynhiggs{} consistently yielded SM Higgs masses $\sim2-4$~GeV above the value yielded by \susyhd{}~\cite{Vega:2015fna}. We found that the differences between the two Higgs sector calculators were enough to present small but noticeable differences in the dark matter observables, particularly for $A$-funnel points. However, as of \feynhiggs{} {\tt 2.12.0}, the differences in the results from \feynhiggs{} and \susyhd{} are far better understood, and it is possible to choose flags in \feynhiggs{} such that the numerical discrepancies are dramatically reduced.  \footnote{We note that the hybrid approach employed by \feynhiggs{} has also been analytically compared to results from pure effective field theory, as employed by \susyhd{}, in \cite{Bahl:2017aev}, which sheds light on the differences between the two approaches.}. Most notable are two changes that yielded large shifts~\cite{Bahl:2016brp,FHchangelog}. The first change is the inclusion of next-to-next-to-leading order (NNLO) $m_t^{\overline{MS}}$, which induces a downward shift in $m_h$ by as much as $\sim2$~GeV relative to when the NLO $m_t^{\overline{MS}}$ is used. The second change is the inclusion of electroweak contributions in evaluating $m_t^{\overline{MS}}$, accounting for a downward shift of about $1$~GeV. 

For the versions employed in our primary pipelines (displayed in Figure~\ref{fig:flow}), \feynhiggs{} and \susyhd{} are in close agreement when \feynhiggs{} includes the resummation of large logs at the 2-loop level; note that this is not the default mode of the calculation, but it is used in this study.  For this reason, we will focus on pipelines that include \feynhiggs{} in the remainder of our analysis, though results from the \susyhd{} pipelines are included in all tables in the Appendix.  We stress that shifts in the value for $m_h$ are significant not just for the calculation of the dark matter observables, but also because they introduce important caveats to previous analyses where the Higgs mass is germane.

\subsubsection{Dark Matter Calculators}

The two dark matter calculators we consider are \darksusy{} and \micromegas{}. Both are sophisticated programs that analyze dark matter observables and relevant collider observables (eg.~$b \rightarrow s\gamma$). We confine our interest to three astrophysical observables:  the neutralino relic density, $\Omega h^2$; the annihilation rate today, $\langle \sigma v \rangle$; and the spin-independent (SI) scattering cross section with nuclei, $\sigma^{\text{SI}}$.

The relic density is calculated by \micromegas{} according to the relation
\begin{equation}
   \Omega h^2\vert_{\text{MCO}} = 2.742 \cdot 10^8 \frac{m_{\tilde{\chi}_{1}^{0}}}{1\,\text{GeV}} Y_0 \;,
\end{equation}
where $Y_0$ is the abundance of dark matter today.  The same relation is used  by \darksusy{} except that the numerical factor is $0.5\%$ larger. The aim of both codes is thus to calculate the abundance of dark matter at the current temperature $Y_0 \equiv Y(T_0)$, where the abundance is defined as the ratio of the number density and entropy density of dark matter $Y = n/s$. Both programs start with the Boltzmann equation~\cite{Griest:1990kh} and follow Reference~\cite{Gondolo:1990dk} to write the differential equation as
\begin{equation} \label{eq:eqstate}
   \frac{dY}{dX} = A(X) \left( Y^2(X) - Y_{eq}^2(X) \right),
\end{equation}
such that
\begin{equation}
  A(X) = \sqrt{\frac{\pi g_{*}(m_{\tilde{\chi}^0_1}/X)}{45}} \frac{m_{\tilde{\chi}^0_1} M_{Pl}}{X^2}\langle \sigma_{eff} v \rangle ,
\end{equation}
where the temperature has been swapped for the dimensionless quantity $X= T/m_{\tilde{\chi}^0_1}$ and $M_{Pl}$ is the Planck mass. $Y_{eq}$ is the thermal equilibrium abundance, and is expressed as
\begin{equation}
   Y_{eq}(T) = \frac{45}{4\pi h_{\text{eff}}^2} \sum_i g_i \frac{m_i}{T} K_2\left (\frac{m_i}{T} \right),
\end{equation}
where $h_{\text{eff}}$ is the number of effective degrees of freedom in the entropy density and $K_n$ is a Bessel function of the second kind. 

The parameter $g_{*}$ is related to the number of degrees of freedom of the system as
\begin{equation}
    g_{*}^{1/2} = \frac{h_{\text{eff}}}{\sqrt{g_{\text{eff}}}} \left( 1 + \frac{T}{3h_{\text{eff}}} \frac{dh_{\text{eff}}}{dT} \right)\;,
\end{equation}
where $g_{\text{eff}}$ is the number of effective degrees of freedom in the energy density. $g_{\text{eff}}$ and $h_{\text{eff}}$ are drawn from hard-coded tables in both programs. In \micromegas{}, the tables come from Olive et al.~\cite{Olive:1980wz,Srednicki:1988ce} by default, but there is an option to use the tables from Hindmarsh \& Philipsen~\cite{Hindmarsh:2005ix}, which is the default in \darksusy{}. The effective degrees of freedom are calculated from their respective extensive quantities, typically by assuming an ideal gas as was done in Olive et al. However, interactions will be significant at in the early universe where temperatures are high, and allowing for interactions requires a modification to the equation of state. As the weak corrections will be suppressed by the W and Z masses, QCD corrections to the effective degrees of freedom will be dominant. Hindmarsh \& Philipsen found that incorporating QCD corrections allows for as much as a $3.5\%$ modification to the relic density~\cite{Hindmarsh:2005ix}.  

From here, the two programs diverge in their treatment of Equation~\ref{eq:eqstate}, as care must be taken due to the stiffness of the ODE~\cite{Gondolo:2004sc,Belanger:2014vza}. To calculate $Y(T_0)$, \micromegas{} employs the freeze-out
 approximation\footnote{This description holds for the case of a single dark matter particle. In general, \micromegas{} can handle models with two component dark matter, which requires a modification of the Boltzmann equation to allow for additional processes and abundances associated with a second dark sector. In the latter case, the Rosenbrock algorithm~\cite{Press:1992,Hairer:2010} is used to avoid the stiffness of the ODE.}
~\cite{Ellis:1998kh,Ellis:1999mm}, writing
\begin{equation}
  \Delta Y = Y - Y_{eq} = \frac{1}{2A} \; :\; \Delta Y \ll Y_{eq} \;.
\end{equation}
Letting $\Delta Y(X_{f1}) = \delta \, Y_{eq} (X_{f1})$ where $\delta$ is a small number chosen to be 1.5, \micromegas{} solves
\begin{equation}
  Y'(X_{f1}) = \delta (\delta + 2) A(X_{f1}) Y_{eq}^2(X_{f1})
\end{equation}
for $X_{f1}$. This point is used as the starting point for the numerical evaluation of Equation~\ref{eq:eqstate} via the Runge-Kutta method, stopping at a point $X_{f2}$. This latter point is chosen such that $Y_{eq}(X_{f2}) < 0.01 Y(X_{f2})$, and allows for the integration of Equation~\ref{eq:eqstate} to solve for $Y_0$:
\begin{equation}
  \frac{1}{Y(X_0)} = \frac{1}{Y(X_{f2})} + \int_{X_{f2}}^{X_0} A(X) dX \; .
\end{equation}
Because $T_0 = 2.725$~K $, X_0 \sim 10^{14}$ and \micromegas{} takes the upper bound to be effectively infinity. Alternatively, \darksusy{} chooses to solve Equation~\ref{eq:eqstate} without applying approximations. Stiffness is still a concern, so \darksusy{} solves the problem by first discretizing the function with trapezoids and then numerically solving the differential equation with an adaptive step-size approach to Euler's method. 

For the computation of the (co-)annihilation of sparticles contributing to the relic density, both \micromegas{} and \darksusy{} follow Reference~\cite{Edsjo:1997bg}. Furthermore, both codes include all 2-body processes\footnote{\micromegas{} does have the option of allowing 2 and 3-body WZ final states, but these are not part of the default calculation~\cite{Belanger:2013oya}.}
 between neutralinos, charginos, sneutrinos, sleptons, and squarks.  \micromegas{} includes processes with gluons, as well.  External programs are incorporated into the distributions of \darksusy{} and \micromegas{} to calculate the relevant cross sections. 
 
 \micromegas{} includes {\tt CalcHEP}~\cite{Pukhov:2004ca} for the evaluation of relevant tree-level annihilation and coannihilation diagrams at run-time for a given model. 
However, some processes are significantly suppressed so as to be insignificant.  \micromegas{} calculates the Boltzmann suppression factor
\begin{equation}
  B_{f} = \frac{K_1( (m_i + m_j)/T_f)}{K_1(2m_{\tilde{\chi}_{1}^{0}}/T_f)} \approx e^{-X_f \frac{m_i+m_j - 2m_{\tilde{\chi}_{1}^{0}}}{m_{\tilde{\chi}_{1}^{0}}}}
\end{equation}
for each channel and, by default, neglects channels where $B_f <10^{-6}$, though this cut-off can be changed by the user~\cite{Belanger:2001fz}.

Within \darksusy{}, on the other hand, the programs {\tt REDUCE}~\cite{hearn:1993} and {\tt FORM}~\cite{Vermaseren:2000nd} are used in the evaluation of annihilation and coannihilation cross sections. {\tt REDUCE} is used for the evaluation of helicity amplitudes for all processes between charginos and neutralinos. This allows for the analytical determination of one type of diagram only once with a numerical sum over all initial and final states performed for the contributing diagrams afterwards. All other processes involving sfermions have their scattering amplitudes evaluated by {\tt FORM}.

One interesting difference between \micromegas{} and \darksusy{} is in the inclusion of internal bremsstrahlung (IB). \micromegas{} includes final states with two SM particles plus an additional photon for the evaluation of the annihilation cross section both in the early Universe and today. While \darksusy{} includes IB when considering the gamma ray signature from annihilation in our Galaxy's halo, it is not incorporated by default in their calculation of the relic density. As we will see below, this will lead to significant differences in the dark matter observables.

Finally we consider the differences in approaches used to calculate the SI scattering cross sections. Both programs utilize loop corrections to the the scattering amplitudes but follow different frameworks. 
\darksusy{} follows the effective Lagrangian framework laid out in Ref.~\cite{Bergstrom:1995cz}, while  \micromegas{} utilizes the framework of Ref.~\cite{Drees:1993bu}. As discussed, for example, in~\cite{Drees:1993bu}, the effective Lagrangian framework can miss crucial QCD effects, though, with modification, it is capable of reliably reproducing the 1-loop result for most cases.

Another important difference between \darksusy{} and \micromegas{} is in the nucleon form factors used to calculate the expected neutralino-nucleon elastic scattering cross section.  The form factors for each calculator are tabulated in Table~\ref{tab:form}. For all quarks, \darksusy{} uses larger form factors than \micromegas{}. As discussed below, this leads to a difference in the predicted SI neutralino-nucleon scattering cross sections, however, as we will demonstrate, these differences alone are not enough to fully explain the discrepancies in the predictions. It is clear that the details of the loop corrections also play an important role in the calculation, and can provide significant contributions to the SI cross sections.

Finally, we mention that there are differences in how each calculator determines the relevant particle widths used in the calculations. When available, \micromegas{} reads in the decay blocks from the SLHA input file, but otherwise employs their own calculation to find any necessary widths. Alternatively, \darksusy{} does not currently read SLHA decay blocks and always performs their own evaluation of the relevant particle widths, though they note that future versions of their SLHA reader should include this functionality\footnote{See \darksusy{} 5.1.2  src/slha/dsfromslha.F, lines 766-769.}. The importance of inlcuding accurate widths in the relic density calculation is particularly relevant for funnel points; previous studies have found $\mathcal{O}(10\%)$ difference in the calculation of the $A$-funnel between \micromegas{} and \darksusy{}~\cite{Cohen:2013kna}.  As we will demonstrate, this difference in how the width is included can lead to seemingly inconsistent results among pipelines that end with \micromegas{}, while results from \darksusy{} pipelines seem more consistent but are less robust. 

\begin{table}
  \centering
  \includegraphics[scale=0.9]{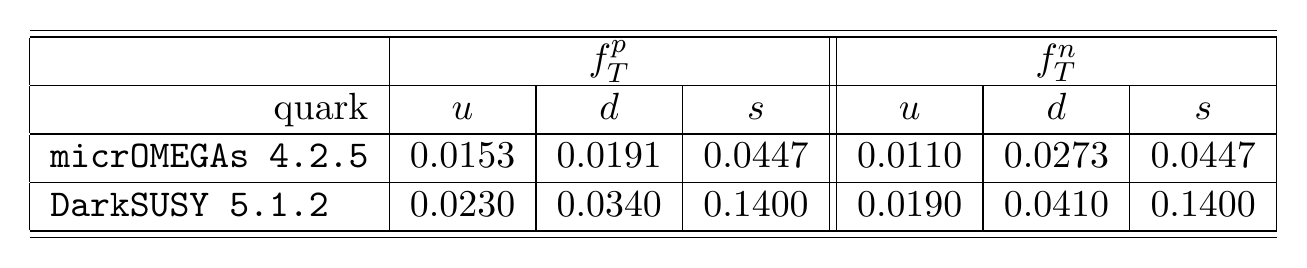}
  \caption{Spin-independent, scattering form-factors used by \micromegas{} and \darksusy{} for protons ($f^p_T$) and neutrons ($f^n_T$).}
  \label{tab:form}
\end{table}

\section{The Physics of Neutralino Dark Matter Benchmarks}
\label{sec:DMphysics}

In this Section, we provide a brief introduction to aspects of the physics of neutralino dark matter that will be relevant for our analysis of the benchmark points. 

The challenge in obtaining the correct relic density of neutralino dark matter observed in the Universe is well-known\footnote{This challenge is alleviated if one considers a non-thermal history for dark matter~\cite{Dutta:2009uf, Kane:2015jia}. In this work, however, we adhere to a thermal history.}.  Specifically, one generically obtains an annihilation cross section at freeze-out that is too small, leading to too much neutralino dark matter in the present epoch. 

The general idea of WIMP dark matter is that, in order to produce the correct relic density, the annihilation cross section of dark matter in the early Universe should have been around 1 pb. This is approximately the value one obtains if dark matter is a new particle with approximately weak-scale mass and with electroweak couplings, a happy accident called the ``WIMP miracle".

However, as has been long appreciated, the details of actual SUSY models are somewhat less attractive than the idea sketched above. Although supersymmetry predicts that the annihilation cross section of two neutralinos should be in the neighborhood of 1 pb, the exact numerical value can span a range that covers orders of magnitude, depending on the composition of the neutralino and the mass spectrum of the other supersymmetric particles. Dark matter that is predominantly higgsino-like ($\tilde{h}$) annihilates to $W$ and $Z$ bosons with a cross section that involves the full strength of the $SU(2)$ gauge coupling, and is moreover enhanced by the presence of spin-one particles in the final state. Higgsino and wino dark matter thus have cross sections that are too large for the observed relic density. On the other hand, binos ($\tilde{B}$) mainly annihilate to quark and lepton pairs, a process that suffers from helicity suppression. Binos therefore typically have a cross section for annihilation that is too low.

The regions of supersymmetric parameter space that are compatible with the dark matter relic density thus tend to be fine-tuned. In the following subsections, we will consider the most well-studied regions, characterized by how the observed dark matter relic density is achieved: coannihilation models, well-tempered dark matter, $A$-funnel annihilation, and pure higgsino composition.

\subsection{Coannihilation of $\tilde{B}$ with light scalars}

The calculation of the relic density for the coannihilation region is very sensitive to the relative masses of the dark matter candidate and the light scalar(s) that coannihilate with it, as detailed eg.~in~\cite{Griest:1990kh}.  Even small discrepancies in the mass spectrum given by different spectrum generators will result in significantly different predictions for the relic density.

Regarding the indirect detection prospects for bino-scalar coannihilation models, the annihilation cross section for bino dark matter in the present Universe occurs mainly through $t$-channel exchange of light scalars.  Bino-nucleon scattering generally proceeds via Higgs- or squark-exchange.  The Higgs exchange diagram is suppressed for models in which bino-squark coannihilations dominate in the early Universe, due to the pure bino nature of the dark matter in these cases. Furthermore, if the first and second generation squarks are heavy, as is the case in our bino-stop coannihilation scenario, the squark-exchange diagram is suppressed, resulting in very small scattering cross sections with nuclei, as we will see.

In the following analysis, three coannihilation scenarios will be relevant: $\tilde{B}-\tilde{\tau}_1$, $\tilde{B}$-$\tilde{t}_1$; and $\tilde{B}$-$\tilde{q}$, where $\tilde{q}$ denotes any first or second generation squark.  In each case, the composition of the neutralino LSP is $\gtrsim 99.9\%$ bino.

\subsection{Well-tempering of Dark Matter}

Well-tempered dark matter has been extensively studied in the context of recent direct, indirect, and collider searches~\cite{Cheung:2012qy, Dutta:2013sta, Badziak:2017the}. The annihilation cross section depends on the mixture of bino and higgsino states, or, equivalently, on the higgsino fraction $|N_{13}|^2 + |N_{14}|^2$. The full expression for the annihilation cross section, as well as various interesting limits, are available for example in Reference~\cite{ArkaniHamed:2006mb}. The predicted relic density and indirect detection signals in the current Universe are both sensitive to the higgsino fraction.

The dominant neutralino-nucleon scattering occurs via CP-even Higgs exchange, which is a bino-higgsino-Higgs coupling. Since a well-tempered neutralino has sizable higgsino and bino fractions, 
the scattering rates with nuclei can be relatively large.

\subsection{$A$-Funnel Annihilation}
A neutralino LSP can annihilate resonantly by exchanging a pseudoscalar Higgs boson $A$ in the $s$-channel, provided $m_{A} \, \sim \, 2 m_{\tilde{\chi}_1^0}$. Prospects of probing the $A$-funnel at colliders and the correlation with the observed Higgs mass have been extensively studied~\cite{Arbey:2012dq, Djouadi:2015jea, Lee:2015uza, Anandakrishnan:2014fia, Anandakrishnan:2013tqa}. Prospects for direct~\cite{Huang:2014xua, Hooper:2013qjx} and indirect~\cite{Erickcek:2015bda} detection have also been studied.

The annihilation cross section can be expressed as~\cite{Griest:1988ma}
\be \label{funnelcs}
\sigma v \, \sim \, \frac{3}{2 \pi}  \frac{y^2_{A \tilde{\chi} \tilde{\chi}}\, y^2_{A b b}\, s}{(m^2_A - s)^2 + m^2_A \Gamma^2_A},
\ee
where
\be \label{mandelstam}
s = 4 m^2_{\tilde{\chi}_1^0} (1- v^2/4)^{-1}
\ee
is the Mandelstam variable, $\Gamma_A = 4.6$ GeV is the width of the pseudoscalar at our benchmark, $m_A = 2042$ GeV is its mass, and the Yukawa couplings to the b-quarks and neutralinos are given by 
\be \label{yukawasafunnel}
y_{A bb} = \frac{i m_b \tan{\beta}}{\sqrt{2} v_{ew}} \,\,\,\,\,\,\,{\rm and} \,\,\,\,\,\,\,
y_{A \tilde{\chi} \tilde{\chi}} = i g_1 N_{11} (N_{14} \cos{\beta} - N_{13} \sin{\beta}).
\ee
To obtain the annihilation cross section, one must take a thermal average of Eq.~\ref{funnelcs} or its more general equivalent.  The resonance occurs in the zero velocity limit in the current Universe, and the annihilation cross section today is given by $\sigma v |_{v \rightarrow 0}$. In the early Universe, the non-zero velocity leads to thermal broadening of the resonance, as is clear from the velocity dependence of the Mandelstam variable, Eq.~\ref{mandelstam}. Small differences in the calculation of $m_A$ and $v$ by different generators can affect the relic density significantly. Note that from Eq.~\ref{yukawasafunnel}, it is clear that to obtain the correct relic density, the $A$-funnel benchmark requires non-zero values of $N_{11}$ and either $N_{13}$ or $N_{14}$. Thus, though the dark matter is typically primarily bino, a higgsino component must be retained, i.e.~the neutralino $\tilde{\chi}_1^0$ must have some bino-higgsino mixture. 

In this case, the SI scattering cross section with nuclei is mediated primarily by Higgs exchange. This is enabled by the non-zero higgsino component $N_{13}$ or $N_{14}$ required to obtain the correct relic density. However, the values of the higgsino fraction required to obtain the correct relic density are typically very small, corresponding to feeble scattering cross sections with nuclei. The direct detection prospects for the $A$-funnel scenario are thus rather challenging, as we will see. This should be contrasted to the case of the well-tempered neutralino, where the large higgsino fraction drives both the relic density and the scattering cross section.

\subsection{Pure Higgsino ($\tilde{h}$) Composition}
Higgsinos with mass of $\sim \, 1$ TeV can satisfy the relic density constraint, with the dominant mechanism being annihilation to gauge bosons. Coannihilation among charged ($\tilde{\chi}^\pm_1$) and neutral ($\tilde{\chi}_1^0$ and $\tilde{\chi}_2^0$) higgsinos is also important in this case. We refer to ~\cite{Cirelli:2005uq} for expressions of the relic density in various limits.

The contribution of Higgs exchange diagrams to the scattering cross section with nuclei is suppressed due to the small gaugino-higgsino mixing. The contribution of squark exchange diagrams is also suppressed at our benchmark point since the squark  masses are at several TeV. Thus, we generally expect small direct detection signals for this benchmark point. We refer to \cite{Cheung:2012qy} for detailed calculations of the scattering cross section of pure higgsinos.

\section{Results: pMSSM Analysis}
\label{sec:pMSSM}

\begin{table}[t]
    \centering
    \includegraphics[scale=0.9]{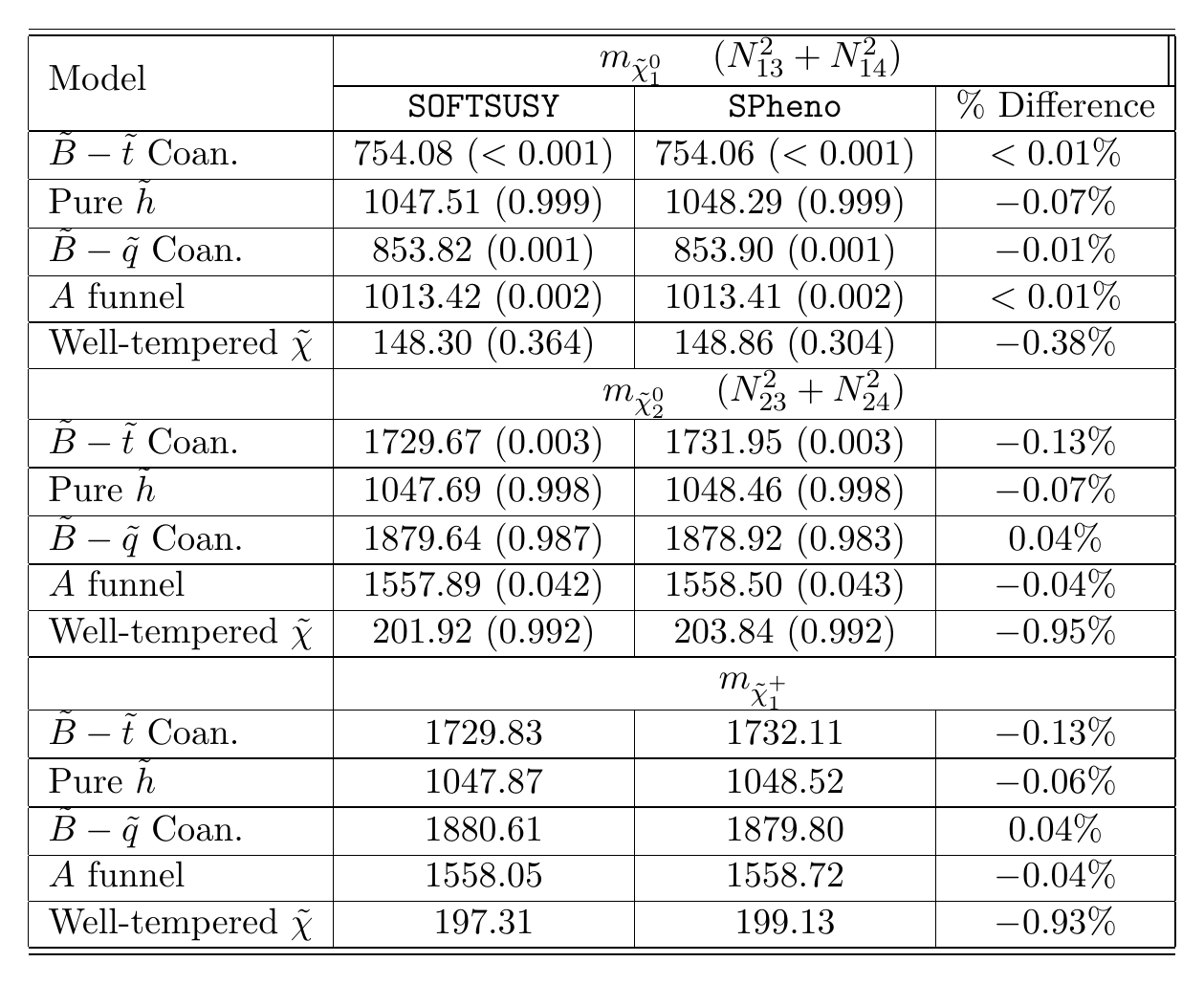}
  \caption{Masses, in GeV, of the lightest and second-lightest neutralinos and the lighter chargino for the pMSSM benchmark points. The corresponding higgsino fraction is given in parentheses. The percent differences are given relative to the \softsusy{} values.
  }
  \label{tab:pneutspectra}
\end{table}

\begin{table}[t]
      \includegraphics[scale=0.9]{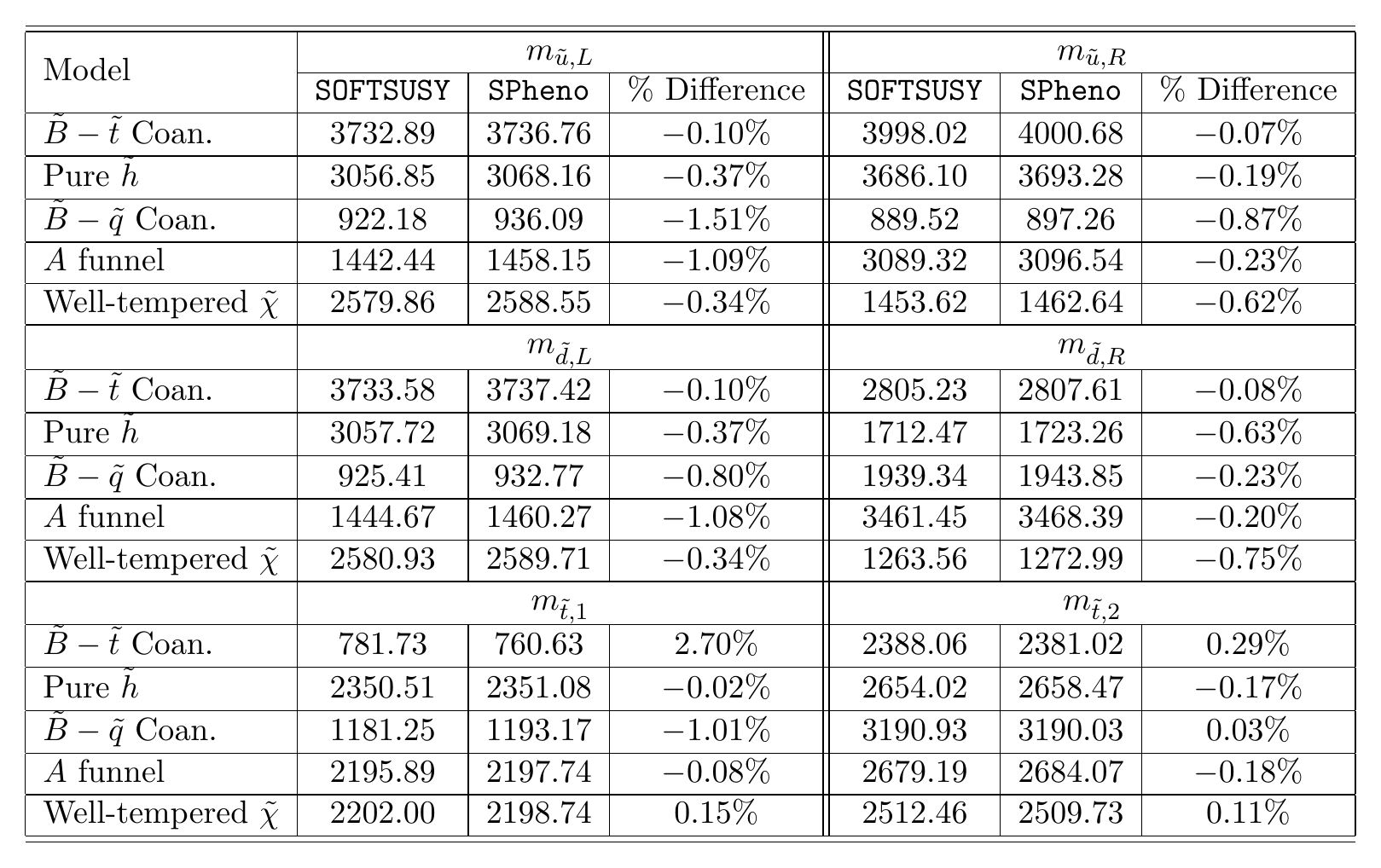}
  \caption{Masses, in GeV, of the squarks for the pMSSM benchmark points. The percent differences are given relative to the \softsusy{} values.}
  \label{tab:psquarkspectra}
\end{table}

\begin{table}
    \centering
    \subfloat[][$m_h$ and $m_H$ as computed via the \feynhiggs{} branch of the pipeline.\label{tab:phiggs1feyn}]{%
      \includegraphics[scale=0.9]{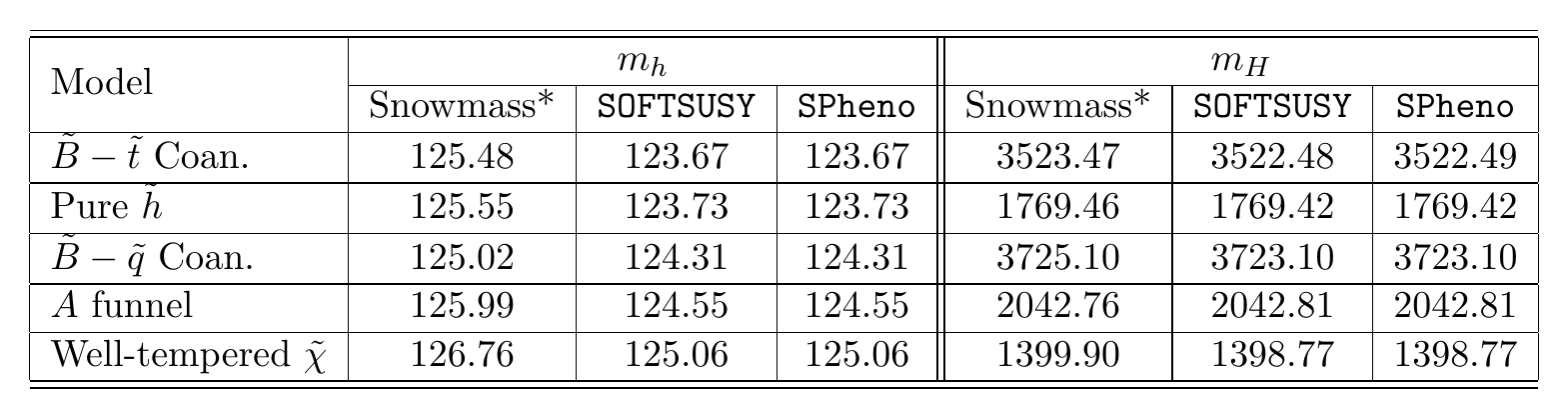}
      }
    \hfill
    \subfloat[][$m_A$ and $m_{H^\pm}$ as computed via the \feynhiggs{} branch of the pipeline.\label{tab:phiggs2feyn}]{%
      \includegraphics[scale=0.9]{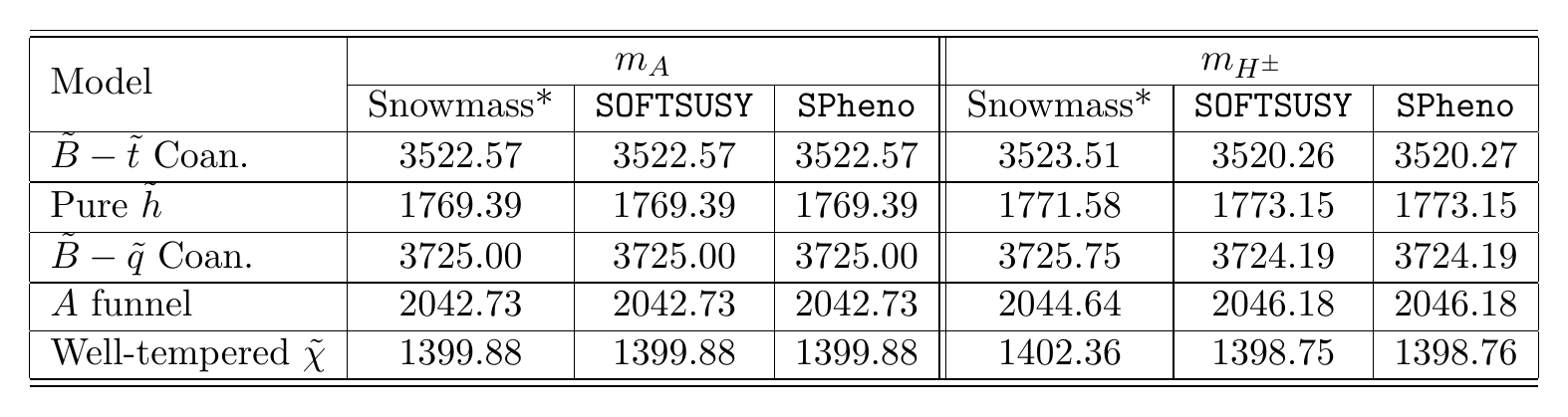}
    }
  \hfill
    \subfloat[][$m_h$ and $m_H$ as computed via the \susyhd{} branch of the pipeline.\label{tab:phiggs1hd}]{%
      \includegraphics[scale=0.9]{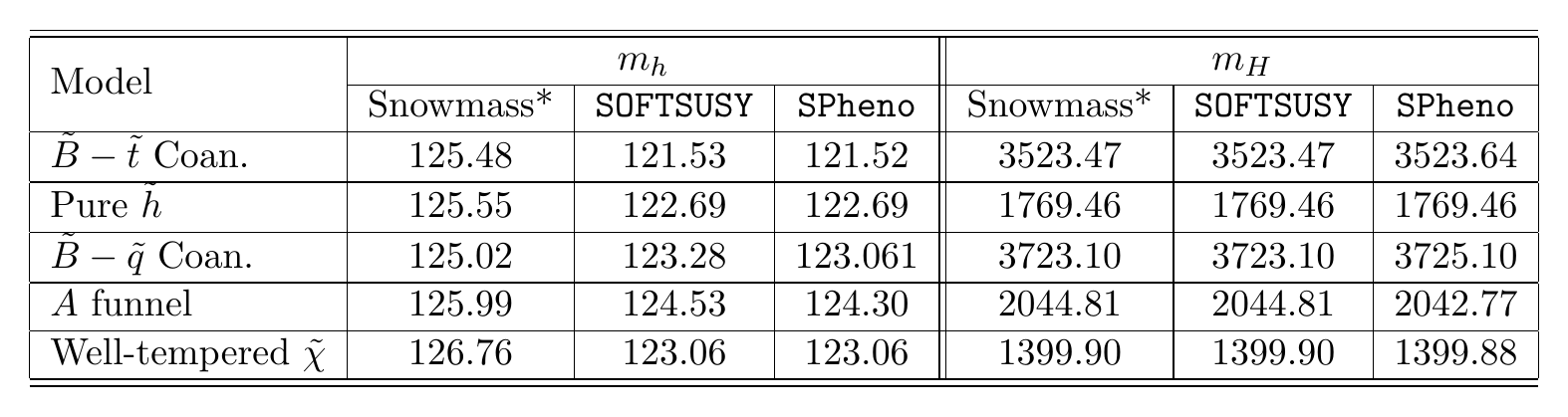}
  }
  \hfill
    \subfloat[][$m_A$ and $m_{H^\pm}$ as computed via the \susyhd{} branch of the pipeline.\label{tab:phiggs2hd}]{%
      \includegraphics[scale=0.9]{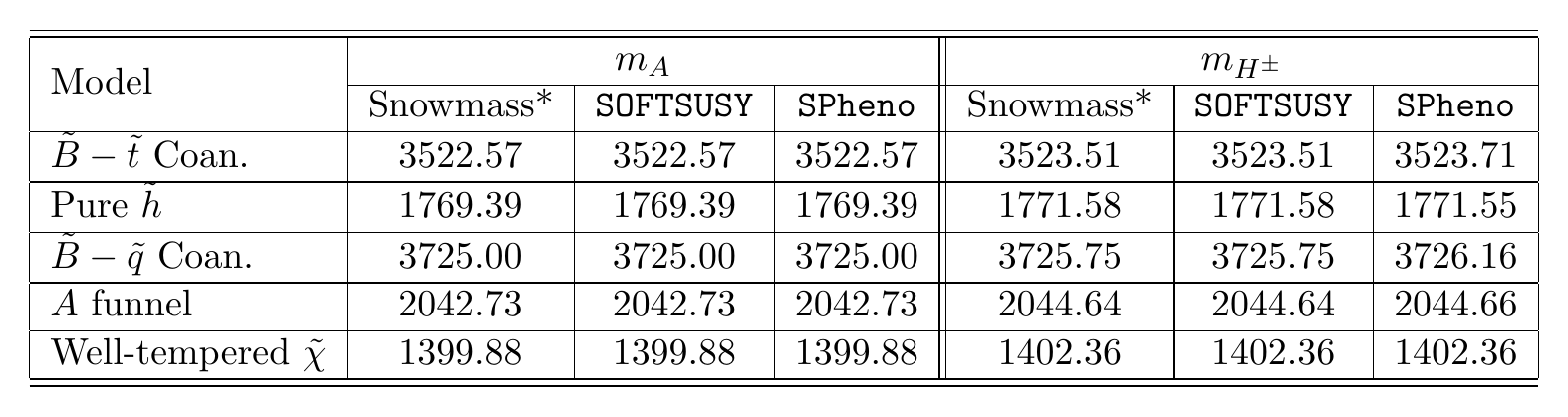}
  }
  \caption{\textbf{pMSSM Higgs masses:} The masses, in GeV, of the CP-even Higgses (Tables~\ref{tab:phiggs1feyn} \& \ref{tab:phiggs1hd}) and the CP-odd and charged Higgs (Tables~\ref{tab:phiggs2feyn} \& \ref{tab:phiggs2hd}). Tables~\ref{tab:phiggs1feyn} \& \ref{tab:phiggs2feyn} employ the \feynhiggs{} pipelines while Tables~\ref{tab:phiggs1hd} \& \ref{tab:phiggs2hd} use \susyhd{}. ``Snowmass*'' refers to the updated Snowmass pipeline which, at this level, amounts to the \softsusy{} spectrum as no Higgs calculator was employed therein. Furthermore, note that \susyhd{} only provides a correction to the SM Higgs mass, and thus $m_H$, $m_A$, and $m_{H^{\pm}}$ in the \susyhd{} tables are the masses as computed by the relevant spectrum generator. }
  \label{tab:phiggs}
\end{table}

Here we present our pMSSM analysis. We consider the five pMSSM points from the Snowmass 2013 benchmarks~\cite{Cahill-Rowley:2013gca} as discussed in Section~\ref{sec:benchmarks}.

The spectra in the neutralino, squark, and Higgs sectors obtained using the different spectrum generators we consider are displayed in Tables~\ref{tab:pneutspectra},~\ref{tab:psquarkspectra}, and ~\ref{tab:phiggs}, respectively. For tables~\ref{tab:pneutspectra} and~\ref{tab:psquarkspectra}, the first column lists the dark matter benchmark scenario, while the next columns display the spectra of relevant sparticles obtained from \softsusy{} and \spheno{}. The final column lists the percent difference in the mass obtained from the two generators. In table~\ref{tab:phiggs}, we compare results for the Higgs sector from \feynhiggs{} and \susyhd{}. As table~\ref{tab:phiggs} demonstrates, as of \feynhiggs{} 2.12.0 and \susyhd{} 1.0.2 these two Higgs sector calculators are in very good agreement.  Hereafter we consider only the \feynhiggs{} branches of the pipelines in Fig.~\ref{fig:flow}.

We pause to elaborate on the Higgs sector, before moving on to analyze the dark matter results. In Tables~\ref{tab:phiggs1feyn}~\&~\ref{tab:phiggs1hd} we show the masses in the Higgs sector for all the dark matter benchmark scenarios studied. The masses corresponding to Table~\ref{tab:phiggs1feyn}~\&~\ref{tab:phiggs2feyn} are computed by \feynhiggs{}, while those corresponding to Tables~\ref{tab:phiggs1hd}~\&~\ref{tab:phiggs2hd} are computed by \susyhd{}. As per convention, Snowmass$^*$ is the updated Snowmass pipeline which amounts to the \softsusy{} spectrum prior to \feynhiggs{}, here. We note that \susyhd{} only corrects $m_h$ and, thus, $m_H$, $m_A$, and $m_{H^{\pm}}$ are identical to those calculated by the relevant spectrum calculator.

From these results, we see that the Higgs sector masses are not significantly affected by the choice of SUSY spectrum generator, with differences amounting to less that $0.01\%$. However, the inclusion of either \feynhiggs{} or \susyhd{} can provide a significant shift in $m_h$ from the Higgs mass calculated by the spectrum generator itself. The Snowmass points were constrained by the Higgs mass ($126\pm1$~GeV), but utilizing \feynhiggs{} moves the mass down by as much as $2$~GeV (1.5\%) which puts the benchmark points in strong tension with the measured value of the Higgs mass ($125\pm0.24$).

Figure~\ref{fig:pmssm} displays our results for the dark matter observables in the pMSSM.  The benchmarks are denoted by different shapes: bino-stop coannihilation (diamonds), pure higgsino (stars), bino-squark coannihilation (circles), $A$-funnel (pentagons), and well-tempered (triangles). Filled/unfilled points correspond to the use of \micromegas{}/\darksusy{}. The pipelines used in generating the results are distinguished by color: we use magenta/cyan to distinguish, \softsusy{}/\spheno{}, respectively, and black/green to delineate between the Snowmass/Snowmass$^*$ pipelines.

The dark matter relic density for the pMSSM benchmarks is shown in the upper left panel of Figure~\ref{fig:pmssm} (and tabulated in Table~\ref{tab:pomega}).  For comparison, we also include the Planck 3-sigma range~\cite{Ade:2015xua} for the relic density, which is highlighted by the red band.
The upper right panel of Fig.~\ref{fig:pmssm} shows the annihilation cross section today (tabulated in Table~\ref{tab:psigv}), with the Fermi-LAT 6-year limits from dwarf spheroidal galaxies~\cite{Ackermann:2015zua} included for comparison (limits on annihilation to $\tau^+\tau^-$ in red and $b\bar{b}$ in blue).  In the lower left panel of Fig.~\ref{fig:pmssm} we show the spin-independent neutralino-nucleon elastic scattering cross section, where we plot the per-nucleon cross section averaged for Xe, computed from the proton and neutron values tabulated in Table~\ref{tab:psigmafeyn} (\feynhiggs{} pipelines), as well as exclusion contours from LUX ~\cite{Akerib:2016vxi} in solid red, PandaX ~\cite{Tan:2016zwf} in solid blue, and LZ (projected)~\cite{Akerib:2015cja} in dashed black lines, for comparison.

We now turn to a description of our results. Throughout our discussion, we will concentrate on the \softsusy{} (magenta) and \spheno{} (cyan) pipelines. In the figures, we also plot the predictions from the Snowmass$^{*}$ and Snowmass$^{\dagger}$ pipelines shown in solid green and solid black, respectively. Since the physics of these pipelines is very similar to the \softsusy{} - \micromegas{} pipeline shown by solid magenta shapes, we will not discuss them separately.

\subsection{Bino-Stop Coannihilation}
The masses relevant for coannihilation are the lightest neutralino and the stop. From Table~\ref{tab:pneutspectra}, we see that there is good agreement  between \softsusy{} and \spheno{} in the neutralino spectrum for these two benchmark scenarios. The dark matter has a mass of $\sim 754$ GeV, with very good agreement between the two spectrum generators, and is almost completely bino.  On the other hand, from Table~\ref{tab:psquarkspectra}, we see that the stop mass differs by 2.7\% between  \softsusy{} and \spheno{}, although it is generally in the range where coannihilation is operational. This has a significant effect on the relic density, which, in these cases, is exponentially sensitive to the mass difference between the dark matter and the relevant coannihilation partner. 

\begin{figure}[h!] 
  \centering 
  \includegraphics[width=0.98\textwidth]{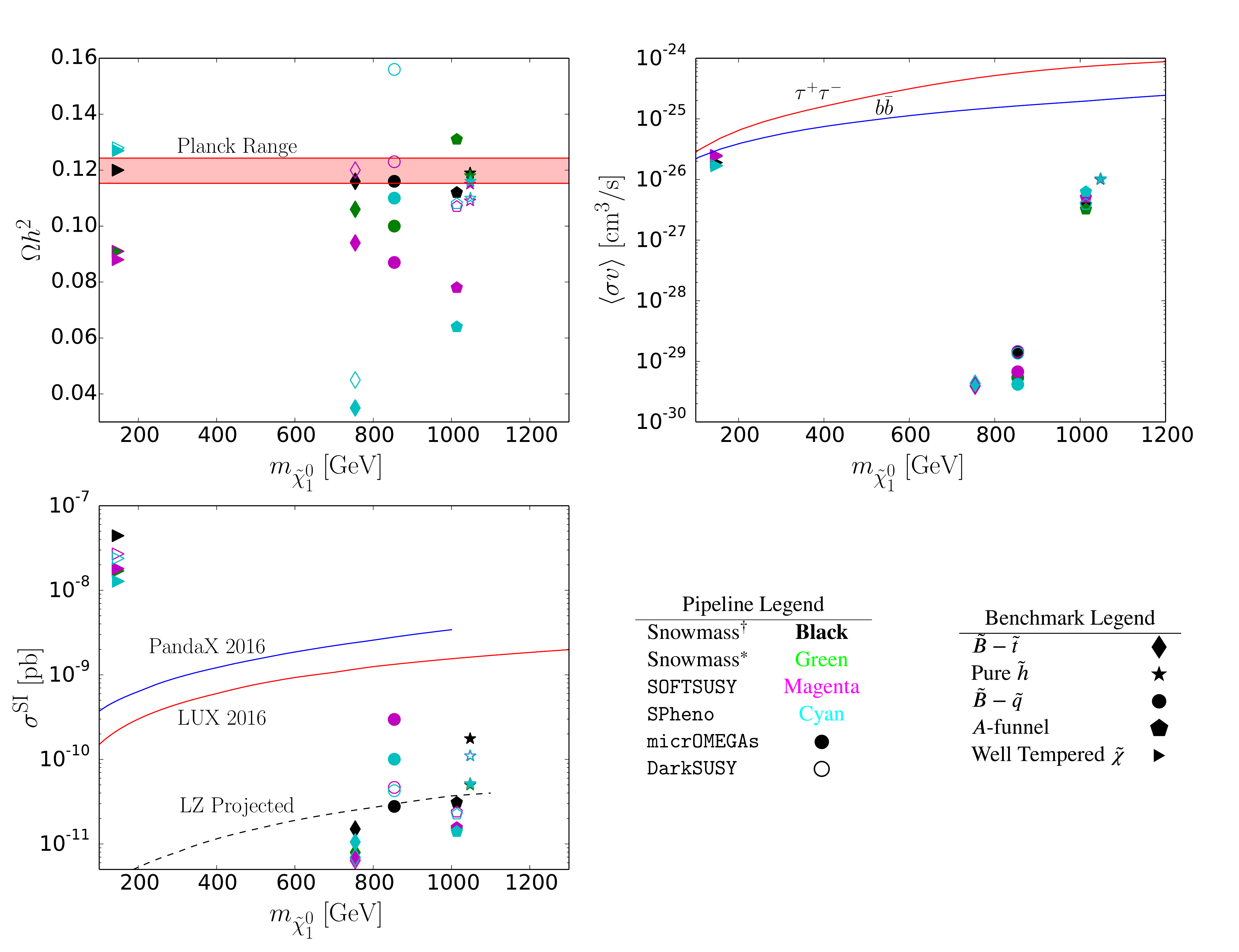}
  \caption{pMSSM results:  The top left, top right, and bottom left panels show the neutralino relic density, annihilation cross section today, and the SI neutralino-nucleon elastic scattering cross section, respectively, all as functions of the dark matter mass, with a common legend in the lower right panel. For comparison, the Planck 3-sigma range for the dark matter abundance~\cite{Ade:2015xua} is highlighted in red in the upper left panel, the limits on the annihilation cross section today from Fermi-LAT's 6-year analysis of dwarf spheroidal galaxies~\cite{Ackermann:2015zua} for annihilation to $\tau^+\tau^-$ (red) and $b\bar{b}$ (blue) are shown in the upper right panel, and exclusion limits from LUX (red)~\cite{Akerib:2016vxi}, PandaX (blue)~\cite{Tan:2016zwf}, and LZ (projected; black)~\cite{Akerib:2015cja} are shown in the lower left panel.}
  \label{fig:pmssm}
\end{figure}

The effect of the variation in the mass of the coannihilation partner can be seen in the upper left panel of Figure~\ref{fig:pmssm}. The solid magenta and cyan diamonds correspond to the relic density values computed by \micromegas{},  for spectra coming from  \softsusy{} and \spheno{}, respectively. While  \softsusy{} yields a value of $\Omega h^2 = 0.094$,  \spheno{} yields a value of $\Omega h^2 = 0.035$, and the difference stems entirely from the difference in stop masses computed by the two generators. Similarly, comparing the hollow magenta and cyan diamonds, which correspond to the relic density values computed by \darksusy{}, we see that while \softsusy{} gives a value of $\Omega h^2 = 0.120$,  \spheno{} yields a value of $\Omega h^2 = 0.045$. 

It is also interesting to compare the values yielded by \micromegas{} and \darksusy{} for the same spectrum. For example, selecting the spectrum from \softsusy{} and comparing the solid and hollow magenta diamonds, we see that \micromegas{} gives  $\Omega h^2 = 0.094$ while \darksusy{} gives $\Omega h^2 = 0.120$. We point out that these theoretical uncertainties exceed the current experimental uncertainty, which, here results in the hollow diamond lying within the Planck-allowed band, while the solid diamond does not. These discrepancies occur due to differences in the calculation of the effective cross section for each annihilation and coannihilation channel and the different relative weights of contributing final states in the coannihilation channels assigned by the calculators.

We note that both \micromegas{} and \darksusy{} give values for the coannihilation cross sections using an effective tree-level calculation. These values can be significantly altered if higher-order SUSY-QCD corrections are taken into account. After including loop diagrams containing a gluon, a gluino, a four-squark vertex, and incorporating gluon radiation, the authors of Reference~\cite{Harz:2012fz} have found a $\sim 20$\% discrepancy with the relic density computed by \micromegas{} in the bino-stop coannihilation region. We will not consider these loop corrections further in this paper, but note that global $\sim20\%$ theoretical uncertainties are expected in all cases.

The annihilation cross section of the bino-stop benchmark model in the current Universe is shown in the upper right panel of Figure~\ref{fig:pmssm}. There is remarkable agreement among the various pipelines. Since coannihilation channels are irrelevant in the present Universe, the sensitivity to the mass difference between the stop and the bino is absent. The annihilation proceeds mainly through the $t$-channel exchange of a stop, and the small difference in the stop mass reported by \softsusy{} and \spheno{} does not affect this diagram as significantly. We find that both \micromegas{} and \darksusy{} give similar values for the different channels in $\tilde{\chi}_1^0 \tilde{\chi}_1^0$ annihilation, the only difference being that \darksusy{} ascribes $\sim5\%$ contribution to the annihilation cross section from $\tilde{\chi}_1^0 \tilde{\chi}_1^0 \rightarrow gg$ final state, while \micromegas{} does not return this channel.

The SI scattering cross section is shown by the diamonds in the lower left panel of Figure~\ref{fig:pmssm}. For very heavy squarks, the leading scattering cross section is mediated by Higgs exchange, which is suppressed if the lightest neutralino is a pure higgsino or gauge state, as discussed in Sec.~\ref{sec:DMphysics}. Thus, the combination of heavy squarks and pure bino eigenstate conspire to give suppressed scattering cross sections for the bino-stop coannihilation benchmark. The cross sections lie below the projected LZ limits, at approximately $10^{-11}$~pb. 

There is a factor of $\sim 5$ discrepancy between the SI scattering cross section yielded by \softsusy{} - \micromegas{} (solid magenta diamond) relative to  \spheno{}-\micromegas{} (solid cyan diamond). From Tables~\ref{tab:pneutspectra},~\ref{tab:psquarkspectra}, and~\ref{tab:phiggs}, we can see that while the higgsino fraction and the Higgs mass match to a high approximation for both \softsusy{} and \spheno{} in the bino-stop coannihilation benchmark, there can be up to a 4 GeV difference in the masses of the squarks. While it is unlikely that this small variation in squark masses can account for the observed variation in the SI scattering cross section, we cannot pinpoint the exact source for the discrepancy. Comparing the dark matter calculators, we see that the solid and hollow magenta diamonds overlap entirely, meaning that after receiving the spectrum from \softsusy{}, both \micromegas{} and \darksusy{} computed the same scattering cross section. The matching is not quite as exact for the spectrum coming from \spheno{} (solid and hollow cyan diamonds), but the values are quite close. While \darksusy{} implements an effective Lagrangian in the heavy squark limit following~\cite{Griest:1988ma} (see Reference~\cite{Bergstrom:1995cz} and references therein for details), \micromegas{} implements the full one-loop Lagrangian following Reference~\cite{Drees:1993bu}.

\clearpage 

\subsection{Bino-Squark Coannihilation}
The physics for this benchmark is similar to that of the bino-stop case. Although there is good agreement in the neutralino spectrum, we see from Table~\ref{tab:psquarkspectra} that squarks can differ by up to 1.51\% between \softsusy{} and \spheno{}. The effect on the relic density can be seen in the upper left panel of Figure~\ref{fig:pmssm}. The solid magenta and cyan circles correspond to the relic density values computed by \micromegas{} for spectra coming from  \softsusy{} and \spheno{}, respectively. While \softsusy{} yields a value of $\Omega h^2 = 0.087$,  \spheno{} yields a value of $\Omega h^2 = 0.110$. The difference can be attributed to the difference in squark masses. The fact that \softsusy{} produces a lower value for the relic density than \spheno{} is due to the fact that squark masses from \softsusy{} are lighter than those from \spheno{}, giving a stronger coannihilation effect. We also note that for a given spectrum generator, say \softsusy{}, \darksusy{} gives a larger value of the relic density than \micromegas. 

The annihilation cross section in the current Universe is shown by the circles in the upper right panel of Figure~\ref{fig:pmssm}. It is evident that there is much more convergence of results among the various pipelines compared to the relic density computation. This can again be ascribed to the fact that coannihilation channels are absent in the current Universe and small differences in squark masses do not translate to large differences in the calculation of the $t$-channel squark exchange diagram.

Comparing the diamonds (stop coannihilation) and circles (squark coannihilation) in the upper right panel of Figure~\ref{fig:pmssm}, we can see that the bino-squark coannihilation points exhibit greater spread in their current annihilation cross sections. This is due to the fact that the cross sections are driven by $t$-channel stop or squark exchange diagrams in the present Universe, and the different pipelines give a greater spread in the squark masses than they do the stop mass for the bino-stop benchmark.

The SI scattering cross section is shown by the circles in the lower left panel of Figure~\ref{fig:pmssm}. As expected, the values are higher than the bino-stop coannihilation case, due to the lighter squarks which contribute to the squark-exchange diagram. However, there is also greater disagreement between the different pipelines. Firstly, comparing the solid magenta circle with the solid cyan circle, we see that the cross section computed with the spectrum coming from \softsusy{} is a factor $\sim 3$ greater than that coming from \spheno{}. This is expected, since the squark masses given by \softsusy{} are lower, from Table~\ref{tab:psquarkspectra}. The same trend can be seen using \darksusy{} (comparing the hollow magenta circle with the hollow cyan circle), although the effect is smaller.

For a given spectrum generator, it is clear that \micromegas{} is giving larger values of the scattering cross section than \darksusy{}, since the solid circles are above the hollow ones. This is partly due to differences in the form factors used by the two calculators (see Table~\ref{tab:form}). Using the the form factors of \darksusy{} in \micromegas{}, we find that the scattering cross sections reported by \micromegas{} reduces by $\sim 30$\%, bringing the two calculators to greater agreement with each other.

\subsection{Pure Higgsino}
The results for the relic density, current annihilation cross section, and scattering cross section of the pure higgsino benchmark are shown by stars in the upper left, upper right, and lower left panels of Figure~\ref{fig:pmssm}, respectively. For the relic density there is good agreement between the different pipelines. The two spectrum calculators produce similar  mass and higgsino fraction of the lightest neutralino for this benchmark, as is evident from Table~\ref{tab:pneutspectra}. Thus, the magenta and cyan stars for a fixed dark matter calculator overlap in the upper left panel of Figure~\ref{fig:pmssm}. There is some discrepancy between the results returned by \micromegas{} and \darksusy{} due to differences in calculation of coannihilation processes between the neutral and charged higgsinos. Since coannihilation is unimportant in the current Universe, these discrepancies disappear and all stars align perfectly in the upper right panel of Figure~\ref{fig:pmssm}. 

For the SI scattering cross section, we notice that the rates for a pure higgsino are suppressed due to the small gaugino-higgsino mixing, similar to the case of a pure bino. As for the relic density, the main difference between pipelines comes from the choice of the dark matter calculator and whether an effective Lagrangian is used or the full one-loop Lagrangian is considered.

\subsection{Well-Tempered Neutralino}
The results for the relic density, current annihilation cross section, and SI scattering cross section of the well-tempered neutralino benchmark are shown by triangles in the upper left, upper right, and lower left panels of Figure~\ref{fig:pmssm}, respectively. 

We first discuss the relic density. The upper left panel of Figure~\ref{fig:pmssm} shows that there are considerable differences between the various pipelines. The \spheno{} pipelines (cyan triangles) give a larger relic density than the \softsusy{} pipelines (magenta triangles). This can be traced to the lower higgsino content of the lightest neutralino given by \spheno{} relative to \softsusy{}. In fact, from Table~\ref{tab:pneutspectra}, the lightest neutralino is 30\% higgsino when calculated by \spheno{}, but 36.4\% higgsino when calculated by \softsusy{}, although the masses agree to better than a percent. 

On the other hand, for a given spectrum generator, there is almost no discrepancy in the relic abundance coming from the dark matter calculators. Thus, the solid and hollow triangles approximately overlap for a given color. What small discrepancy that is evident can be attributed to differences in the computation of the effective annihilation and coannihilation cross sections between the bino, neutral higgsino, and charged higgsinos, as well as the computation of the $t$-channel chargino exchange diagram. The relic density in the present Universe presents far fewer differences, since the coannihilation channels are absent. The triangles thus overlap in the upper right panel of Figure~\ref{fig:pmssm}.

For the SI scattering cross section with nuclei, we see that the well-tempered benchmark is the only one of our pMSSM benchmarks that is constrained by current experiments, regardless of the pipeline adopted. This is because of the non-negligible higgsino fraction. The different higgsino fractions reported by \softsusy{} and \spheno{} affect the relative positions of the magenta and cyan triangles in the lower left panel of Figure~\ref{fig:pmssm}. Since \softsusy{} reports the larger higgsino fraction, the scattering cross section for magenta triangles are larger than those for the cyan triangles corresponding to the \spheno{} pipelines. 

For a given spectrum generator, say \softsusy{}, there is some disagreement between the SI scattering cross sections computed by \micromegas{} (solid magenta triangle) versus \darksusy{} (hollow magenta triangle). This can be attributed to the different form factors, as well as the way in which loop effects are incorporated. To check the effect of the different form factors, we used \darksusy{}'s form factors from Table~\ref{tab:form} in the \micromegas{} code. For most points, doing so brings \micromegas{}'s values for the scattering cross sections into better agreement with those of \darksusy{}. However, some differences remain, suggesting that other details of the calculation are also important. We also carried out the tree level calculation for SI scattering, following
 the discussion of Appendix C in Reference~\cite{Bertone:2004pz}\footnote{We note that there are minor errors in Equation 204 of Reference~\cite{Bertone:2004pz} related to the squark and Higgs masses.  To correct these minor errors, we followed References~\cite{Falk:1999mq}~\&~\cite{Ellis:2000ds} and References~\cite{Jungman:1995df},~\cite{Falk:1998xj},~\&~\cite{Falk:1999mq}.} and using the values in Table~\ref{tab:form}. Surprisingly, we found $\mathcal{O}(10\%)$ differences between our tree-level calculation and that of the codes', implying substantial loop contributions that are different between the two codes.

\subsection{A-funnel}
The results for the relic density, current annihilation cross section, and scattering cross section of the well-tempered neutralino benchmark are shown by pentagons in the upper left, upper right, and lower left panels of Figure~\ref{fig:pmssm}, respectively. 

We first discuss the relic density. From the upper left panel of Figure~\ref{fig:pmssm} and Table~\ref{tab:pomega}, it is evident that there is a large variation, of more than factor of two, among the different pipelines, though the variation is $\lesssim 2$ if one neglects the  Snowmass and Snowmass* pipelines. We note that for a given spectrum generator, say \softsusy{}, there is some difference in the calculation performed by \micromegas{} (solid magenta pentagon) versus \darksusy{} (hollow magenta pentagon). This difference amounts to an uncertainty of 69\% and 37\% for \spheno{} and \softsusy{}, respectively, both of which far exceed the experimental uncertainty in the measurement of the dark matter abundance. 

The resonance region is notoriously sensitive to the approximations used to compute the relic density, especially for sharp peaks. The sharpness parameter in our case (following the same notation as Reference~\cite{Griest:1990kh}) is 
\be
\epsilon \,\, \equiv \,\, \left( \frac{\Gamma_A}{m_A} \right)^2 \,\, \sim 5 \times 10^{-6} \,\,.
\ee
Reference~\cite{Griest:1990kh} compared various approximation schemes in the relic density calculation (such as Taylor expansion in $v$) to a full numerical computation for values of $\epsilon$ near this value. Depending on the approximation, the relic density can vary over several orders of magnitude. Even for a full numerical computation, the resonance region is sharp enough that a factor of $\sim 2$ can easily appear, unless there is an exact matching of calculation. 

Finally, we note that there is a significant difference between the \softsusy{}-\feynhiggs{}-\micromegas{} and \softsusy{}-\susyhd{}-\micromegas{} pipelines for the relic density of the $A$-funnel benchmark point, as can be seen by comparing Tables~\ref{tab:pomegafeyn} and~\ref{tab:pomegahd}.  This discrepancy is due to the inclusion of the pseudoscalar width, as in Eq.~\ref{funnelcs}.  \feynhiggs{} calculates the pseudoscalar width, which is read in by \micromegas{} (but not \darksusy{}, as discussed below).  However, \susyhd{} does not calculate the width, so any program further down the pipeline either takes the width approximation from the spectrum generator or calculates the width itself.  \spheno{} does estimate a width with reasonable agreement between its width and that calculated by \feynhiggs{}, yielding decent agreement between the \spheno{}-\feynhiggs{}-\micromegas{} and \spheno{}-\susyhd{}-\micromegas{} pipelines.  But \softsusy{} does not report a width that can be used by \micromegas{}, so for the \softsusy{}-\susyhd{}-\micromegas{} pipeline, \micromegas{} calculates its own width, which differs by nearly a factor of 1.7 from that calculated by \feynhiggs{}.  This leads to a discrepancy of nearly a factor of 2 in the relic abundances, as can be verified, for example, by evaluating Eq.~41 of Ref.~\cite{Griest:1990kh}. We do not see a discrepancy in the relic densities 
of any of the \darksusy{} pipelines because \darksusy{} always calculates all relevant particle widths, since, as mentioned above, up to version 5.1.2 \darksusy{} does not read in SLHA decay blocks.

The annihilation cross section today is given by an even sharper peak in $\langle\sigma v\rangle$ centered at $m_A = 2 m_{\tilde{\chi}_1^0}$. From the upper right panel of Fig.~\ref{fig:pmssm}, and comparing Tables~\ref{tab:pomegafeyn} and~\ref{tab:psigvfeyn}, we see that the values reported by the various pipelines are in agreement with what we expect from the upper left panel; that is, there is a factor of $\sim2$ spread in the annihilation cross sections, similar to the factor of $\sim2$ spread in the relic densities (with the largest annihilation cross section corresponding to the smallest relic abundance, and so forth).
We note that the current annihilation cross sections in the upper right panel of  Fig.~\ref{fig:pmssm} are a factor $\sim 5$ reduced compared to their values in the early Universe, 
used to calculate the abundances in the upper left panel of Fig.~\ref{fig:pmssm}. This happens due to the thermal broadening of the resonance region in the early Universe.

The SI scattering cross sections reported in the lower left panel of Fig.~\ref{fig:pmssm} match very closely due to the close agreement in the relevant masses and higgsino fractions. For a given spectrum generator, say \softsusy{}, there is some disagreement between the scattering cross section computed by \micromegas{} (solid magenta pentagon) versus \darksusy{} (hollow magenta pentagon). As for other benchmarks, this can be attributed to the different form factors, as in Table~\ref{tab:form} and the way in which loop effects are incorporated.

\subsection{Summary: Broad Trends in the pMSSM Analysis}

It is instructive to look back at our analysis and draw some broad conclusions. The dark matter models studied in this Section are a well-known subset of pMSSM benchmarks that satisfy the observed dark matter relic density. These benchmarks have been used in numerous studies, typically relying on one of the pipelines described in our work. Furthermore, connections between supersymmetric model building and cosmology often concern regions of parameter space based around one of the (co)annihilation mechanisms that our benchmarks capture.

It should be clear from our work that the theoretical uncertainty in the relic density calculation using standard pipelines far exceeds the experimental uncertainty. From the upper left panel of Figure~\ref{fig:pmssm}, it is apparent that only a minority of pipeline choices for any given benchmark actually fall within the red band that delineates the Planck range. For the coannihilation and funnel models, this spread is especially broad, with theoretical calculations yielding results that can vary by as much as 300\%. The spread is somewhat lower for the well-tempered neutralino and pure higgsino benchmarks, but even in those cases it far exceeds the experimental uncertainty. We note also that there can be significant spread in the relic density due to updates to the software packages even for the same sequence of calculators.

There are several underlying reasons for the discrepancies in the relic density and other dark matter observables among the pipeline choices:

\begin{itemize}

\item \textbf{Small variations in the spectrum} -- Coannihilations and $A$-resonance models are critically dependent on the relative masses of the dark matter particle and other light supersymmetric states. Within a pMSSM framework, the spectrum generators can easily produce 1\% - 2\% variation in the low energy spectrum of squarks, stops, or the pseudoscalar Higgs, leading to a large variation of the relic density, evident in the upper left panel of Figure~\ref{fig:pmssm}. However, the annihilation cross section in the present Universe is far less dependent on the masses of other light states, since coannihilation channels become irrelevant. This is reflected in the much greater convergence among pipelines in the upper right panel of Figure~\ref{fig:pmssm}.

\item \textbf{Composition of the lightest neutralino} -- The well-tempered neutralino framework depends critically on the higgsino fraction of the dark matter for its relic density. We see from Table~\ref{tab:pneutspectra} that the spectrum generators can vary by up to 20\% in their calculation of the higgsino fraction relative to each other. Furthermore, the higgsino fraction plays a crucial role in the SI scattering cross section, most evident in the well-tempered benchmark. 

\item \textbf{LO and NLO calculation of coannihilation channels} -- For a given spectrum, there is wide variation ($\sim 50$\%) in the relic density reported by \micromegas{} and \darksusy{} (discrepancies between solid and hollow points of the same shape and color in the upper left panel of Fig.~\ref{fig:pmssm}), especially in scenarios where coannihilation channels become important. This stems from differences in the tree level computation implemented in these programs. Moreover, as we have pointed out, the incorporation of NLO SUSY-QCD will further change the relic density calculation, by up to as much as 20\%.

\item \textbf{Differences in form factors} -- The form factors employed by \micromegas{} and \darksusy{} are displayed in Table~\ref{tab:form}. These differences affect the SI scattering cross sections reported in the lower left panel of Figure~\ref{fig:pmssm} for a given spectrum calculator (discrepancies between solid and hollow points of the same shape and color). For a given spectrum, using the the form factors of \darksusy{} in \micromegas{}, we find a definitive shift in the SI cross sections, bringing them into closer agreement.

\item \textbf{NLO effects in scattering cross section} -- Even accounting for the differences in spectrum and form factors, we see that different dark matter calculators report different SI scattering cross sections, especially for very low cross sections. These differences are likely coming from the fact that \darksusy{} implements an effective Lagrangian in the heavy squark limit following Reference~\cite{Griest:1988ma} (see Reference~\cite{Bergstrom:1995cz} and references therein for details) while \micromegas{} implements the full one-loop Lagrangian following Reference~\cite{Drees:1993bu}. For example, the pure higgsino or pure bino benchmarks in the lower left panel of Figure~\ref{fig:pmssm} show a lot of variation.   The theory calculations for pure higgsino and wino scattering cross sections have only converged recently~\cite{Cheung:2012qy}. The discrepancies among the pipelines is likely to become a pressing issue in the future, when experimental sensitivity reaches the relevant cross sections.

\end{itemize}

\begin{table}
  \centering
  \includegraphics[scale=0.9]{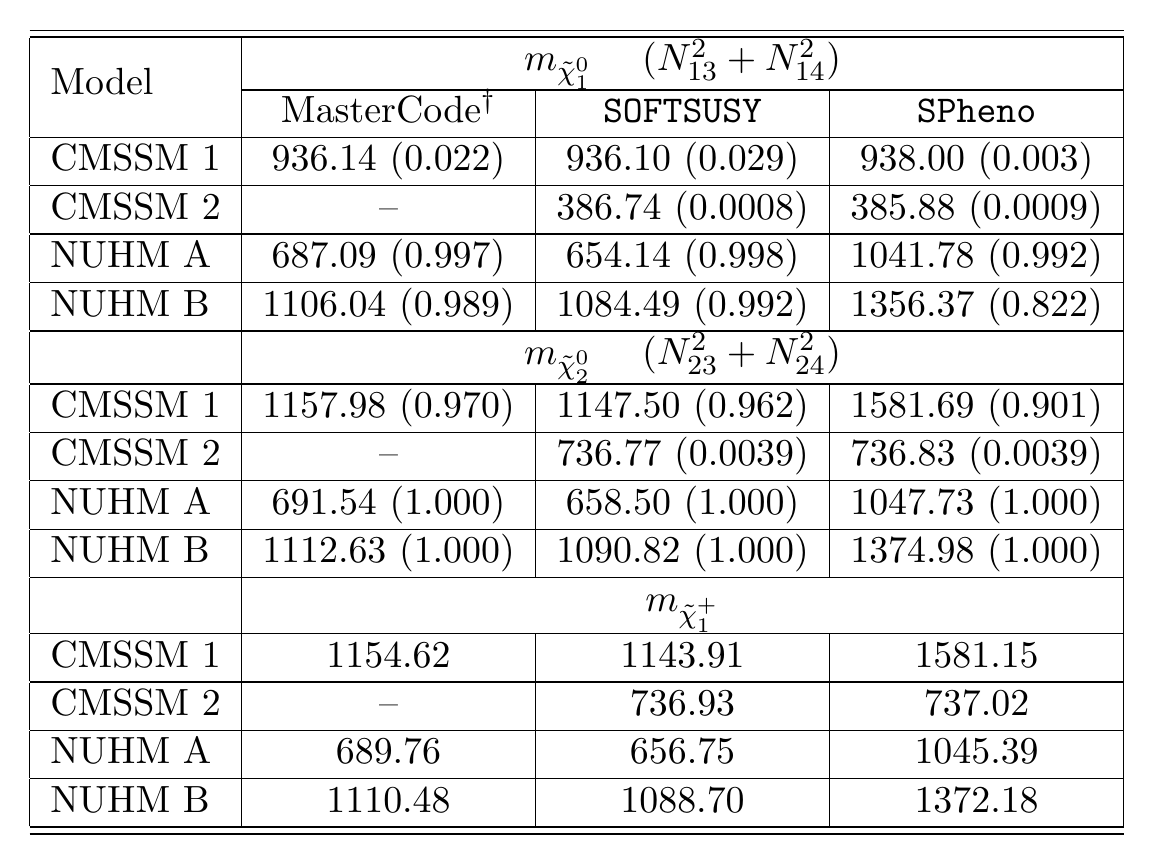}
  \caption{Masses, in GeV, of the lightest and second-lightest neutralinos and the lighter chargino for the GUT benchmark points. The corresponding higgsino fraction is given in parentheses.}
  \label{tab:mneutspectra}
\end{table}

\section{Results: GUT Analysis}
\label{sec:GUT}

In the previous sections, we have presented our analysis of several standard pMSSM benchmark scenarios. For electroweak-scale models like the pMSSM, the supersymmetric mass spectrum is used as an input for the spectrum generators at low energies, and the final sparticle spectrum is used as an input for the dark matter calculators. Thus, there is very little running of the sparticle masses and the results reported by different spectrum generators are generally in good agreement. In this section, we will analyze dark matter benchmark points in the context of supersymmetric models with boundary conditions for soft terms specified at the GUT scale. In particular, we will study two models: CMSSM and NUHM. In this case, the effects of RG running performed by the two spectrum calculators are expected to become more important, and can substantially affect the dark matter observables.
We begin this Section by first discussing the sparticle spectra of the GUT benchmark models.

\begin{table}
  \centering
      \includegraphics[scale=0.9]{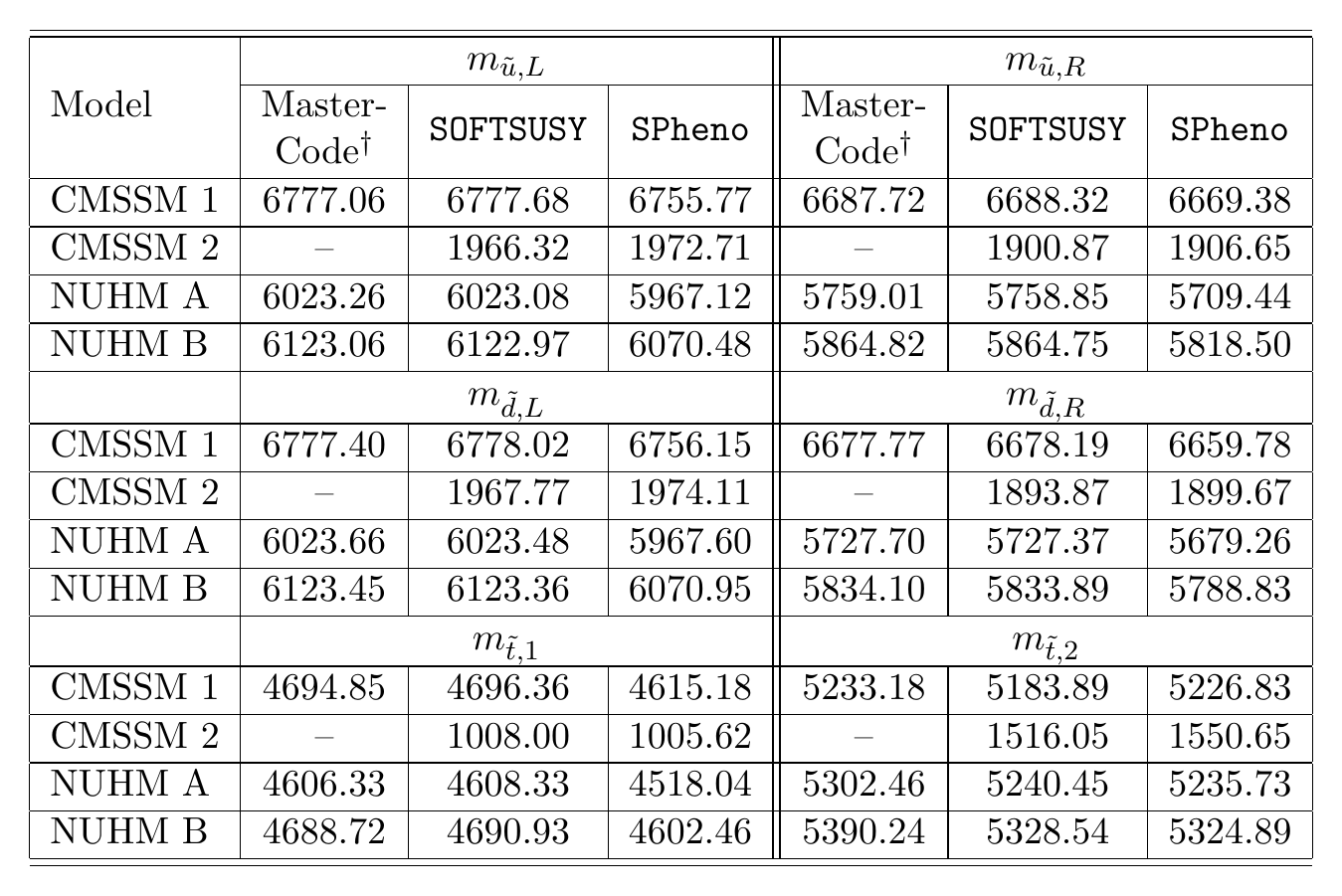}
  \caption{Masses, in GeV, of the squarks for the GUT benchmark points.}
  \label{tab:msquarks}
\end{table}


The GUT models are defined in Table~\ref{tab:gutpoints}. The low energy spectrum of neutralinos and squarks generated by \softsusy{} and \spheno{} are shown in Tables~\ref{tab:mneutspectra} and \ref{tab:msquarks}, and Higgs masses are shown in Table~\ref{tab:mhiggs}. Masses presented in Tables~\ref{tab:mhiggs1feyn}~\&~\ref{tab:mhiggs2feyn} are computed by \feynhiggs{} while those in Tables~\ref{tab:mhiggs1hd}~\&~\ref{tab:mhiggs2hd} \susyhd{}. 
The higgsino mass parameter $\mu$ is shown in Table~\ref{tab:mufeyn}. We also display in Tables~\ref{tab:mneutspectra}-\ref{tab:mufeyn} the values obtained via the MasterCode$^\dag$ pipeline, as described in Table~\ref{tab:pipelines}. 

From Table~\ref{tab:mneutspectra}, we see that for the CMSSM points, the lightest neutralino is mainly bino, while for the NUHM points, it is mainly higgsino. The higgsino fraction calculated by the different spectrum generators has appreciable differences in the cases of the CMSSM 1 and NUHM B, which will significantly affect the dark matter observables reported for the two pipelines, as we shall see. Even in the case of the NUHM A point, the small difference in higgsino fraction ($<1\%$ difference between \softsusy{} and \spheno{}) will be important.

There is good agreement between \softsusy{} and \spheno{} for the mass of the dark matter in the CMSSM cases. However, there is a vast disagreement in the mass of dark matter between \softsusy{} and \spheno{} for both NUHM points. There are also large discrepancies in the mass of the second lightest neutralino and the charginos in all cases except CMSSM 2. These discrepancies all stem from differences in the value of $\mu$, as is evident from Table~\ref{tab:mufeyn}. From Table~\ref{tab:mhiggs}, we see that there are also large discrepancies in the mass of the pseudoscalar Higgs $A$, which is critical for the dark matter relic density computation in the $A$-funnel region of parameter space. On the other hand, the squark spectrum agrees among different generators quite well, as is evident from Table~\ref{tab:msquarks}, although there can be variations of up to $\sim 2$ \% in the calculation of squark masses. Even these $\sim 2$ \% discrepancies will be important in the following analysis.

  \begin{table}
  \centering
  \subfloat[][$m_h$ and $m_H$ as computed via the \feynhiggs{} branch of the pipeline.\label{tab:mhiggs1feyn}]{%
      \includegraphics[scale=0.9]{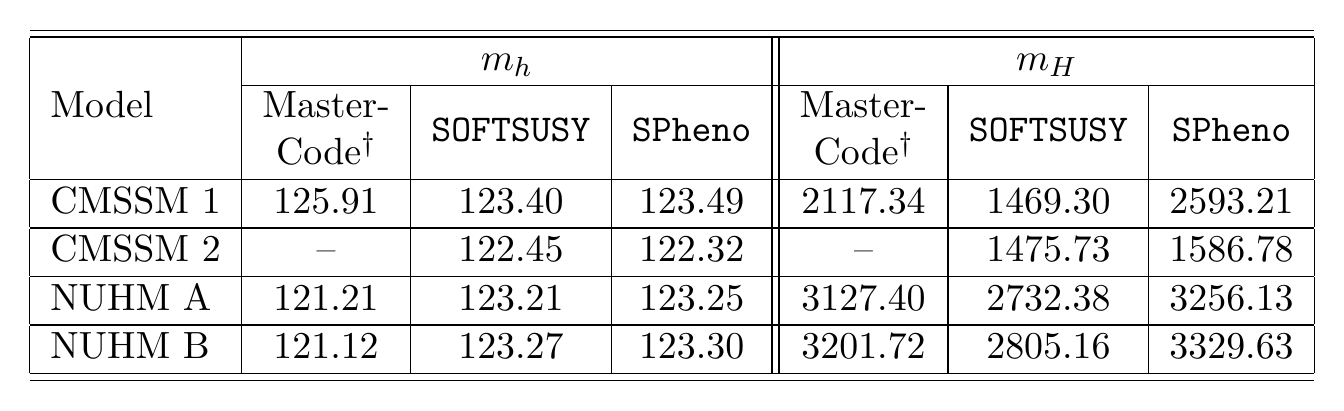}
  } \hfill
  \subfloat[][$m_A$ and $m_{H^{\pm}}$ as computed via the \feynhiggs{} branch of the pipeline.\label{tab:mhiggs2feyn}]{%
      \includegraphics[scale=0.9]{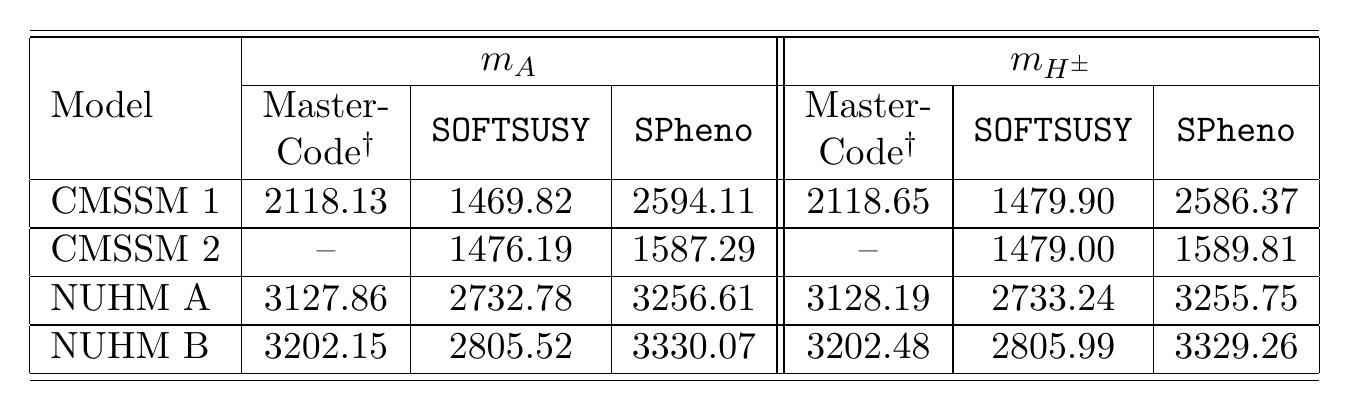}
  } \hfill
  \subfloat[][$m_h$ and $m_H$ as computed via the \susyhd{} branch of the pipeline.\label{tab:mhiggs1hd}]{%
      \includegraphics[scale=0.9]{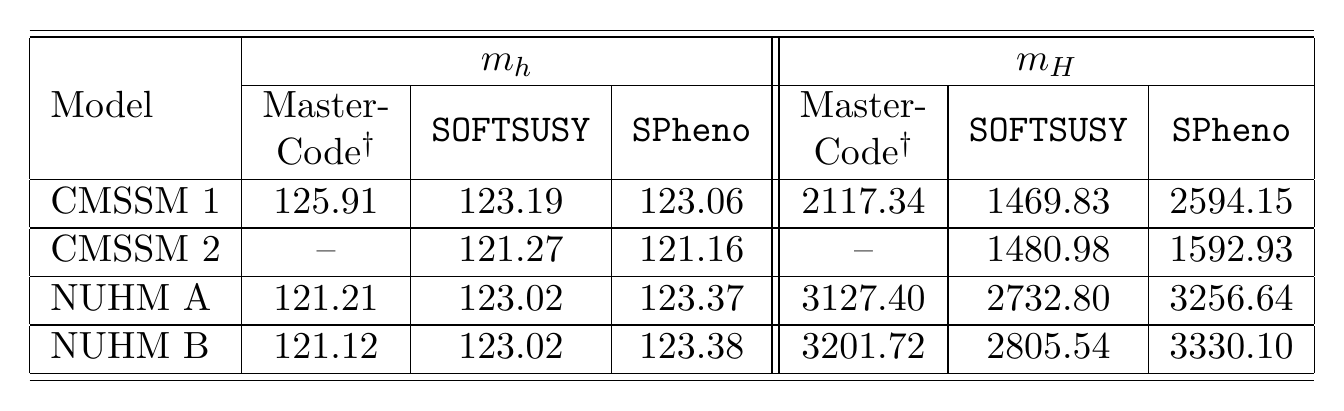}
  } \hfill
  \subfloat[][$m_A$ and $m_{H^{\pm}}$ as computed via the \susyhd{} branches of the pipeline.\label{tab:mhiggs2hd}]{%
      \includegraphics[scale=0.9]{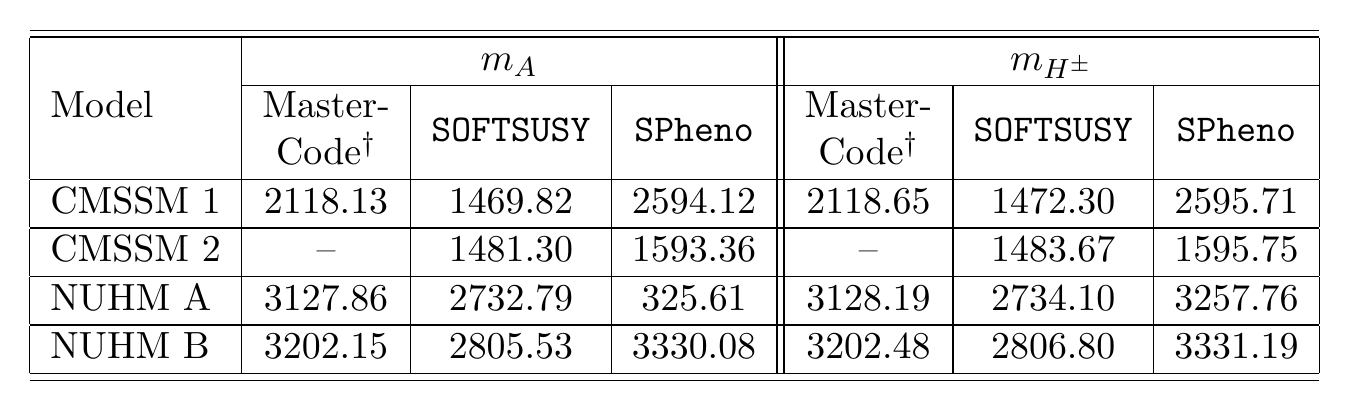}
  }
  \caption{Masses, in GeV, of the CP-even Higgses (Tables~\ref{tab:mhiggs1feyn} and~\ref{tab:mhiggs1hd}) and the CP-odd and charged Higgses (Tables~\ref{tab:mhiggs2feyn} and~\ref{tab:mhiggs2hd}) for the GUT scale points. Masses presented in Tables~\ref{tab:mhiggs1feyn}~\&~\ref{tab:mhiggs2feyn} are computed by \feynhiggs{} while those in Tables~\ref{tab:mhiggs1hd}~\&~\ref{tab:mhiggs2hd} \susyhd{}.}
  \label{tab:mhiggs}
\end{table}

We thus see that the largest discrepancies seem to occur in the calculation of the electroweak symmetry breaking (EWSB) sector between different spectrum generators. We now turn to a more detailed study of these differences: as a prelude to our investigation of dark matter observables, we discuss in detail the differences between the calculation of the higgsino mass parameter $\mu$ and the pseudoscalar Higgs mass $m_A$ by the different spectrum generators. Finally, we discuss the dark matter observables obtained from the different pipelines.

\subsection{Comparison of EWSB Sectors} \label{sec:ewsb}

Here we explore the EWSB calculations performed by the different spectrum generators. The differences in the EWSB calculations are particularly exacerbated at large values of $\tan{\beta}$ and $m_0$, which is the regime where our GUT benchmark models CMSSM 1 and NUHM A/B lie. We discuss these issues in this section.

The RGE's for the higgsino mass parameters in the MSSM are, following the notation of Reference~\cite{Martin:1997ns},
\bea \label{muone}
16 \pi^2 \frac{d}{dt}m^2_{H_u}  &=& 3 X_t - 6 g^2_2|M_2|^2 - \frac{6}{5} g^2_1|M_1|^2 + \frac{3}{5}g^2_1 S \nonumber \\ 
16 \pi^2 \frac{d}{dt}m^2_{H_d}  &=& 3 X_b + X_{\tau} - 6 g^2_2|M_2|^2 - \frac{6}{5} g^2_1|M_1|^2 - \frac{3}{5}g^2_1 S \,\,.
\eea
In the above equations, 
\bea
X_{(t,b,\tau)} &=& 2 \left| y^2_{(t,b,\tau)} \right| \, \left( m^2_{H_{(u,d,d)}} + m^2_{(Q_3,Q_3,L_3)} + m^2_{(\bar{u}_3,\bar{d}_3,\bar{e}_3)} + A^2_{(t,b,\tau)} \right) \nonumber \\ 
S &=& m^2_{H_u} - m^2_{H_d} + Tr \left(\mathbf{m^2_{Q}} - \mathbf{m^2_{L}} - 2 \mathbf{m^2_{\bar{u}}} + \mathbf{m^2_{\bar{d}}} + \mathbf{m^2_{\bar{e}}}\right),
\eea
in standard notation. The higgsino mass parameter is given, in the large $\tan{\beta}$ limit, by
\be \label{mutwo}
\mu^2 \, \sim \, \frac{m^2_{H_d} - m^2_{H_u} \tan^2{\beta}}{\tan^2{\beta} - 1} - \frac{1}{2}m^2_Z , \,\,
\ee
where all quantities are defined at $m_Z$. The pseudoscalar Higgs mass is given at tree level by
\be \label{mAexprth}
m^2_A \, = 2 |\mu |^2 + m^2_{H_u} + m^2_{H_d},
\ee
with all quantities defined at the scale $M_{SUSY}$, and the quantities on the right hand side of Eq.~\ref{mAexprth} related to those at $m_Z$ by radiative corrections.
Obviously, the calculated value of $m_A$ depends on the computation of $\mu, \, m^2_{H_u}$, and $m^2_{H_d}$.


  \begin{table}
    \centering
    \includegraphics[scale=0.9]{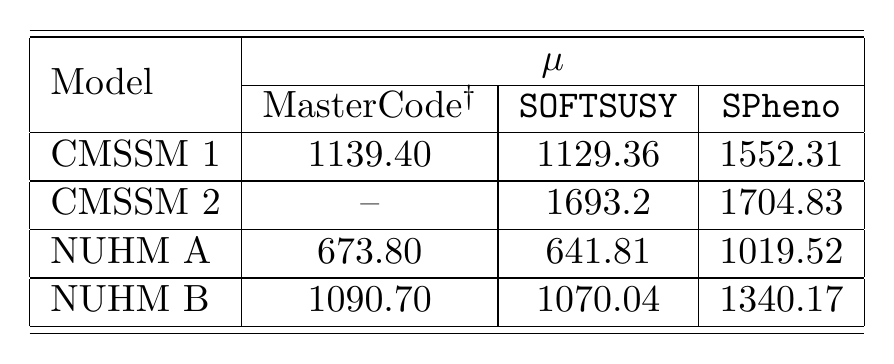}
    \caption{Higgsino mass parameter, $\mu$, after \feynhiggs{}.}
    \label{tab:mufeyn}
 \end{table}

From Eq.~\ref{muone}, Eq.~\ref{mutwo} and Eq.~\ref{mAexprth}, it is clear that the values of $\mu$  and $m_A$ reported by the programs will be greatly affected by their calculation of the top and bottom Yukawas $y_t$  and $y_b$, as well as the calculation of squark and stop masses that enter into $X_t$ and $X_b$. In Table~\ref{tab:msquarks}, there are variations of $\sim 2$\% in the stop masses between \softsusy{} and \spheno{}. In fact, \softsusy{} consistently reports higher values of the stop masses than \spheno{} across benchmark models. There are also variations in $y_t$ and $y_b$ between the programs, as studied in~\cite{Allanach:2003jw}. These factors result in vastly different values of $\mu$ and $m_A$, especially for large values of $m_0$ where the squark and Yukawa calculations differ substantially.

\begin{figure}
  \centering
  \includegraphics[width=0.98\textwidth]{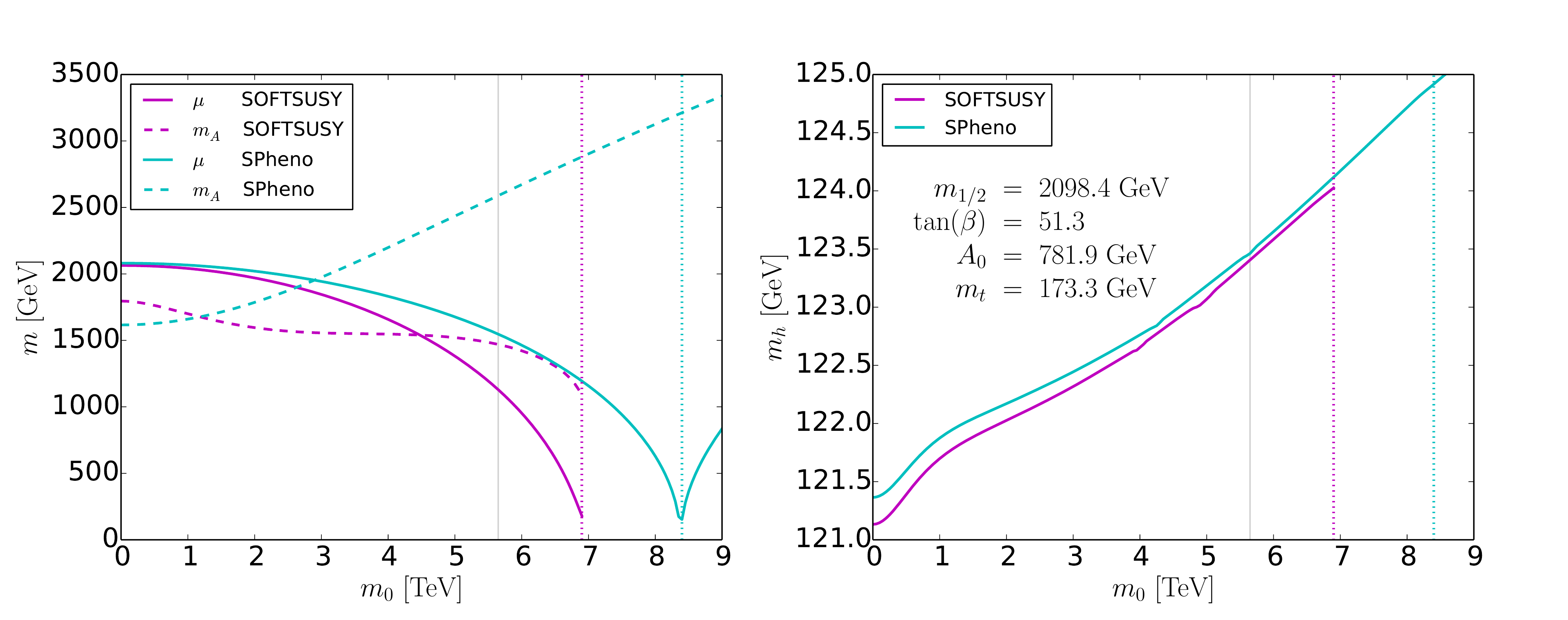}
  
  \caption{We show the higgsino mass parameter, $\mu$, and the pseudoscalar mass, $m_A$, (left panel), as well as $m_h$ (right panel), each as functions of $m_0$, as computed by \feynhiggs{} with spectral input from \softsusy{} (magenta) or \spheno{} (cyan) for the CMSSM.  The CMSSM 1 benchmark is denoted with a solid grey line.  Vertical dotted lines indicate the value of $m_0$ above which $\mu$ becomes unphysical ($\mu^2<0$). Data near the benchmarks is presented in Table~\ref{tab:cmssmscan}.}
  \label{fig:cmssm_mu_m0}
\end{figure}

\begin{figure}
  \centering
  \includegraphics[width=0.98\textwidth]{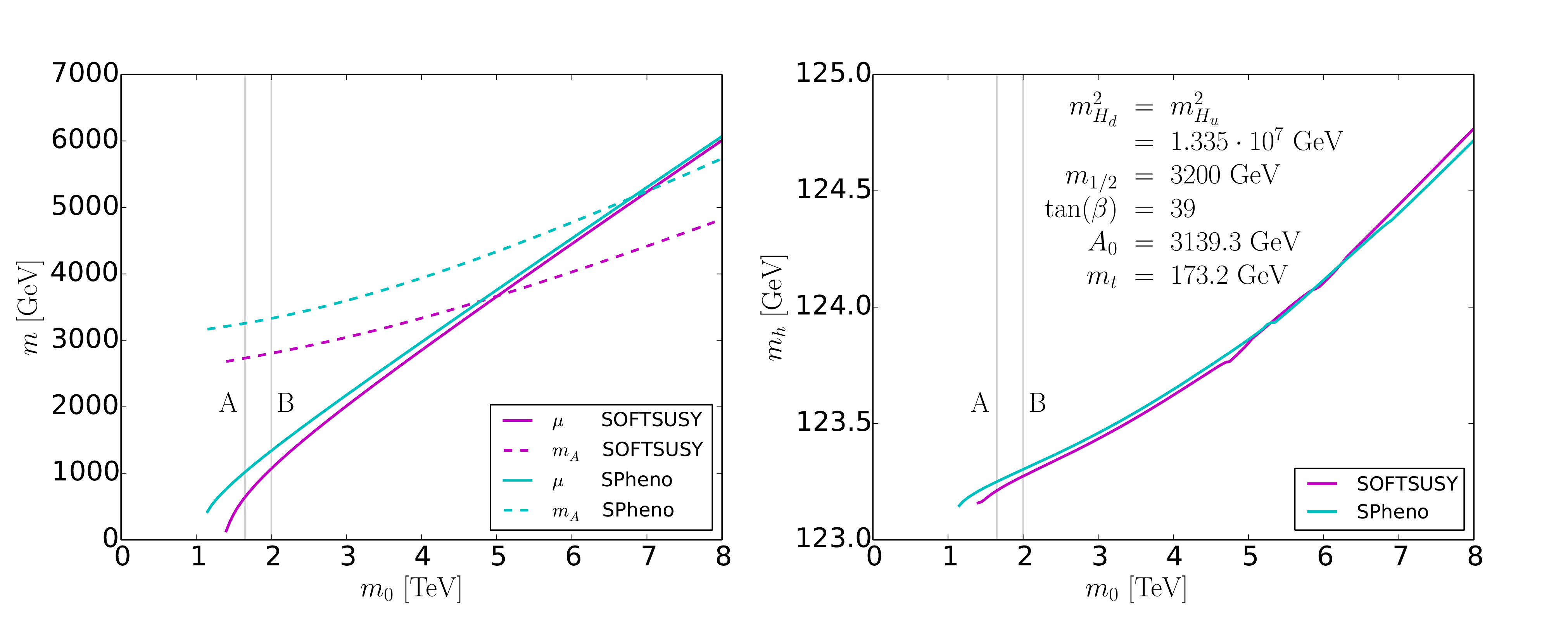}
  \caption{We show the higgsino mass parameter, $\mu$, and the pseudoscalar mass, $m_A$, (left panel), as well as $m_h$ (right panel), each as functions of $m_0$, as computed by \feynhiggs{} with spectral input from \softsusy{} (magenta) or \spheno{} (cyan) for the NUHM.  The NUHM A and B benchmarks are denoted by solid grey vertical lines, as labeled.}
  \label{fig:nuhm_mu_m0}
\end{figure}

In Figures~\ref{fig:cmssm_mu_m0} and~\ref{fig:nuhm_mu_m0}, we show the resulting variations the higgsino mass parameter, $\mu$, and the pseudoscalar mass, $m_A$, (left panels), as well as $m_h$ (right panels), each as functions of $m_0$, as computed by \feynhiggs{} with spectral input from \softsusy{} (magenta) and \spheno{} (cyan) near the CMSSM 1 and NUHM benchmarks. Our benchmark points are denoted with a solid grey line in each panel.  Vertical dotted lines indicate the values of $m_0$ above which $\mu$ becomes unphysical ($\mu^2<0$). Data near the benchmarks is presented in Table~\ref{tab:cmssmscan}.

We now discuss some general features of the Figures. From the left panel of Figure~\ref{fig:cmssm_mu_m0}, we see that for both generators, $\mu$ decreases as $m_0$ increases. We can understand this as follows. From Eq.~\ref{muone}, we see that increasing the scalar masses makes the RG running of both $m^2_{H_u}$
 and $m^2_{H_d}$ steeper, decreasing their values at low scales. On the other hand, increasing the scalar masses within the CMSSM also increases the boundary values of $m^2_{H_u}$ and $m^2_{H_d}$ at the GUT scale. The low-scale values of $m^2_{H_u}$ and $m^2_{H_d}$ are thus determined by these two competing effects. For our selection of $m_{1/2}$ and $\tan{\beta}$, we have checked that the cumulative effect is to decrease $m^2_{H_u}$ with increasing $m_0$ near the benchmark point, for both  \softsusy{} and \spheno{}. We have also verified this using the approximate relations for the renormalization group running given in Reference~\cite{Drees:1995hj}.
 
 On the other hand, we have found that near the benchmark, the effect of increasing $m_0$ is to increase $m^2_{H_d}$. This is because while increasing $m_0$ increases the slope of the renormalization group running from Eq.~\ref{muone},  this increase is suppressed compared to the $m^2_{H_u}$ case by the small value of the bottom Yukawa. The cumulative effect is that the increase in the boundary value of $m^2_{H_d}$ for increasing $m_0$ dominates, so $m^2_{H_d}$ increases with increasing $m_0$.

We see from Eq.~\ref{mutwo} that $m^2_{H_u}$ and $m^2_{H_d}$ contribute with opposite signs to $\mu$. However, the dominant contribution is from $m^2_{H_u}$, since $m^2_{H_d}$ is suppressed by the large value of $\tan{\beta}$. Thus, following the behavior of $m^2_{H_u}$, $\mu$ too decreases with increasing $m_0$.

The values of $\mu$ (solid curves) from \softsusy{} and \spheno{} start to diverge radically for $m_0 > 2$ TeV in the left panel of Figure~\ref{fig:cmssm_mu_m0}. This is the regime where differences in the squark masses and the top Yukawa calculated by the two spectrum generators start to become important in determining $\mu$. From Table~\ref{tab:msquarks}, we see that \softsusy{} produces heavier stop masses than \spheno{} for the same CMSSM model point. Thus, $\mu$ runs to smaller values faster in \softsusy{} compared to \spheno{}. The values are $\mu = 1129$ GeV for \softsusy{} and $\mu = 1552$ GeV for \spheno{} at the CMSSM 1 benchmark. 

From the right panel of Figure~\ref{fig:cmssm_mu_m0}, we see that for both programs the mass of the lightest CP-even Higgs increases with increasing $m_{0}$. This is expected, due to the usual loop corrections to the Higgs mass. We also note that \spheno{} reports a slightly larger Higgs mass than \softsusy{} due to a combination of the low energy values of the stop mass and the trilinear coupling. Finally, we point out that the Higgs mass calculation is relatively robust to uncertainties in the EWSB calculations, since $m_h$ is sensitive only to $m_A$ (not $\mu$ independently) at tree level, which is reflected in the behavior of $m_h$ at very large $m_0$ near where $\mu$ becomes unphysical.  We note that the uncertainties in the EWSB calculations tend to cancel each other in the calculations of $m_h$, while, in contrast, they do not cancel each other in the calculation of $m_H$, which tracks $m_A$ quite closely.

We now move on to a discussion of the pseudoscalar Higgs mass $m_{A}$ (dashed curves) in the left panel of Figure~\ref{fig:cmssm_mu_m0}. From the tree level relation for $m_A$ in Eq.~\ref{mAexprth}, we expect that the value of $m_A$ reported will depend on the relative magnitudes of $\mu$, $m^2_{H_u}$, and $m^2_{H_d}$ reported by the spectrum generators. The values of $m_A$ returned by \softsusy{} \emph{decrease} steadily with increasing $m_0$. However, the values of $m_A$ given by \spheno{} instead \emph{increase} with increasing $m_0$. At the benchmark value of $m_0$, we have $m_A = 1469$ GeV given by \softsusy{} and $m_A = 2595$ GeV given by \spheno{}.

The NUHM benchmarks are somewhat different from the CMSSM case discussed above. From the left panel of Figure~\ref{fig:nuhm_mu_m0}, we see that with increasing $m_0$, the value of $\mu$ increases. This is due to the fact that although $m^2_{H_u}$ and $m^2_{H_d}$ individually decrease with increasing $m_0$, their difference increases with increasing $m_0$. For the NUHM A point, the values are $\mu = 642$ GeV from \softsusy{} and $\mu = 1020$ GeV from \spheno{}. For the NUHM B point, the values are $\mu = 1070$ GeV from \softsusy{} and $\mu = 1340$ GeV from \spheno{}. The values of the pseudoscalar and lightest CP-even Higgs masses increase with increasing $m_0$ for both spectrum generators, as can be seen from the left and right panel, respectively, of Figure~\ref{fig:nuhm_mu_m0}.

Finally, the CMSSM 2 point, which is a stau coannihilation model, is largely insensitive to uncertainties in the EWSB sector.  The dark matter physics of the CMSSM 2 benchmark primarily concerns the LSP, which is strongly bino-like with good agreement among the pipelines, and the lighter stau,  with masses of 538.29 GeV and 390.01 GeV from \softsusy{} and \spheno{}, respectively.  Since the lightest neutralino mass for the CMSSM 2 is $\sim 386$ GeV for both \softsusy{} and \spheno{}, this means that only \spheno{} pipelines represent true coannihilation models, while \softsusy{} pipelines do not coannihilate efficiently enough to achieve the correct relic abundance.

\subsection{Dark Matter Observables}

\begin{figure} 

    \includegraphics[width=0.98\textwidth]{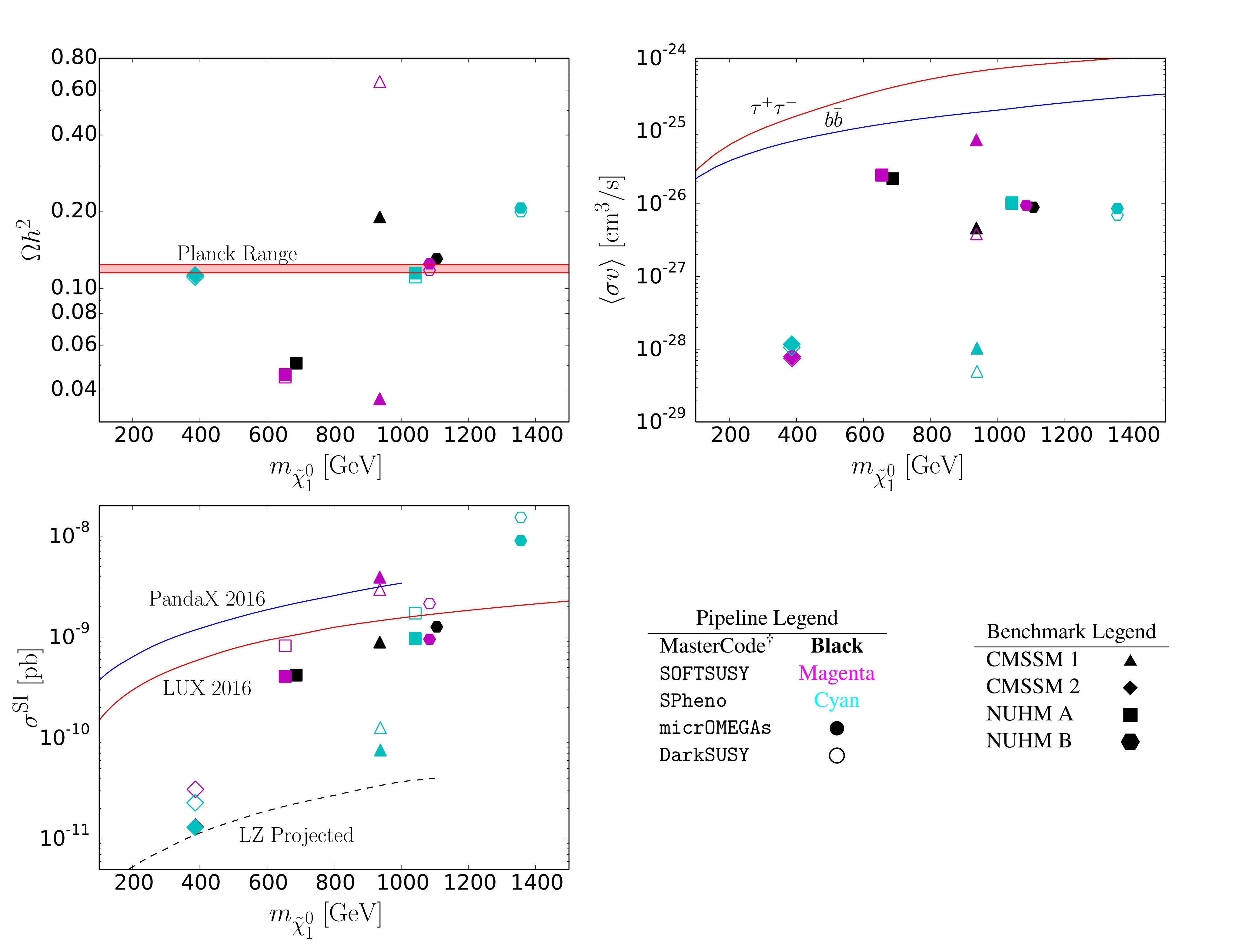}
  
  \caption{GUT model results.  The upper left, upper right, and lower left panels show the neutralino relic density, annihilation cross section today, and the SI neutralino-nucleon elastic scattering cross section, respectively, all as functions of the dark matter mass, with a common legend in the lower right panel. 
  We note that relic density for some pipelines yields values far larger than those plotted here, and for this reason the CMSSM 1 \spheno{} points and the CMSSM 2 \softsusy{} points do not appear in the upper left panel.  For comparison, the Planck 3-sigma range for the dark matter abundance~\cite{Ade:2015xua} is highlighted in red in the upper left panel, the limits on the annihilation cross section today from Fermi-LAT's 6-year analysis of dwarf spheroidal galaxies~\cite{Ackermann:2015zua} for annihilation to $\tau^+\tau^-$ (red) and $b\bar{b}$ (blue) are shown in the upper right panel, and exclusion limits from LUX (red)~\cite{Akerib:2016vxi}, PandaX (blue)~\cite{Tan:2016zwf}, and LZ (projected; black)~\cite{Akerib:2015cja} are shown in the lower left panel.}
  \label{fig:msugra}
\end{figure}

In this Section, we study the dark matter observables for the GUT benchmarks. For each of the CMSSM and NUHM benchmark models, we discuss the relic density, the annihilation cross section today, and the predicted scattering cross section, all of which are plotted in Figure~\ref{fig:msugra}.  As in Figure~\ref{fig:pmssm}, for comparison, the Planck 3-sigma range for the dark matter abundance~\cite{Ade:2015xua} is highlighted in red in the upper left panel, the limits on the annihilation cross section today from Fermi-LAT's 6-year analysis of dwarf spheroidal galaxies~\cite{Ackermann:2015zua} for annihilation to $\tau^+\tau^-$ (red) and $b\bar{b}$ (blue) are shown in the upper right panel, and exclusion limits from LUX (red)~\cite{Akerib:2016vxi}, PandaX (blue)~\cite{Tan:2016zwf}, and LZ (projected; black)~\cite{Akerib:2015cja} are shown in the lower left panel. Here, we introduce a new set of unique shapes to denote the various pipelines, but follow the same colour scheme as in Figure~\ref{fig:pmssm} -- black is used again for our implementation of the original pipeline (MasterCode$^\dag$, rather than Snowmass$^\dag$)\footnote{There are no green markers to indicate an updated MasterCode pipeline, since the updated MasterCode pipeline is the same as our \softsusy{}-\feynhiggs{}-\micromegas{} pipeline.}.

\subsubsection{CMSSM Benchmarks}

We first examine the relic density as plotted in the upper left panel of  Figure~\ref{fig:msugra} and
tabulated in Table~\ref{tab:momega}.

For the CMSSM 1 point, the mass of the lightest neutralino from Table~\ref{tab:mneutspectra} is $m_{\neu{1}} = 936.10$ GeV from \softsusy{}, and $m_{\neu{1}} = 938.0$ GeV from \spheno{}. The dark matter is mostly bino in both cases. It is, however, the mass of the second lightest neutralino that differs radically between the two programs. For \softsusy{}, we have $m_{\neu{2}} = 1147.50$ GeV, while for \spheno{}, we have $m_{\neu{2}} = 1581.69$ GeV. The second lightest neutralino is mostly higgsino. The huge discrepancy in masses is due to the very different values of $\mu$ reported by the two programs, as discussed above and shown in Figure~\ref{fig:cmssm_mu_m0}. The large difference in $\mu$ is also reflected in the different higgsino fractions of the lightest neutralino. From \softsusy{}, the higgsino fraction in $\neu{1}$ is around $\sim 3$\%, while from \spheno{}, the higgsino fraction is only $0.3$\%. We note another large discrepancy among the results from the different pipelines is that the values of $m_A$ obtained from \softsusy{} and \spheno{} are 1470 GeV and 2594 GeV, respectively.

These differences in the spectra have a profound effect on the relic density. From the upper left panel of Figure~\ref{fig:msugra}, we find that the  \softsusy{}-\micromegas{} pipeline (solid magenta triangles) and \softsusy{}-\darksusy{} pipeline (hollow magenta triangles) give values for the relic density $\Omega h^2 = 0.037$ and $\Omega h^2 = 0.648$, respectively, while the pipelines that involve \spheno{} as the spectrum calculator give relic densities that are  $\Omega h^2 \gtrsim 10$, with the value returned from the \darksusy{} pipeline larger than that from the \micromegas{} pipeline by a factor of 2. 

In fact, the CMSSM 1 benchmark model is not even a cosmologically-favored point (at least within a thermal history) if one uses \spheno{} as the spectrum calculator. The lightest neutralino in that case is an almost pure bino, far away from the $A$-resonance, and without any possible contributions from coannihilation channels. The pipelines involving \spheno{} give a relic density that is $\gtrsim 2$ orders of magnitude larger than those given by \softsusy{} pipelines. 

For pipelines involving \softsusy{}, the CMSSM 1 benchmark falls approximately into the category of well-tempered dark matter\footnote{The original MasterCode CMSSM best fit point from~\cite{Buchmueller:2013rsa} was primarily an $A$-funnel point, as is our MasterCode$^\dag$ point.  From updated pipelines, the value of $m_A$ is much farther from $2m_{\tilde{\chi}_1^0}$ such that the $A$-funnel resonance does not have a significant impact.}. 
With the \softsusy{} spectrum, we see a factor of $\sim 20$ difference in $\Omega h^2$ between \micromegas{} (solid magenta triangles) and \darksusy{} (hollow magenta triangles), with neither giving a value within the limits of experimental uncertainty. As in the pMSSM, for well-tempered dark matter, \darksusy{} tends to give a relic density value that is larger than the one given by \micromegas{} for the same spectrum. This could arise due to differences in the way the effective annihilation cross section between the dark matter and the neutral and charged higgsinos is computed by the two programs, as well as differences in the calculation of the $t$-channel chargino exchange diagram. While in the case of the pMSSM the difference was small, in the case of the GUT model benchmark the difference is enormous, possibly due to the proximity to the $A$-funnel region. With \micromegas{} we obtain a dark matter candidate that annihilates too efficiently in the early Universe, while with \darksusy{} we obtain a candidate that does not annihilate efficiently enough.

Regarding the relic density for the CMSSM 2 point, as mentioned above, only \spheno{} pipelines represent true coannihilation models, while \softsusy{} pipelines do not coannihilate efficiently enough to achieve the correct relic abundance.  Thus, the relic abundance from \softsusy{} pipelines appears at large values of $\Omega h^2$ beyond the range shown in the upper left panel of Figure~\ref{fig:msugra}.

We now turn to the annihilation cross section in the current Universe for the CMSSM benchmarks. The results are plotted in the upper right panel of Figure~\ref{fig:msugra} and tabulated in Table~\ref{tab:msigv}. We see that the enormous difference in the relic density computation between spectra coming from \softsusy{} and \spheno{}  continues to persist in the computation of the current annihilation cross section for the CMSSM 1. For this benchmark we see the largest discrepancies in the calucation of the annihilation cross section today, nearly three orders of magnitude. Since the charged higgsinos are much heavier in the spectrum generated by \spheno{}, the $t-$channel chargino exchange diagram is suppressed in this case. This leads to a much smaller annihilation cross section, $\sim 10^{-28}$ cm$^{3}$s$^{-1}$. For a given spectrum, the difference between \micromegas{} and \darksusy{} persists, with \darksusy{} giving smaller annihilation cross section as before.  For the CMSSM 2 benchmark, neutralino-stau coannihilations play no role in the annihilation today, so we see reasonably good agreement among the pipelines, albeit with a very low annihilation cross section. 

Finally, we consider the neutralino-nucleon scattering cross sections, which are presented in the lower left panel of Figure~\ref{fig:msugra} (as per-nucleon scattering cross sections) and tabulated in Table~\ref{tab:msigmafeyn} and~\ref{tab:msigmahd}, where the subtables are organized by proton then neutron scattering cross sections, first for the spin-independent then for spin-dependent scattering.

From the lower left panel of Figure~\ref{fig:msugra}, we see that the scattering cross sections for the CMSSM 1 spectrum coming from \softsusy{} are much larger than those for the corresponding spectrum coming from \spheno{} for both \darksusy{} and \micromegas{}.  In fact, the \softsusy{} model is already being constrained by current experiments. This is due to the larger higgsino content of the dark matter in the \softsusy{} case, which leads to an enhancement of the Higgs exchange diagram. Moreover, as was observed in the pMSSM cases, there are discrepancies between the scattering cross sections reported by \micromegas{} versus \darksusy{}, even for the same spectrum, from the differences in form factors employed by each code (see Table~\ref{tab:form}) and the way in which loop effects are incorporated.  For the CMSSM 2, \micromegas{} gives the same SI scattering cross section no matter which spectrum generator is employed, while \darksusy{} yields somewhat larger SI scattering cross sections that do depend somewhat on the details of the spectrum that differ.  Since the LSP for the CMSSM 2 benchmark is nearly pure bino and the SI cross sections are strongly suppressed, differences in the SI scattering cross sections from the \spheno{}-\darksusy{} versus \softsusy{}-\darksusy{} pipelines likely come from loop corrections.

\subsubsection{NUHM Benchmarks} 
As discussed in Section~\ref{sec:benchmarks}, since the original MasterCode NUHM best fit point has a nearly pure higgsino LSP and is therefore very sensitive to the details of the spectrum, our NUHM A and B benchmarks were chosen with the requirements that a valid relic density would be achieved by NUHM A via the \spheno{} pipelines and by NUHM B via the \softsusy{} pipelines.

From Table~\ref{tab:mneutspectra}, the mass of the lightest neutralino obtained by \softsusy{} is far smaller that that obtained by \spheno{} for both NUHM A and B.
The dark matter is mostly higgsino in all cases, and the radically different masses for the LSP (and other light -inos) are due to the very different values of $\mu$ reported by the two programs, as discussed previously in Section~\ref{sec:ewsb}.

The analyses of the dark matter observables for both NUHM benchmarks follow the trends of the pure higgsino case discussed previously for the pMSSM. For both benchmarks, we expect that the relic density for pipelines involving \softsusy{} should be lower than that given by pipelines involving \spheno{}, due to the smaller higgsino mass in the former case. This expectation is borne out in the upper left panel of Figure~\ref{fig:msugra}.  For a given spectrum, the relic densities computed by \micromegas{} and \darksusy{} match quite well, as evidenced by the fact that solid and hollow NUHM markers more or less overlap. 

For the annihilation cross section in the current Universe, we expect that the lower higgsino mass of the \softsusy{} pipelines should correspond to a larger annihilation cross section than the \spheno{} pipelines. This is borne out by the relative positions of the magenta and cyan NUHM markers in the upper right panel of Figure~\ref{fig:msugra}, though the difference is less pronounced for \micromegas{} pipelines for the NUHM B benchmark.

The scattering cross sections with nuclei are shown in the lower left panel of Figure~\ref{fig:msugra}. We first note that the SI scattering cross sections for  higgsino dark matter in this case are within a factor of a few of $\sim 10^{-9}$ pb, which is much larger than the cross section for the pure higgsino case in the pMSSM analysis. This can be attributed to the fact that the dark matter in the NUHM benchmarks is less pure higgsino than in the ``pure higgsino'' pMSSM benchmark, as can be seen by comparing Tables~\ref{tab:mneutspectra} and~\ref{tab:pneutspectra}.  The more pure the LSP, the smaller the SI scattering cross section. We see also that \darksusy{} gives larger scattering cross sections than \micromegas{}, consistent with results discussed above. As before, we can attribute this to the difference in form factors between the two programs.

\label{sec:generic}

\section{Conclusions} 
\label{sec:conclusions}

In this paper, we have performed comparative studies of the physics of supersymmetric dark matter calculated with a sequence of spectrum generators (\softsusy{} and \spheno{}), Higgs sector calculators (\feynhiggs{} and \susyhd{}) and to dark matter observable calculators (\micromegas{} and \darksusy{}).  We placed our study in the context of several SUSY benchmark models that are interesting in  light of LHC Run-1 and null results from recent dark matter searches as studied previously by~\cite{Cahill-Rowley:2013gca, Buchmueller:2013rsa, Cohen:2013kna}.  We have compared calculations for the sparticle spectra, the Higgs sectors, and the dark matter observables for each benchmark, and we have incorporated the various generators and calculators into comprehensive pipelines to study not only the effects of the choice of an individual package, but also all downstream effects of those choices on subsequent calculations.

This study was conducted in two parts. In the first part, we investigated a set of pMSSM models from Ref.~\cite{Cahill-Rowley:2013gca}. The dark matter scenarios we considered were coannihilation (bino-stop and bino-squark), $A$-funnel, well-tempered neutralinos, and pure higgsinos. We discovered that the spectrum generators can differ by up to 1 - 2 \% in their predicted masses for the stop and the first two generations of squarks, and by up to 20\% in the gauge composition of the lightest neutralino, for a given pMSSM model. As for the dark matter observables, differences of up to a factor of $\sim 3 - 5$ in the relic density and current annihilation cross section, and up to a factor of $\sim 10$ in the predicted scattering cross section, were found to exist between the different pipelines. These discrepancies are already pressing in the case of the relic abundance of dark matter, for which the uncertainty in the experimental value is small compared to the theoretical uncertainty in pMSSM predictions.  For the annihilation cross section today and the SI scattering cross section, the discrepancies will become important if/when future dark matter indirect and direct detection experimental sensitivities reach the predicted levels. 

In the second part of our study, we considered four interesting benchmark models defined at the GUT scale -- two CMSSM points and two NUHM points.  For GUT-scale models, we found that discrepancies among the various pipelines are often amplified by the renormalization group running.  For our CMSSM and NUHM benchmarks, we found that the spectrum generators can give low energy values of the higgsino mass parameter $\mu$ and the pseudoscalar Higgs mass $m_A$ that differ by up to 150\% - 200\% (though the differences can be much greater at larger $m_0$). This can lead to large differences in the annihilation and scattering cross sections computed by the dark matter calculators.

As a community, we have made tremendous progress in predicting signals of the particle nature of dark matter, made possible by pioneering work in theory and computation and closely related to increasing experimental sophistication.  Though no definitive signals have yet emerged, ongoing attention to the theoretical calculations related to MSSM neutralino dark matter, and dark matter observables in general, is now more important than ever as experiments begin to probe the canonical SUSY WIMP parameter space.

\acknowledgments
{}
We thank Paolo Gondolo, Werner Porod, and Alexander Pukhov for clarifications related to the implementation of calculations in \darksusy{}, \spheno{}, and \micromegas{}, respectively, and Keith Olive for information on the MasterCode best fit points. PS would also like to thank Nordita, the Nordic Institute for Theoretical Physics, for their hospitality.
The work of PS is supported in part by NSF Grant No. PHY-1720282.

\renewcommand{\thetable}{A\arabic{table}}
\setcounter{table}{0}

\begin{appendix}

\section{Data Tables for pMSSM Models}
\label{app:pmssm}

In this Appendix, we collect data tables for the pMSSM section of our paper.

\begin{table}[h]
  \centering
  \subfloat[][$\Omega h^2$ as computed via the \feynhiggs{} branch of the pipeline.\label{tab:pomegafeyn}]{%
      \includegraphics[scale=0.9]{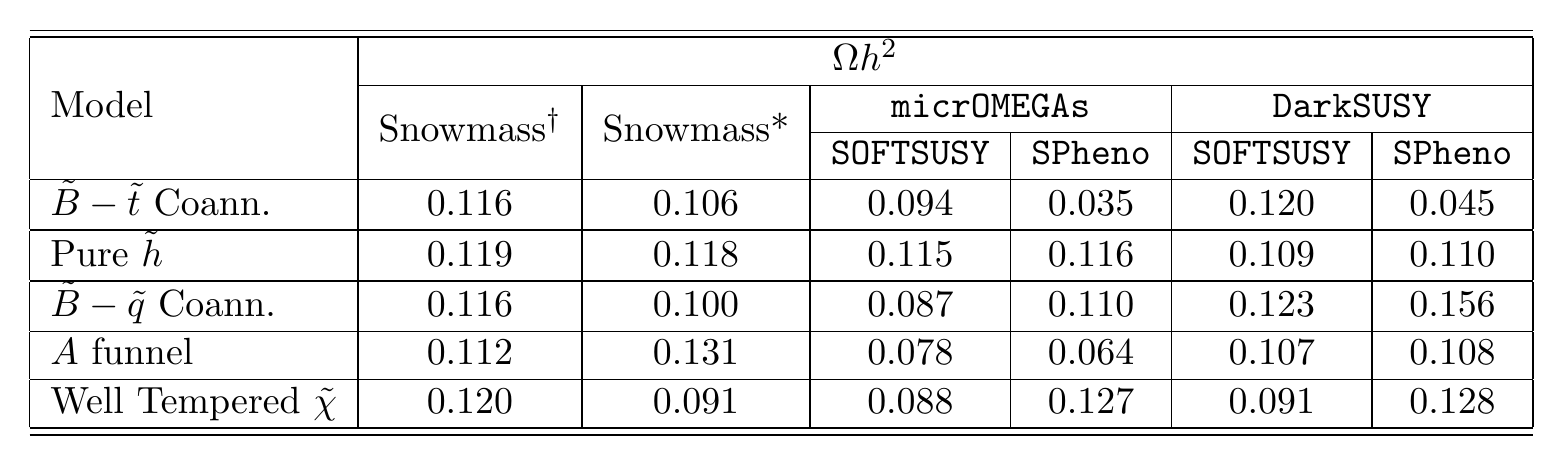}
  }
  \hfill
  \subfloat[][$\Omega h^2$ as computed via the \susyhd{} branch of the pipeline.\label{tab:pomegahd}]{%
      \includegraphics[scale=0.9]{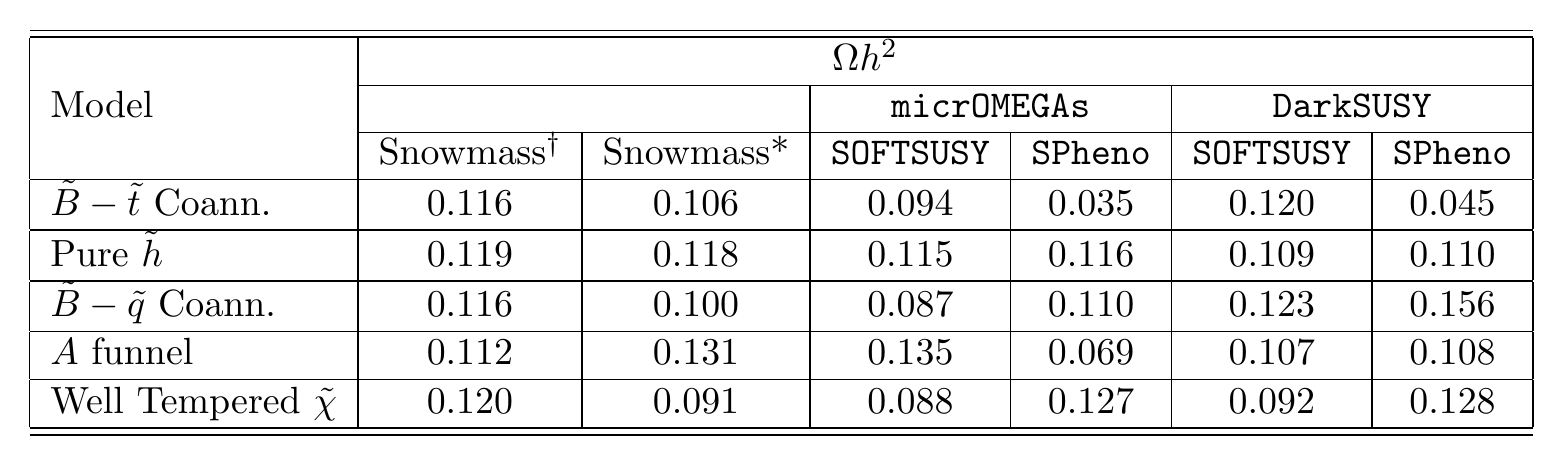}
  }
  \caption{\textbf{pMSSM dark matter relic density:} The relic density as computed by the various pipelines. Table~\ref{tab:pomegafeyn}'s pipelines make use of \feynhiggs{}, while Table~\ref{tab:pomegahd} uses \susyhd{}. Here ``Snowmass$^*$ refers to the updated version of the Snowmass pipeline and ``Snowmass$^\dag$'' refers to our incarnation of the Snowmass pipeline with \micromegas{} {\tt 2.4.5} (rather than {\tt v2.4}; see Table~\ref{tab:pipelines}). Values for the \feynhiggs{} pipelines (Table~\ref{tab:pomegafeyn}) are plotted in the upper left panel of Figure~\ref{fig:pmssm} and percent differences for both pipelines may be found in Table~\ref{tab:omegaperc}.}
  \label{tab:pomega}
  \end{table}

\begin{table}[b]
    \centering
    \subfloat[][Percent differences in $\Omega h^2$ as computed via the \feynhiggs{} branch of the pipeline.\label{tab:omegapercfeyn}]{%
      \includegraphics[scale=0.9]{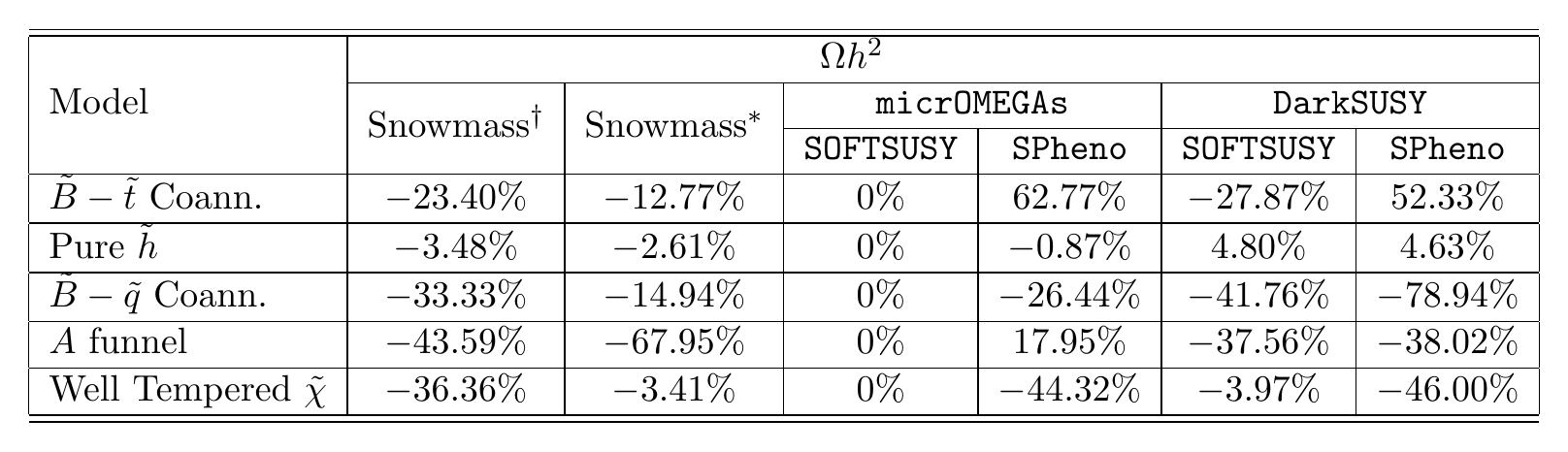}
  }
  \hfill
  \subfloat[][Percent differences in $\Omega h^2$ as computed via the \susyhd{} Branch of the pipeline.\label{tab:omegaperchd}]{%
      \includegraphics[scale=0.9]{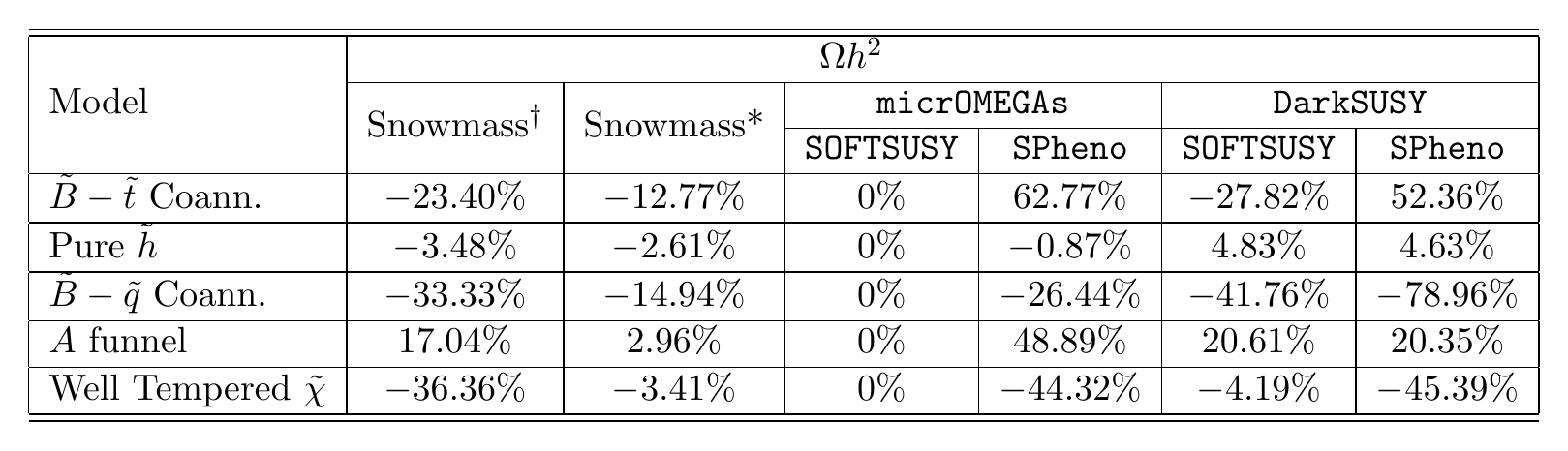}
  }
  \hfill
  \caption{\textbf{pMSSM dark matter relic density:} Percent differences $\Omega h^2$, relative to the \softsusy{}-\micromegas{} pipelines.}
  \label{tab:omegaperc}
\end{table}

\begin{table}[h]
  \centering
  \subfloat[][$\langle \sigma v \rangle$ as computed via the \feynhiggs{} branch of the pipeline.\label{tab:psigvfeyn}]{%
      \includegraphics[scale=0.9]{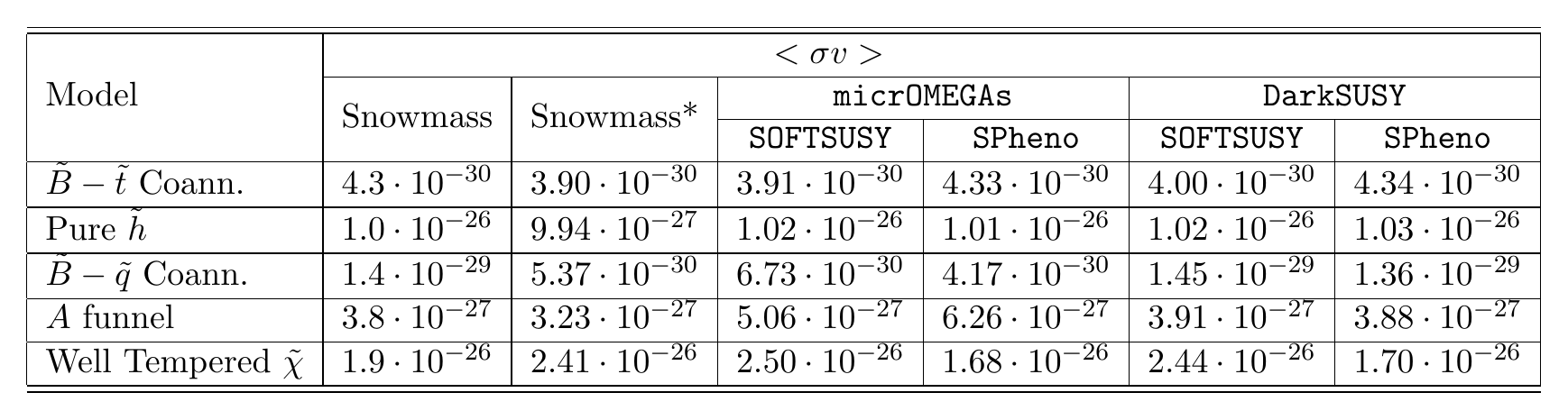}  
    }
    \hfill
    \subfloat[][$\langle \sigma v \rangle$ as computed via the \susyhd{} branch of the pipeline.\label{tab:psigvhd}]{%
          \includegraphics[scale=0.9]{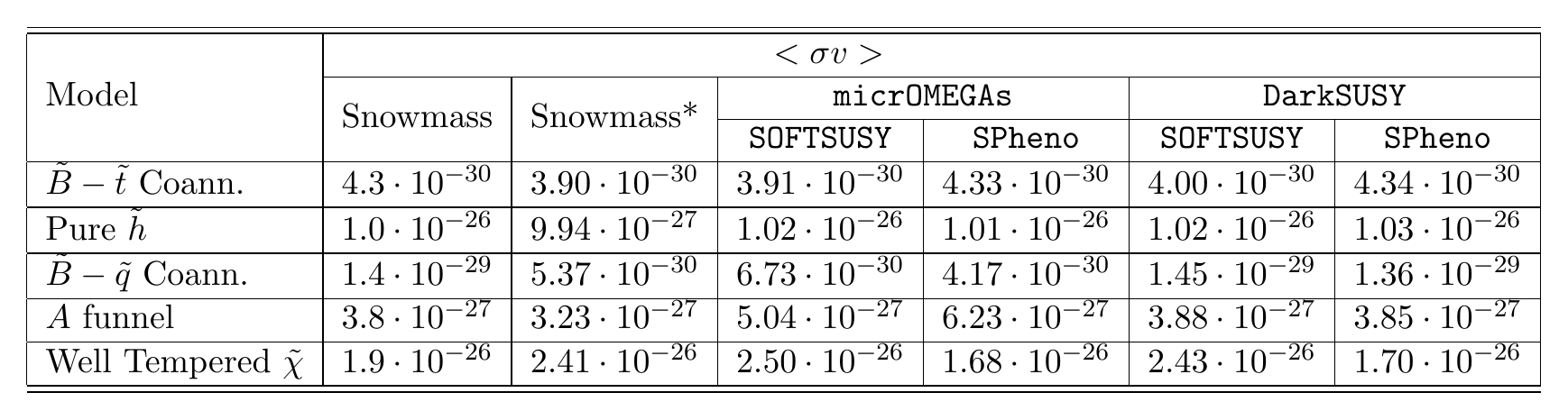}
    }
    \caption{\textbf{pMSSM dark matter annihilation cross section:} Dark matter annihilation cross section today, in cm$^3$s$^{-1}$, for pMSSM benchmarks as computed by the various pipelines. Table~\ref{tab:psigvfeyn} displays results from the pipelines using \feynhiggs{} while Table~\ref{tab:psigvhd}'s pipelines use \susyhd{}. Values for the \feynhiggs{} pipelines (Table~\ref{tab:psigvfeyn}) are plotted in the upper right panel of Figure~\ref{fig:pmssm} and the percent differences are given in Table~\ref{tab:sigvperc}.}
    \label{tab:psigv}
  \end{table}
  
  \begin{table}[b]
  \centering
  \subfloat[][Percent differences for $\langle \sigma v \rangle$ as computed via the \feynhiggs{} branch of the pipeline.\label{tab:sigvpercfeyn}]{%
      \includegraphics[scale=0.9]{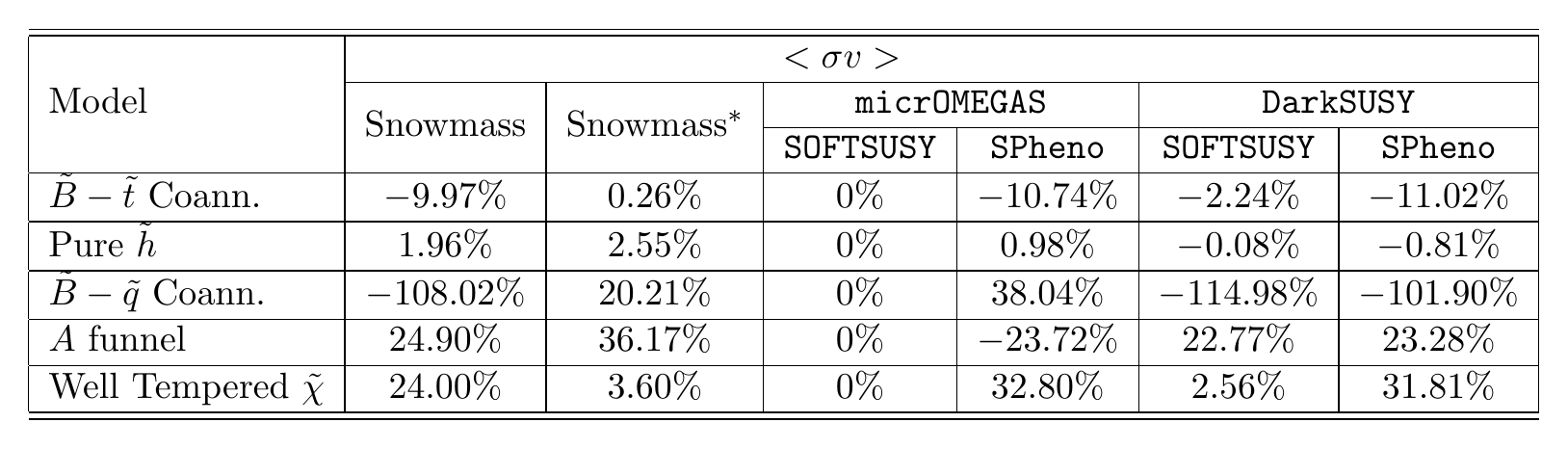}
  }
  \hfill
  \subfloat[][Percent differences for $\langle \sigma v \rangle$ as computed via the \susyhd{} branch of the pipeline.\label{tab:sigvperchd}]{%
      \includegraphics[scale=0.9]{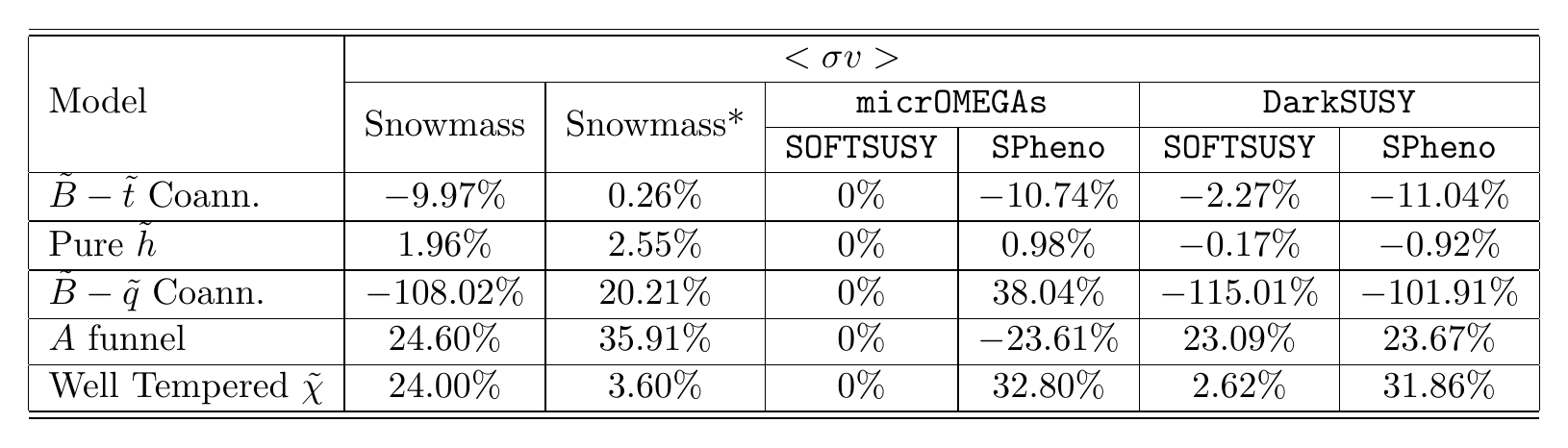}
  }
  \caption{\textbf{pMSSM dark matter annihilation cross section:} Percent differences for $\langle \sigma v \rangle$, relative to the \softsusy{}-\micromegas{} pipelines.}
  \label{tab:sigvperc}
\end{table}

\begin{table}[h]
  \centering
  \subfloat[][$\sigma^{SI}_p$ as computed in the \feynhiggs{} branch of the pipeline.\label{tab:pprotindepfeyn}]{%
      \includegraphics[scale=0.9]{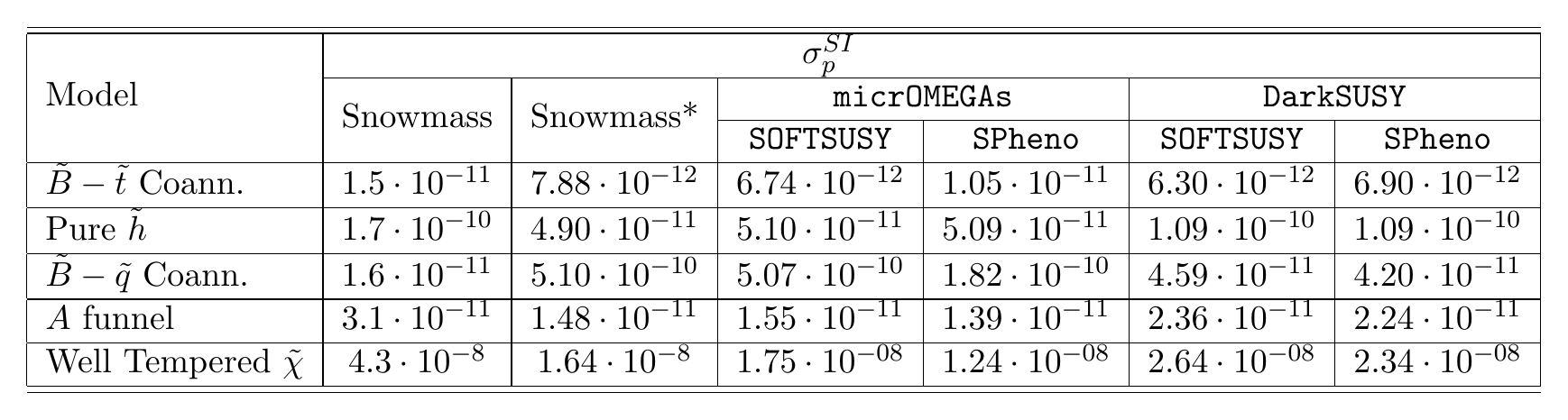}
  }
  \hfill
  \subfloat[][$\sigma^{SI}_n$ as computed in the \feynhiggs{} branch of the pipeline.\label{tab:pneutindepfeyn}]{%
      \includegraphics[scale=0.9]{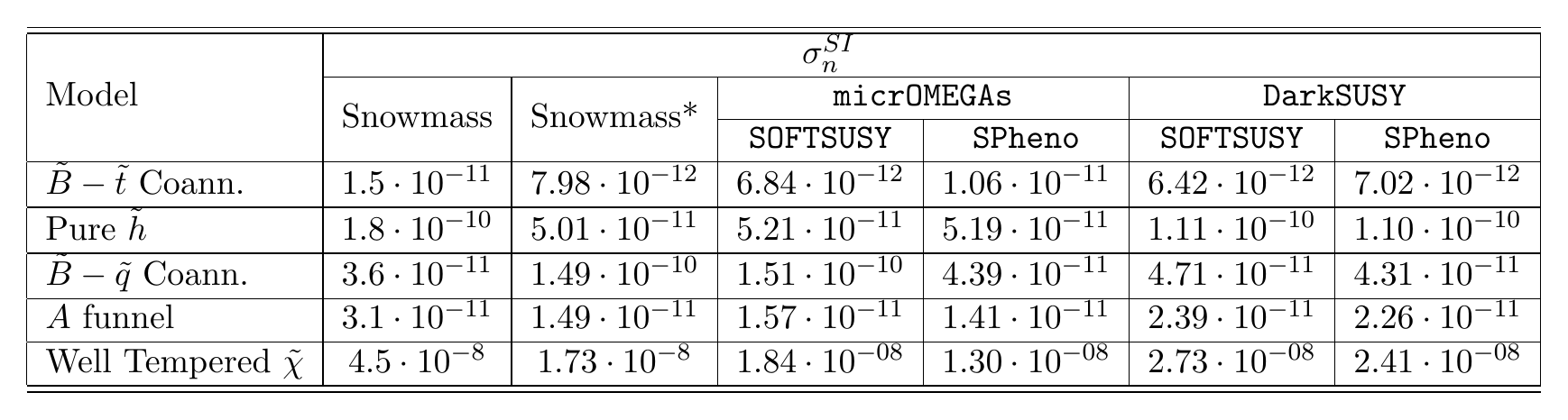}
  }
  \hfill
  \subfloat[][$\sigma^{SD}_p$ as computed in the \feynhiggs{} branch of the pipeline.\label{tab:pprotdepfeyn}]{%
      \includegraphics[scale=0.9]{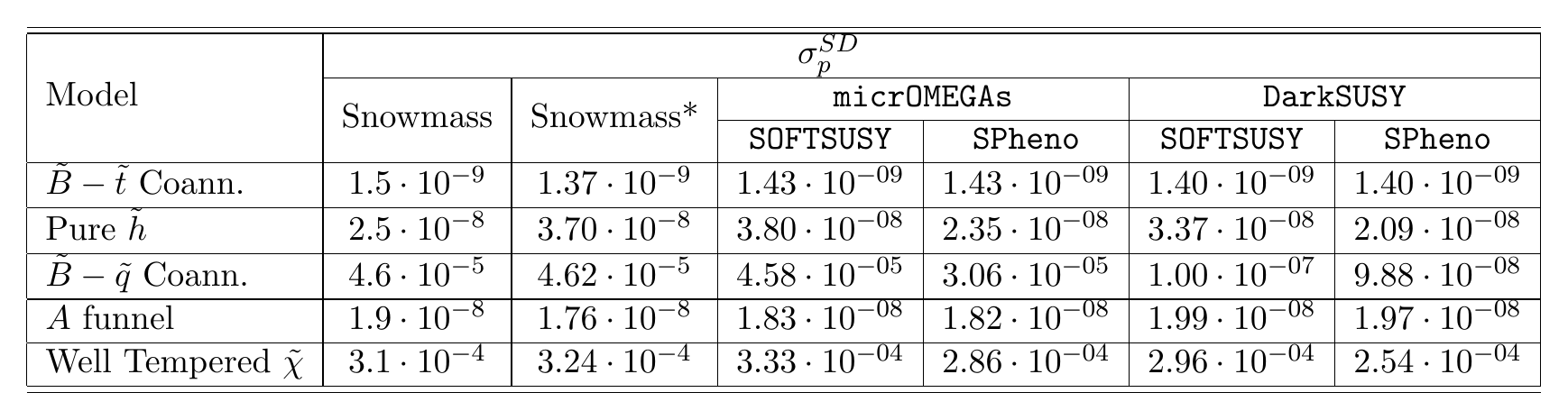}
  }
  \hfill
  \subfloat[][$\sigma^{SD}_n$ as computed in the \feynhiggs{} branch of the pipeline.\label{tab:pneutdepfeyn}]{%
      \includegraphics[scale=0.9]{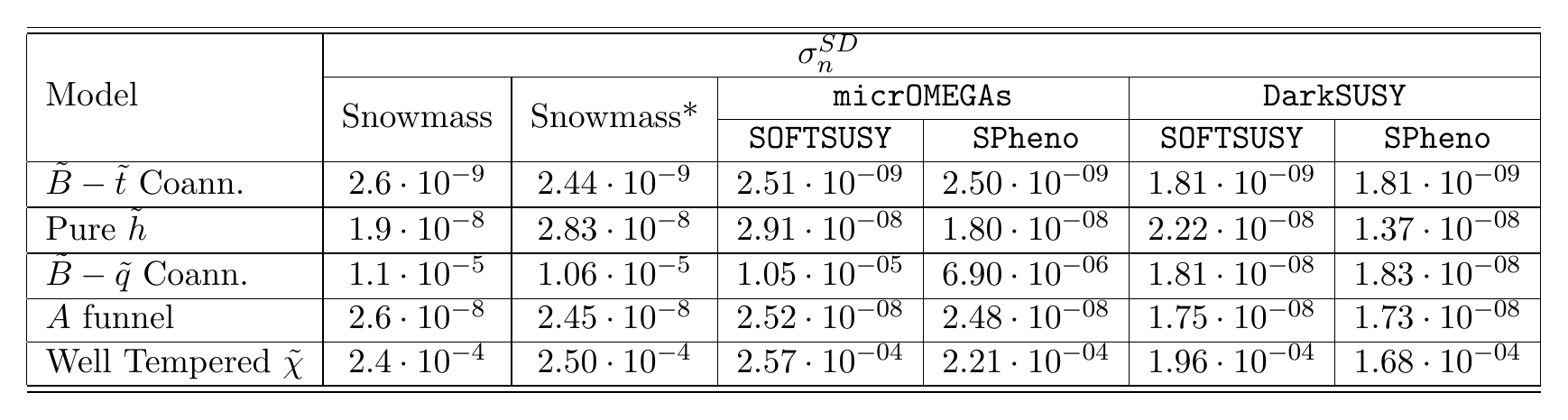}
  }
  \caption{\textbf{pMSSM dark matter scattering cross section:} Spin independent neutralino-nucleon elastic scattering cross sections, in pb, as computed by the various \feynhiggs{} pipelines. The spin-independent per-nucleon average for Xe is plotted in the lower left panel of Figure~\ref{fig:pmssm} and percent differences are given in Table~\ref{tab:sigmapercfeyn}.}
  \label{tab:psigmafeyn}
\end{table}
\begin{table}
  \centering
  \subfloat[][$\sigma^{SI}_p$ as computed in the \susyhd{} branch of the pipeline.\label{tab:pprotindephd}]{%
      \includegraphics[scale=0.9]{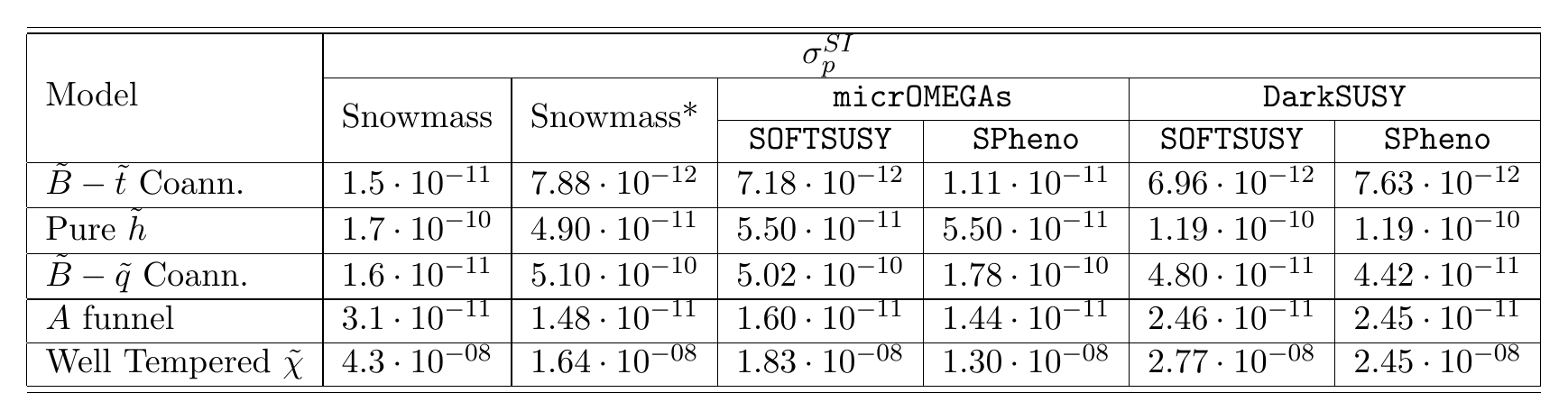}
  }
  \hfill
  \subfloat[][$\sigma^{SI}_n$ as computed in the \susyhd{} branch of the pipeline.\label{tab:pneutindephd}]{%
      \includegraphics[scale=0.9]{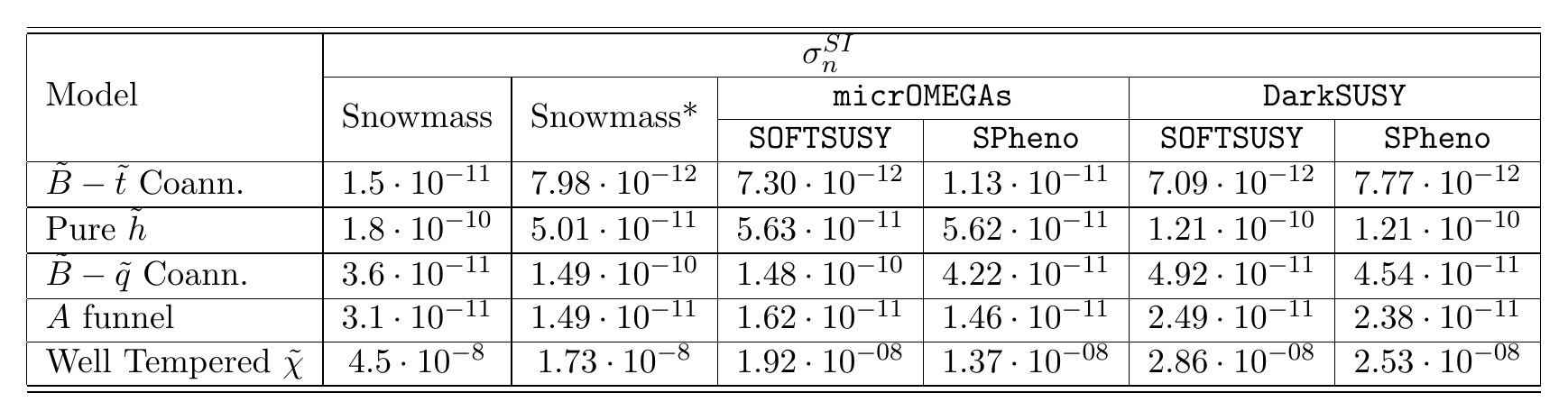}
  }
  \hfill
  \subfloat[][$\sigma^{SD}_p$ as computed in the \susyhd{} branch of the pipeline.\label{tab:pprotdephd}]{%
      \includegraphics[scale=0.9]{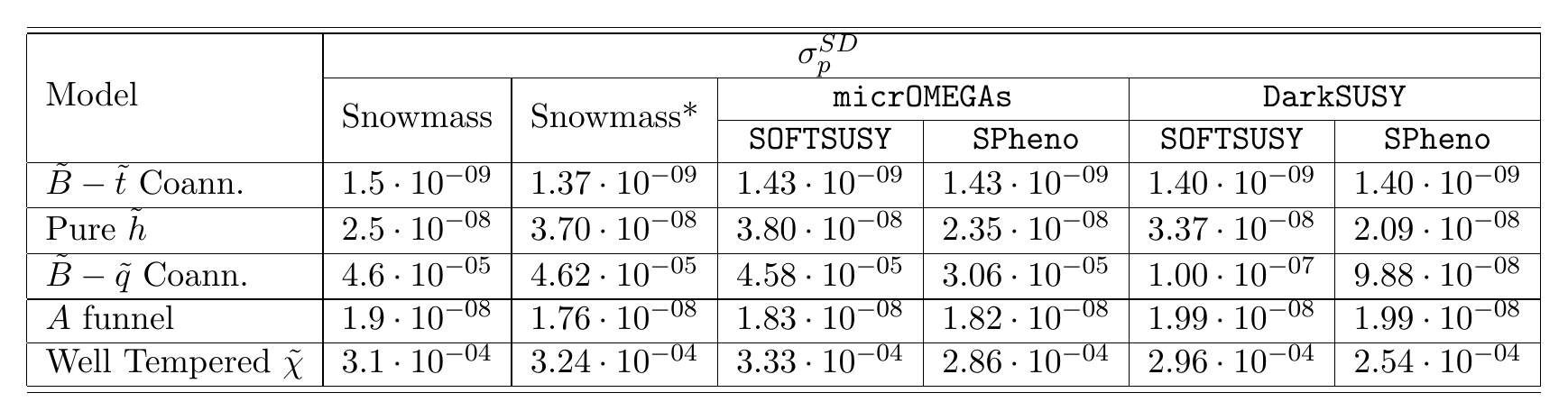}
  }
  \hfill
  \subfloat[][$\sigma^{SD}_n$ as computed in the \susyhd{} branch of the pipeline.\label{tab:pneutdephd}]{%
      \includegraphics[scale=0.9]{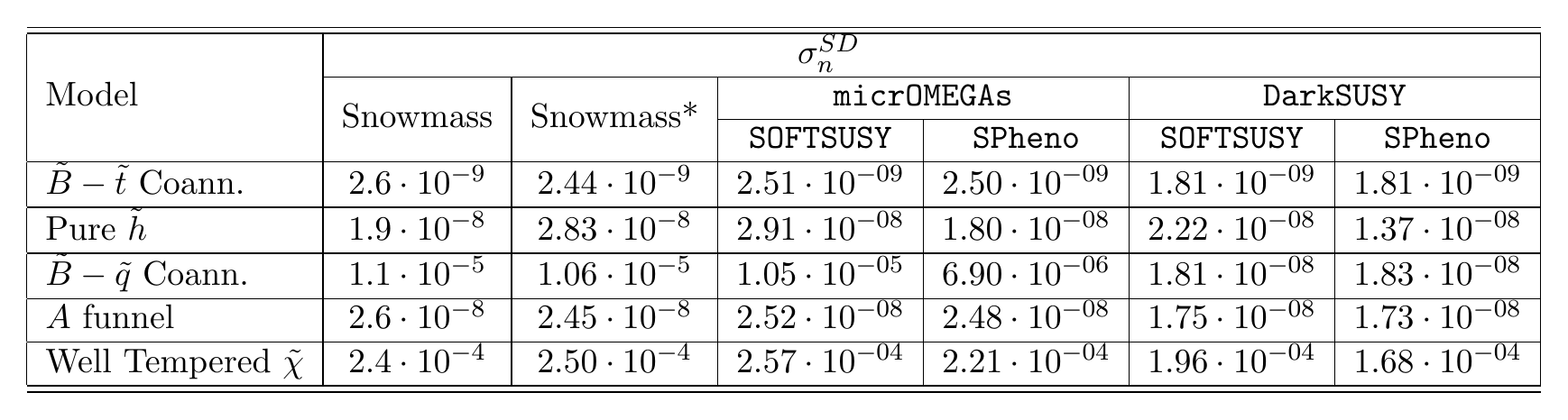}
  }
  \caption{\textbf{pMSSM dark matter scattering cross section:} Spin independent neutralino-nucleon elastic scattering cross sections, in pb, where pipelines not labeled as ``Snowmass'' are computed by the various \susyhd{} pipelines. Percent differences for the values presented here may be found in Table~\ref{tab:sigmaperchd}.}
  \label{tab:psigmahd}
\end{table}


\begin{table}
  \centering
  \subfloat[][Percent differences for $\sigma^{SI}_p$ as computed in the \feynhiggs{} branch of the pipeline.\label{tab:protindeppercfeyn}]{%
      \includegraphics[scale=0.9]{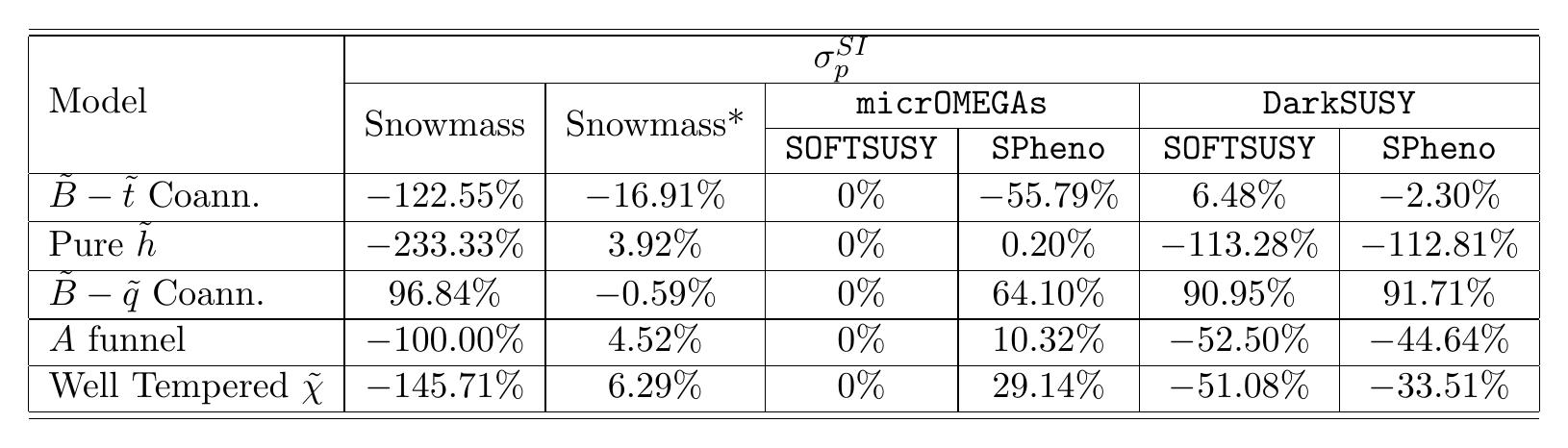}
  }
  \hfill
  \subfloat[][Percent differences for $\sigma^{SI}_n$ as computed in the \feynhiggs{} branch of the pipeline.\label{tab:neutindeppercfeyn}]{%
      \includegraphics[scale=0.9]{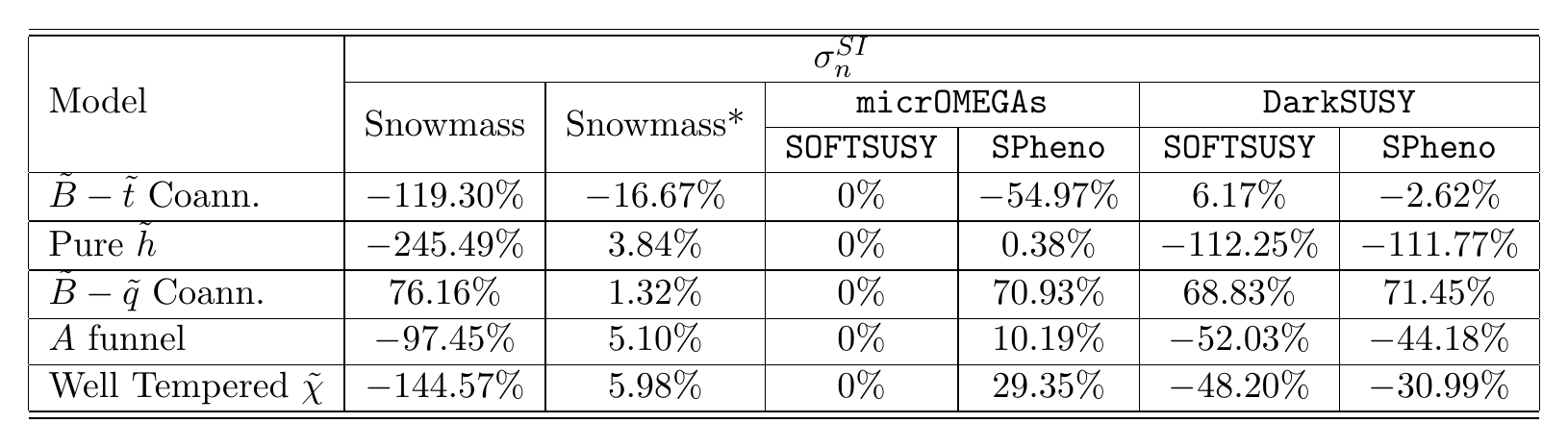}
  }
  \hfill
  \subfloat[][Percent differences for $\sigma^{SD}_p$ as computed in the \feynhiggs{} branch of the pipeline.\label{tab:protdeppercfeyn}]{%
      \includegraphics[scale=0.9]{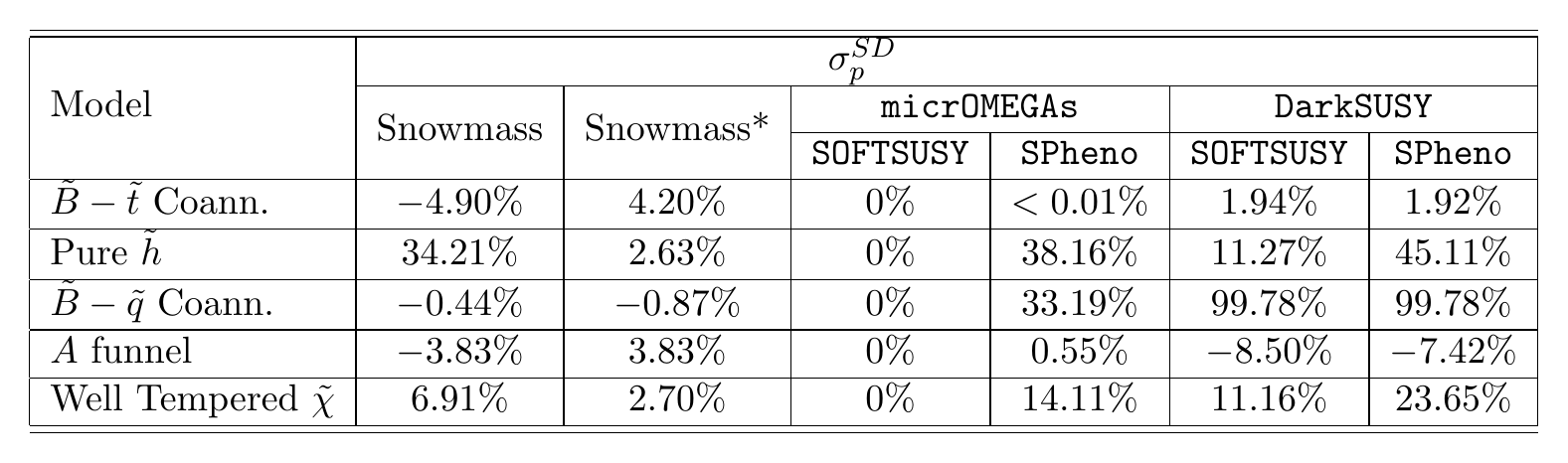}
  }
  \hfill
  \subfloat[][Percent differences for $\sigma^{SD}_n$ as computed in the \feynhiggs{} branch of the pipeline.\label{tab:neutdeppercfeyn}]{%
      \includegraphics[scale=0.9]{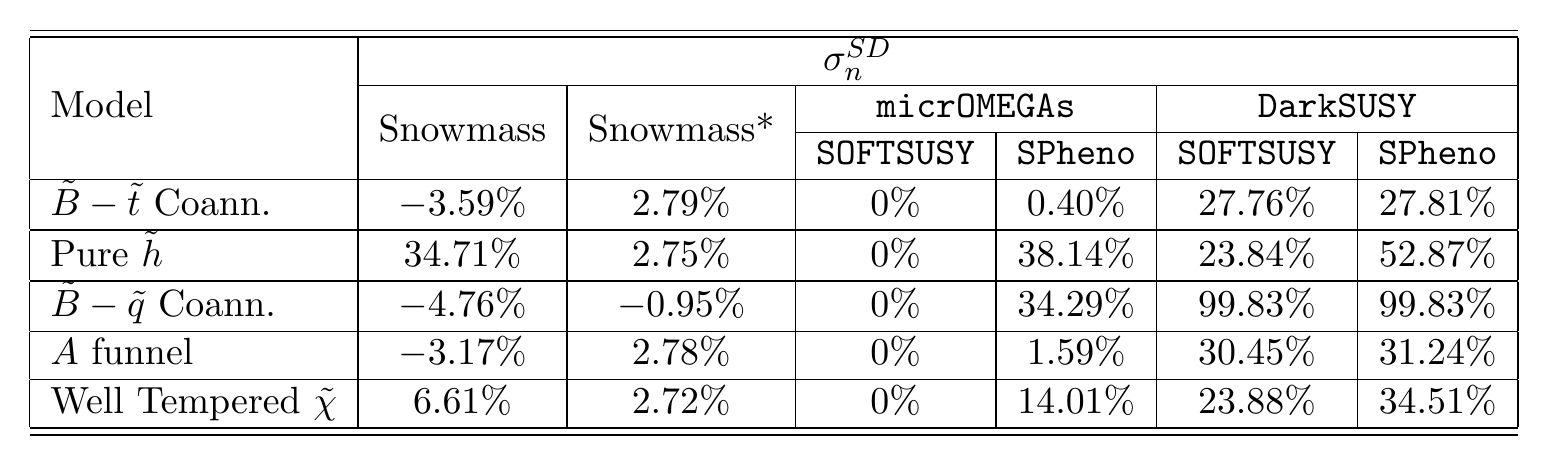}
  }
  \caption{\textbf{pMSSM dark matter scattering cross section:} Percent difference of the values found in Table~\ref{tab:psigmafeyn} (pipelines with \feynhiggs{}) relative to the \softsusy{}-\micromegas{} pipelines.}
  \label{tab:sigmapercfeyn}
\end{table}
\begin{table}
  \centering
  \subfloat[][Percent differences for $\sigma^{SI}_p$ as computed in the \susyhd{} branch of the pipeline.\label{tab:protindepperchd}]{%
      \includegraphics[scale=0.9]{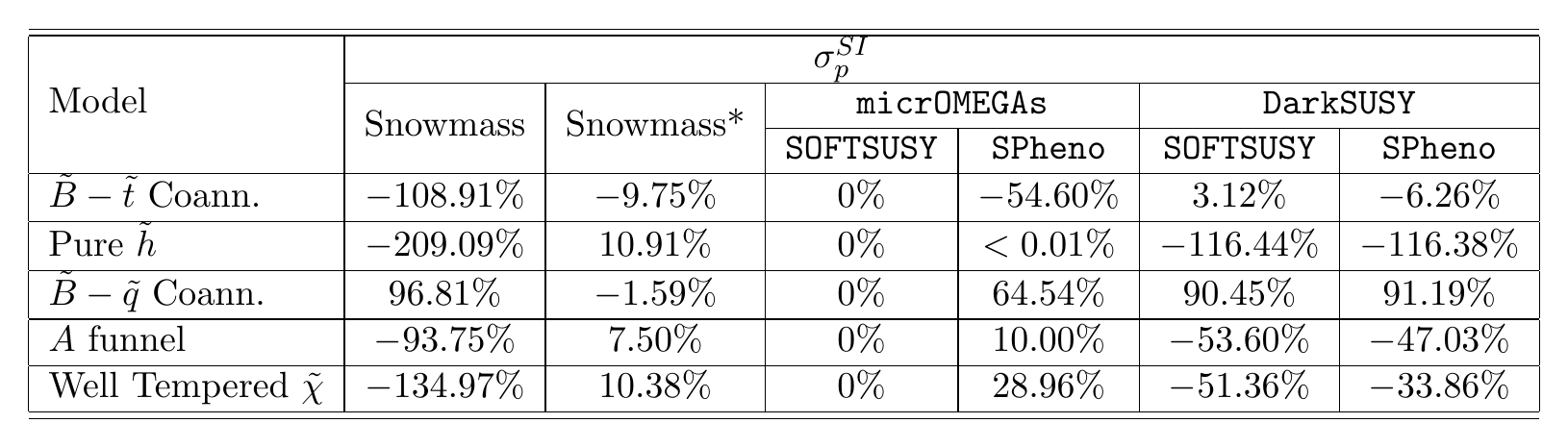}
  }
  \hfill
  \subfloat[][Percent differences for $\sigma^{SI}_n$ as computed in the \susyhd{} branch of the pipeline.\label{tab:neutindepperchd}]{%
      \includegraphics[scale=0.9]{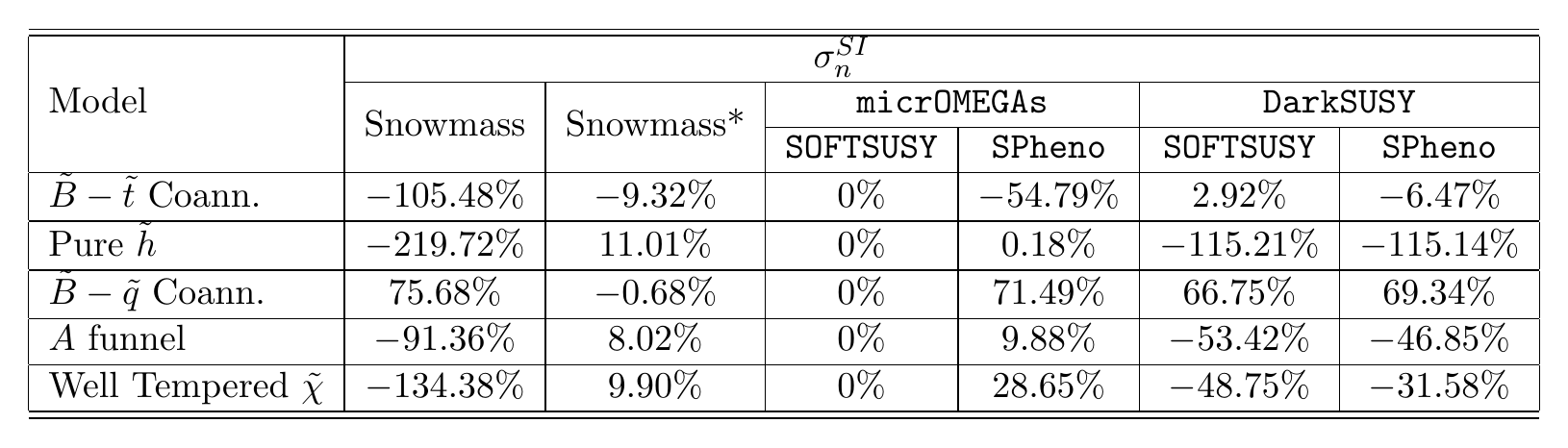} 
  }
  \hfill
  \subfloat[][Percent differences for $\sigma^{SD}_p$ as computed in the \susyhd{} branch of the pipeline.\label{tab:protdepperchd}]{%
      \includegraphics[scale=0.9]{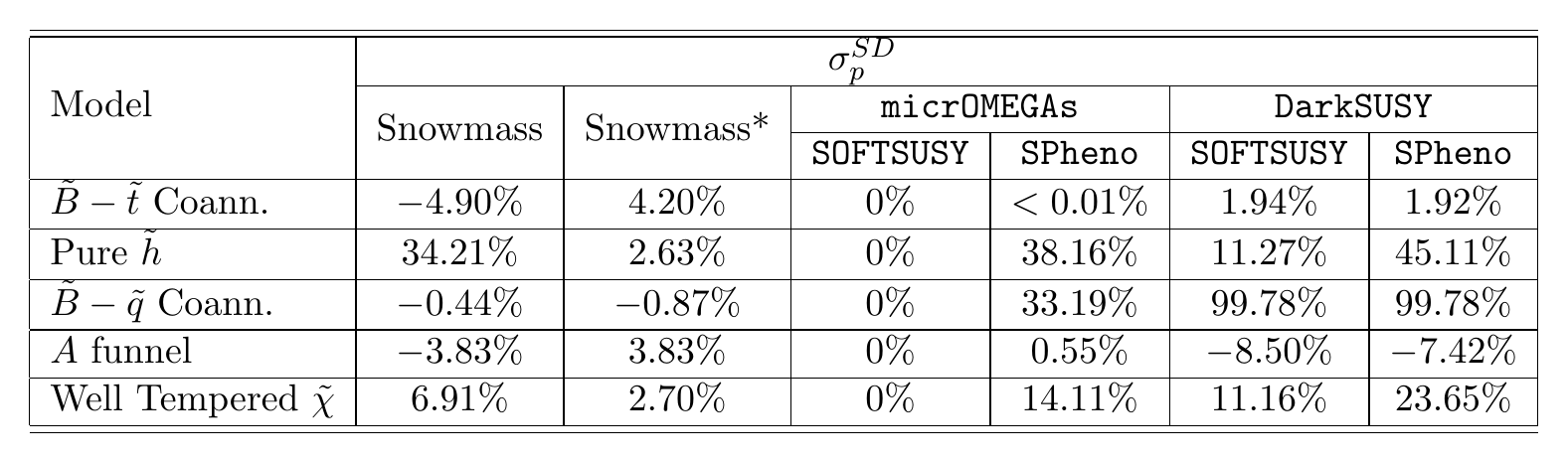}
  }\hfill
  \subfloat[][Percent differences for $\sigma^{SD}_n$ as computed in the \susyhd{} branch of the pipeline.\label{tab:neutdepperchd}]{%
      \includegraphics[scale=0.9]{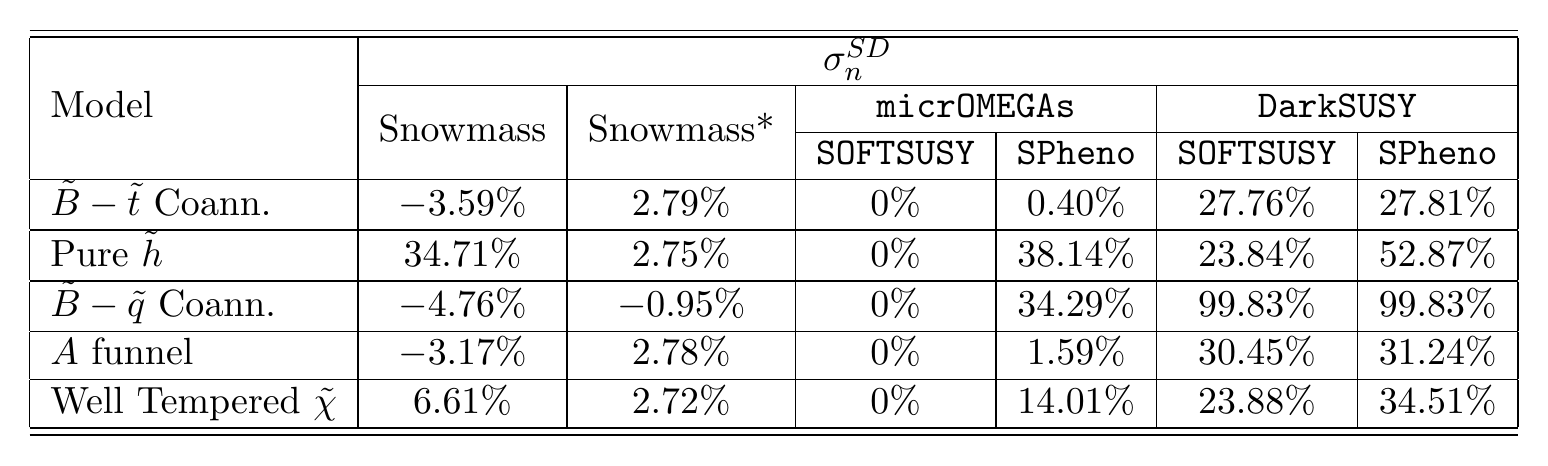}
  }
  \caption{\textbf{pMSSM dark matter scattering cross section:} Percent difference of the values found in Table~\ref{tab:psigmahd} (pipelines with \susyhd{}) relative to the \softsusy{}-\micromegas{} pipelines.}
  \label{tab:sigmaperchd}
\end{table}

\clearpage
\section{Data Tables for GUT Models}
\label{app:msugra}

In this Appendix, we collect data tables for the GUT model analysis of our paper.

\begin{table}
  \centering
  \subfloat[][$\Omega h^2$ as computed via the \feynhiggs{} branch of the pipeline.\label{tab:momegafeyn}]{%
      \includegraphics[scale=0.9]{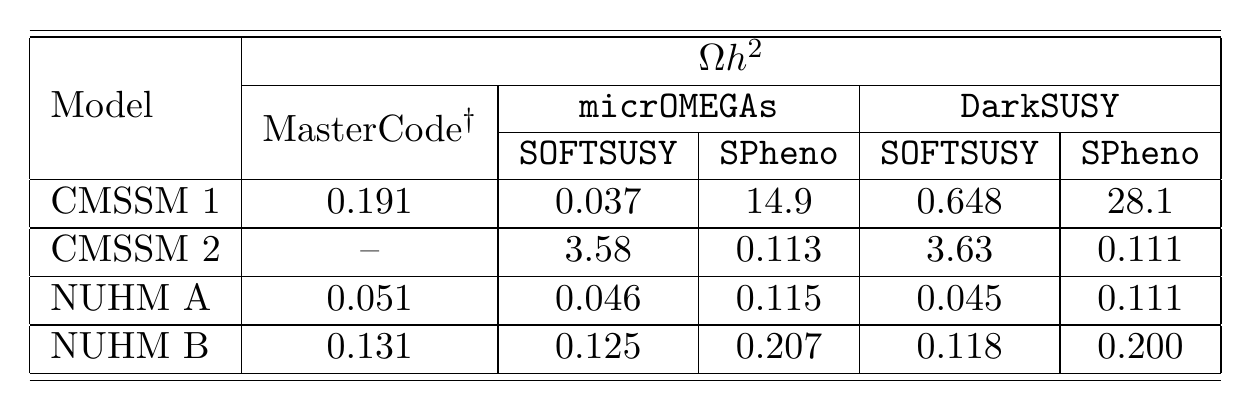}
  } \hfill
  \subfloat[][$\Omega h^2$ as computed via the \susyhd{} branch of the pipeline.\label{tab:momegahd}]{%
      \includegraphics[scale=0.9]{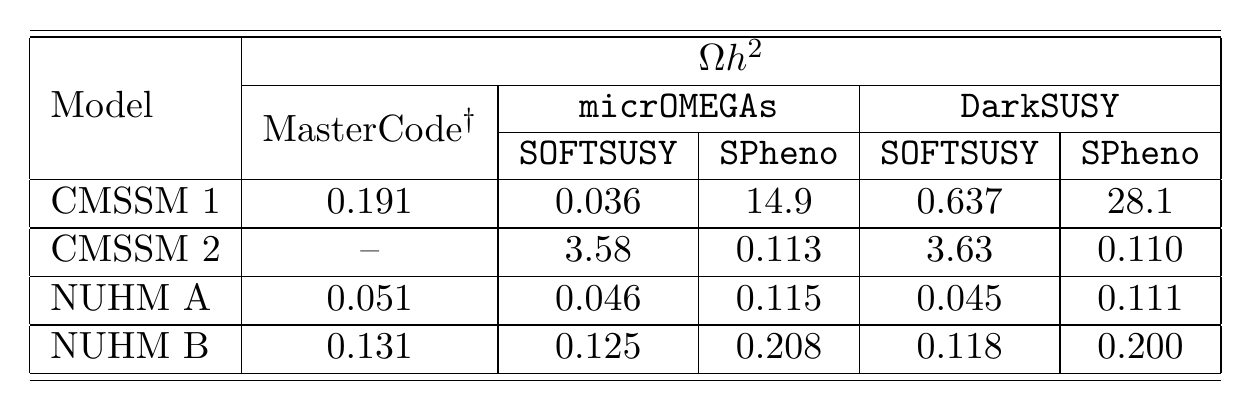}
  }
  \caption{\textbf{GUT model dark matter relic density:} Dark matter relic density as computed by the various pipelines. Table~\ref{tab:momegafeyn}'s pipelines make use of \feynhiggs{}, while Table~\ref{tab:momegahd} uses \susyhd{}. The values from Table~\ref{tab:momegafeyn} are plotted in the upper left panel of Figure~\ref{fig:msugra}.}
  \label{tab:momega}
\end{table}

\begin{table}
  \centering
  \subfloat[][$< \sigma v >$ today as computed in the \feynhiggs{} pipelines.\label{tab:msigvfeyn}]{%
      \includegraphics[scale=0.9]{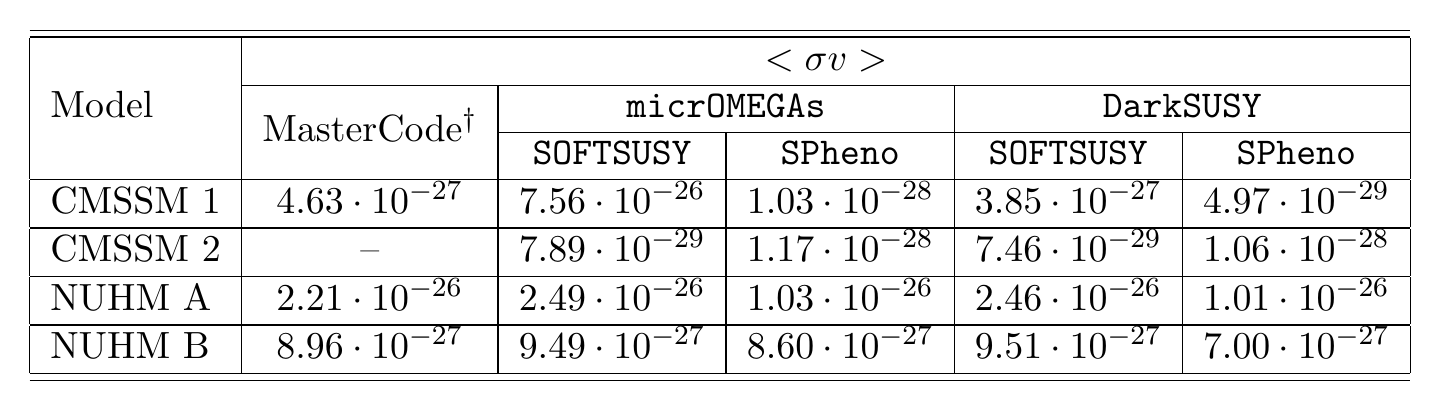}
  } \hfill
  \subfloat[][$< \sigma v >$ today as computed in the \susyhd{} pipelines.\label{tab:msigvhd}]{%
      \includegraphics[scale=0.9]{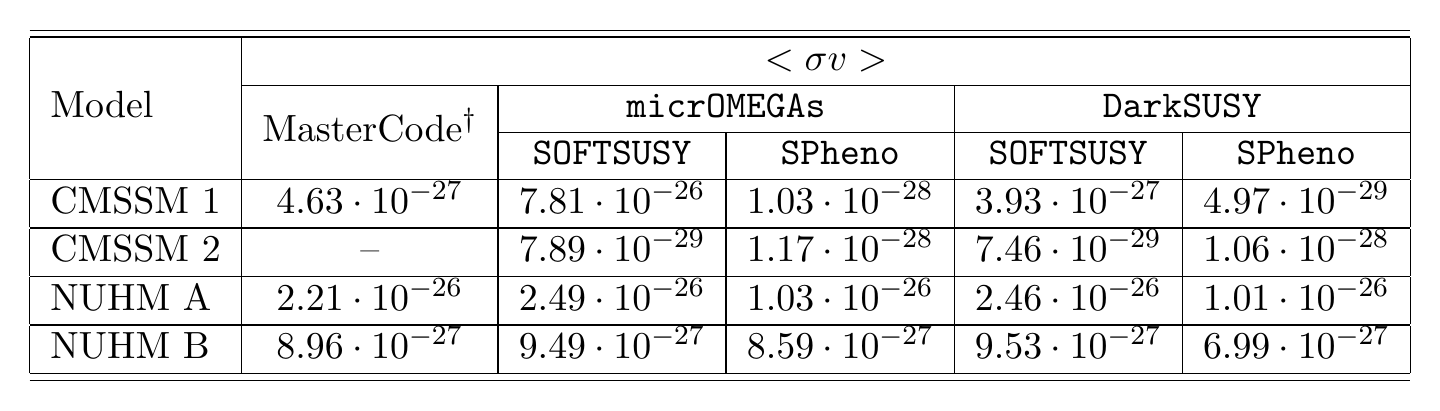}
    }
    \caption{\textbf{GUT model dark matter annihilation cross section:} Dark matter annihilation cross section today, in cm$^3$s$^{-1}$, for GUT models as computed by the various pipelines. Table~\ref{tab:msigvfeyn} shows the results from the pipelines that make use of \feynhiggs{}, while Table~\ref{tab:msigvhd} shows those that use  \susyhd{}. The values in Table~\ref{tab:msigvfeyn} are plotted in the upper right panel of Figure~\ref{fig:msugra}.}
  \label{tab:msigv}
\end{table}

\begin{table}
  \centering
  \subfloat[][$\sigma^{SI}_p$ as computed in the \feynhiggs{} branch of the pipeline.\label{tab:mprotindepfeyn}]{%
      \includegraphics[scale=0.9]{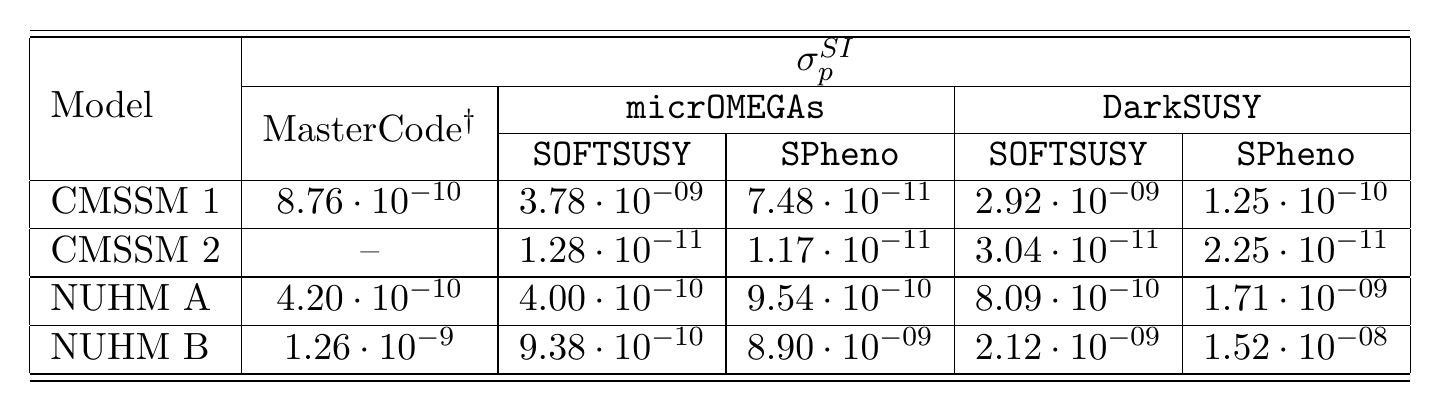}
  } \hfill
  \subfloat[][$\sigma^{SI}_n$ as computed in the \feynhiggs{} branch of the pipeline.\label{tab:mneutindepfeyn}]{%
      \includegraphics[scale=0.9]{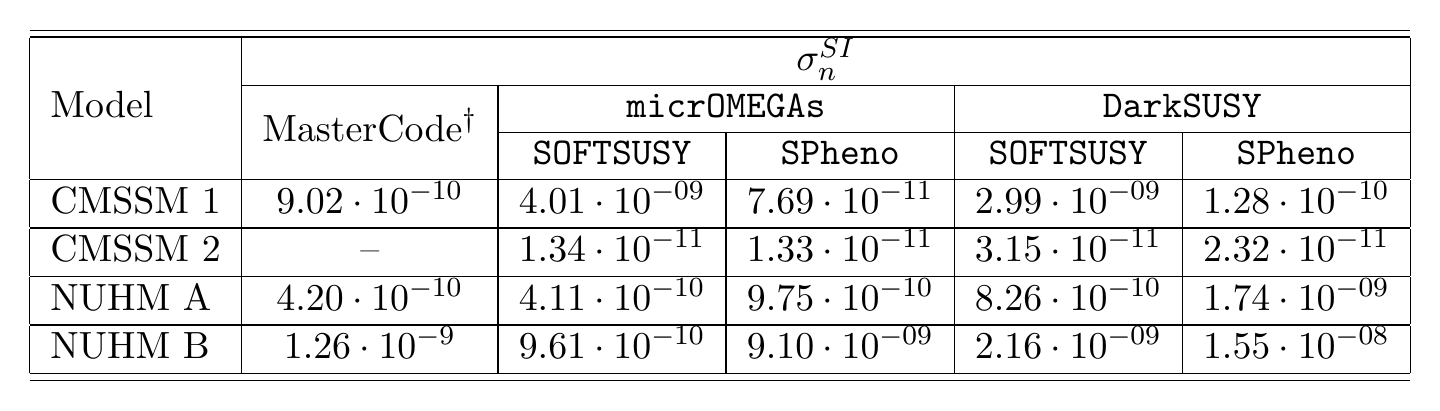}
  } \hfill
  \subfloat[][$\sigma^{SD}_p$ as computed in the \feynhiggs{} branch of the pipeline.\label{tab:mprotdepfeyn}]{%
      \includegraphics[scale=0.9]{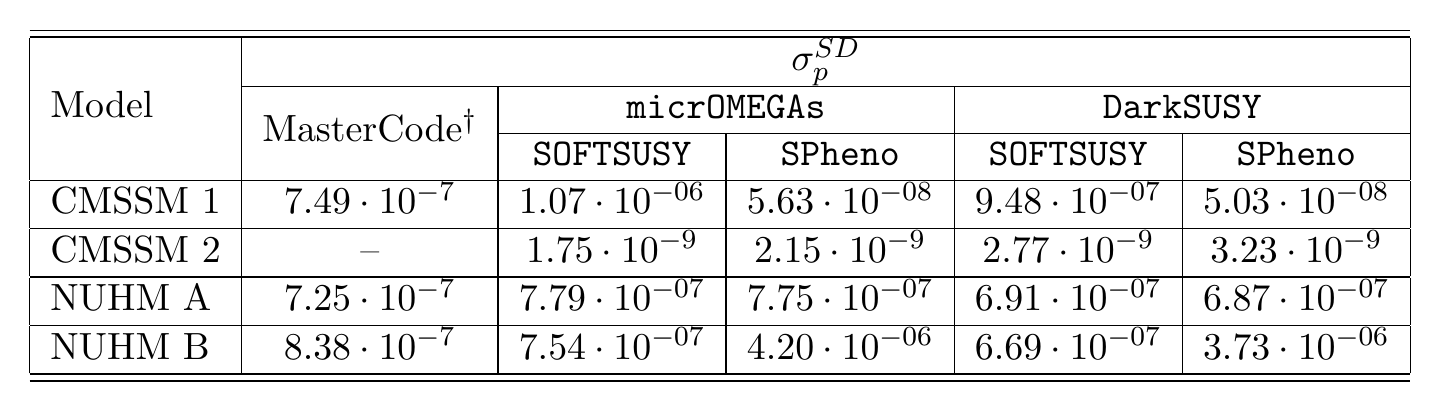}
  } \hfill
  \subfloat[][$\sigma^{SD}_n$ as computed in the \feynhiggs{} branch of the pipeline.\label{tab:mneutdepfeyn}]{%
      \includegraphics[scale=0.9]{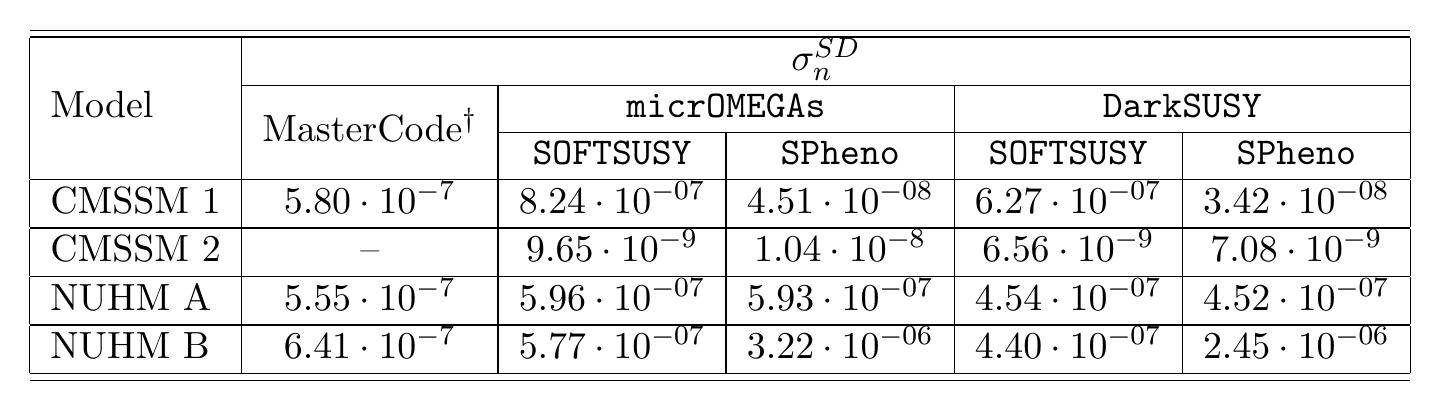}
  }
  \caption{\textbf{GUT model dark matter scattering cross section}: Spin independent neutralino-nucleon elastic scattering cross sections, in pb, as computed by the various \feynhiggs{} branches of the pipelines. The per nucleon average for Xe is plotted in the lower left panel of Figure~\ref{fig:msugra} using the \feynhiggs{} pipelines' values (Tables~\ref{tab:mprotindepfeyn}~\&~\ref{tab:mneutindepfeyn}).}
  \label{tab:msigmafeyn}
\end{table}

\begin{table}
  \centering
  \subfloat[][$\sigma^{SI}_p$ as computed in the \susyhd{} branch of the pipeline.\label{tab:mprotindephd}]{%
      \includegraphics[scale=0.9]{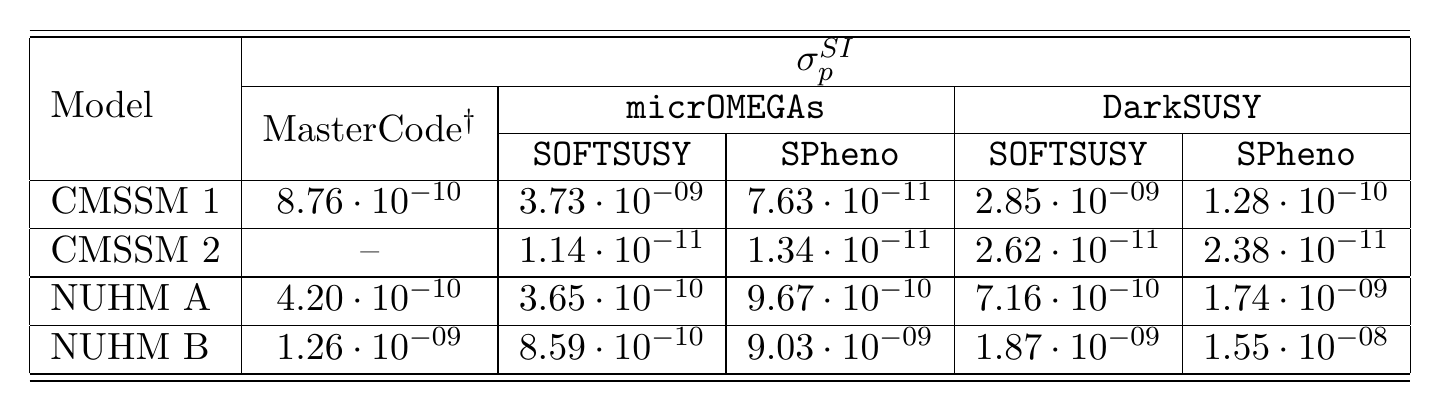}
  } \hfill
  \subfloat[][$\sigma^{SI}_n$ as computed in the \susyhd{} branch of the pipeline.\label{tab:mneutindephd}]{%
      \includegraphics[scale=0.9]{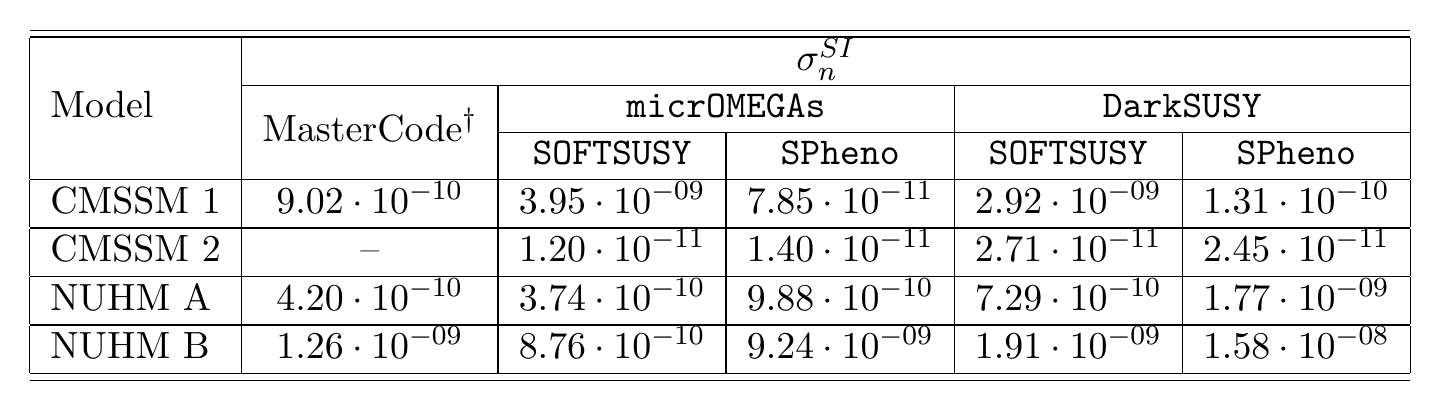}
  } \hfill
  \subfloat[][$\sigma^{SD}_n$ as computed in the \susyhd{} branch of the pipeline.\label{tab:mprotdephd}]{%
      \includegraphics[scale=0.9]{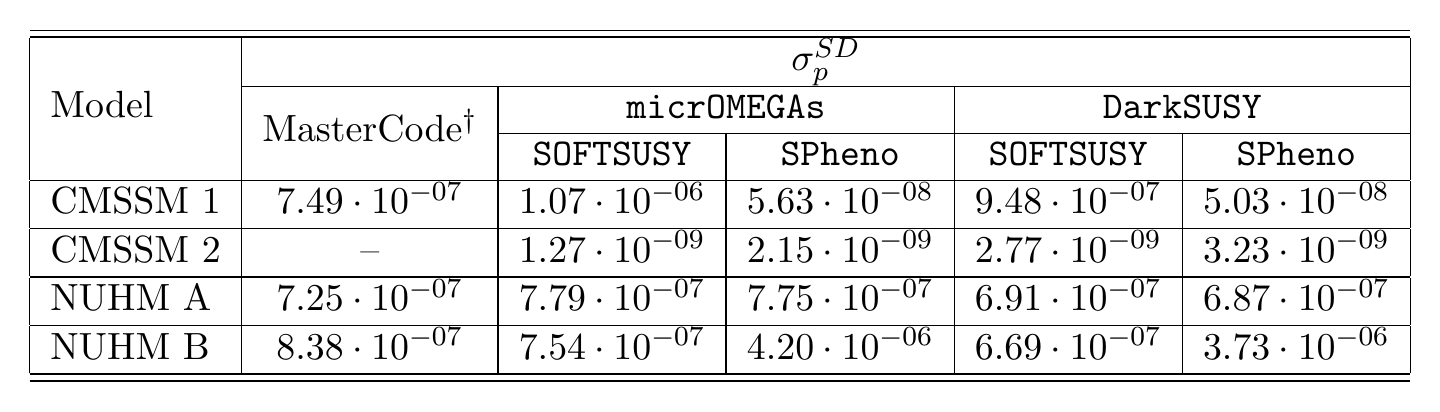}
  } \hfill
  \subfloat[][$\sigma^{SD}_n$ as computed in the \susyhd{} branch of the pipeline.\label{tab:mneutdephd}]{%
      \includegraphics[scale=0.9]{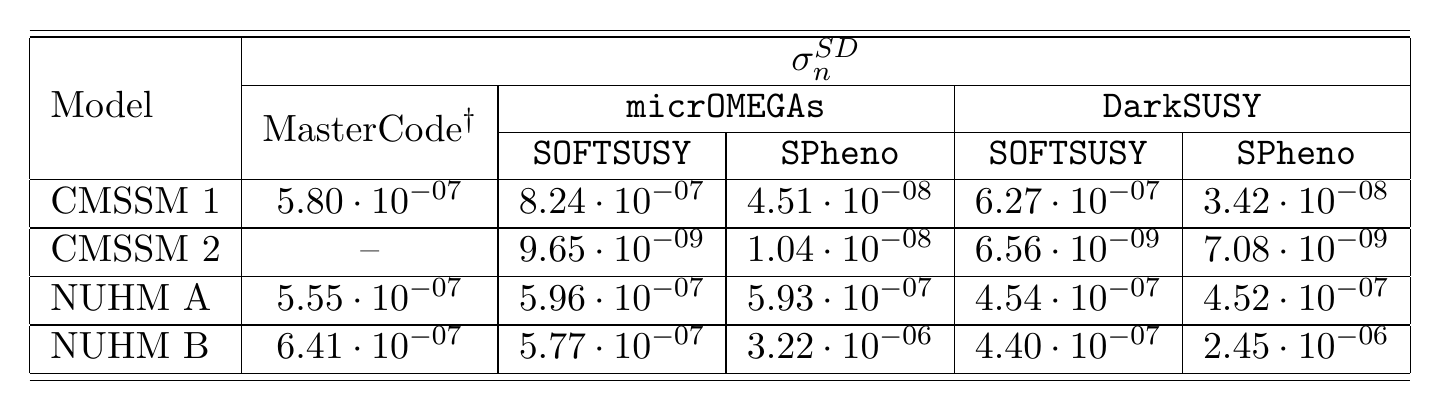}
  }
  \caption{\textbf{GUT model dark matter scattering cross section:} Spin independent neutralino-nucleon elastic scattering cross sections, in pb, as computed by the various \susyhd{} branches of the pipelines (except for the MasterCode$^\dag$ pipeline, which uses \feynhiggs{}.).}
  \label{tab:msigmahd}
\end{table}

\begin{table}
  \centering
    \subfloat[][CMSSM points in the neighborhood of the CMSSM 1 benchmark point (here denoted by the star) as computed via \softsusy{}-\feynhiggs{}.\label{tab:cmssmsfs}]{%
      \includegraphics[scale=0.9]{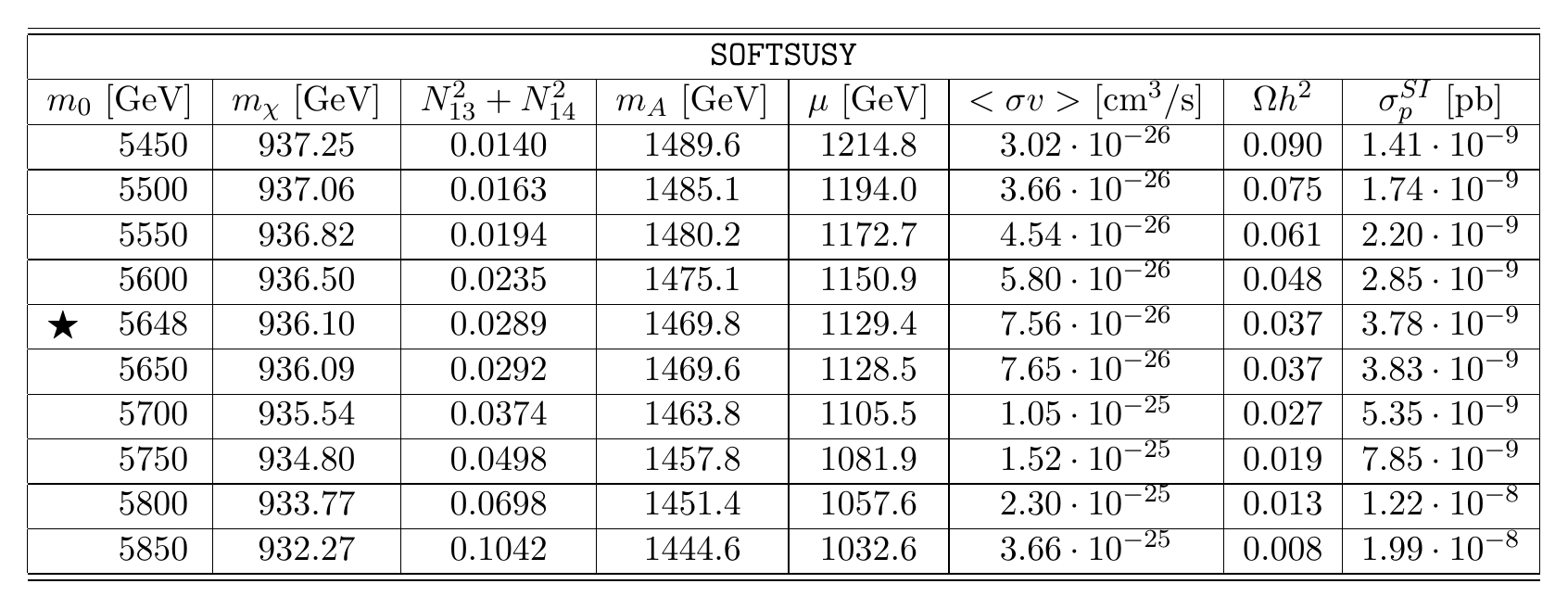}
    } \hfill
    \subfloat[][CMSSM points in the neighborhood of the CMSSM 1 benchmark point (here denoted by the star) as computed via \spheno{}-\feynhiggs{}.\label{tab:cmssmsph}]{%
      \includegraphics[scale=0.9]{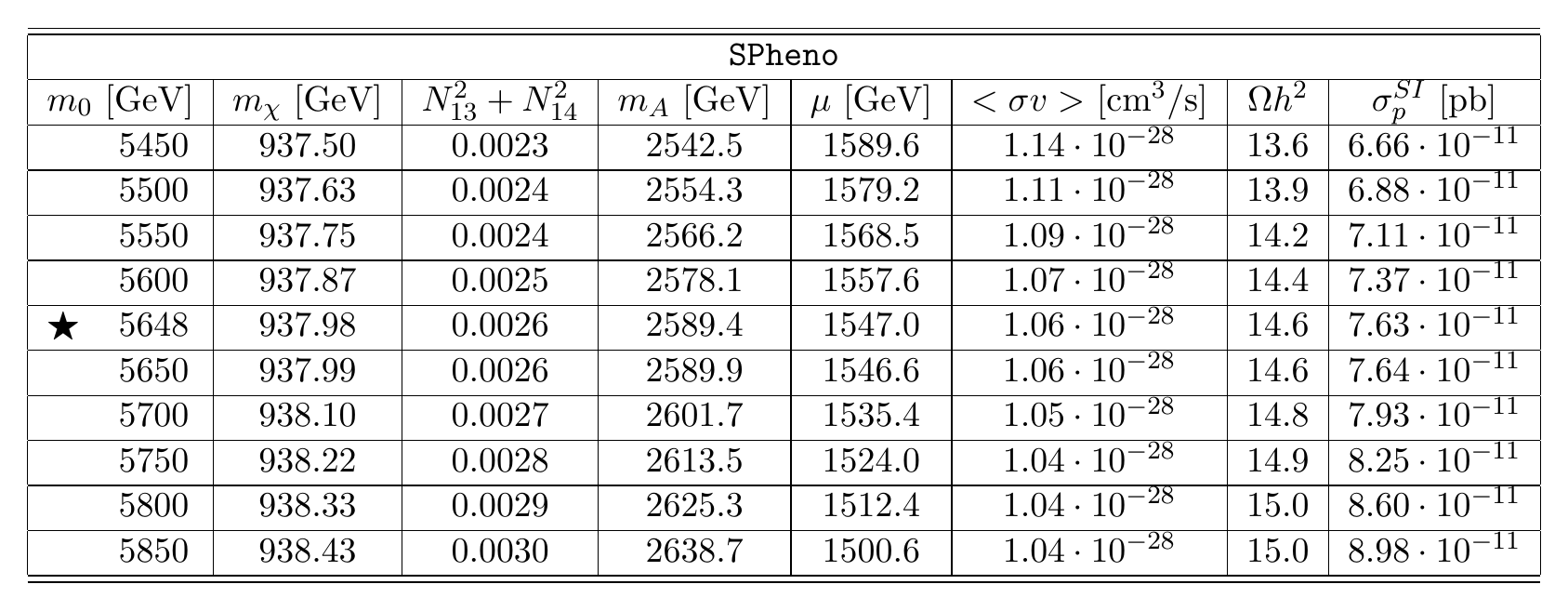}
    } \hfill
  \caption{\textbf{CMSSM EWSB sector:} The variation of $\mu$, $m_A$, and dark matter observables with universal scalar mass $m_0$ in the neighborhood of the CMSSM 1 benchmark point (denoted by the star), i.e.~$m_{1/2}=2098.41$~GeV, $A_0 =781.89$~GeV, $\tan(\beta)=51.28$, $\mu>0$, $m_t=173.30$~GeV. These results are also plotted in Figure~\ref{fig:cmssm_mu_m0}. The dark matter observables are computed by \micromegas{} for both Table~\ref{tab:cmssmsfs} and Table~\ref{tab:cmssmsph}. }
  \label{tab:cmssmscan}
\end{table}

\end{appendix}

\clearpage

\end{document}